\documentclass[10pt,twocolumn,amsmath,amsfonts,
amssymb,notitlepage,showpacs,pra,aps]{revtex4-2}
\usepackage{color}
\usepackage{tikz}
\usepackage{quantikz}
\usetikzlibrary{positioning}
\date{\today}

\usepackage{hyperref}
\usepackage[utf8x]{inputenc}
\usepackage{amsmath}
\usepackage{amsfonts}
\usepackage{amssymb}
\usepackage{graphicx}
\usepackage{braket}
\usepackage[normalem]{ulem}

\begin{document}
\title{Resonating Kagome Dimer coverings
in Rydberg atom arrays}
%: From the Thin Cylinder Limit to General Cases}
%in the thin cylinder limit and beyond}

%\author{Xicheng Wang}
%\email{xw727@cornell.edu}
%\affiliation{Zhili College, Tsinghua University,  Beijing 100084, China.}
%\affiliation{Department of Physics, Cornell University, Ithaca, New York 14853, USA}
%\author{Erich J Mueller}
%\email{em256@cornell.edu}
%\affiliation{Laboratory of Atomic and Solid State Physics, Cornell University, Ithaca, New York}

\author{Xicheng Wang}
\email{xw727@cornell.edu}
\affiliation{Zhili College, Tsinghua University, Beijing 100084, China}
\affiliation{Department of Physics, Cornell University, Ithaca, NY 14853, USA}

\author{Erich J. Mueller}
\email{em256@cornell.edu}
\affiliation{Laboratory of Atomic and Solid State Physics, Cornell University, Ithaca, NY 14853, USA}

\begin{abstract}
Motivated by experiments on Rydberg atom arrays, we explore the properties of  uniform quantum superpositions of kagome dimer configurations and construct an efficient algorithm for experimentally producing them.  We begin by considering the thin cylinder limit, where these states 
have simple descriptions.
%correspond to dimer crystals and topological valence bond solids{\color{red} Is it solid for XC4?}.  
We then develop a matrix product representation of the states on arbitrary cylinders, which leads to a natural protocol to efficiently grow them.  We  explain how our approach can be adapted to other quantum computing hardware.
%with particular attention to the thin cylinder limit.  For the thinnest cylinders these are resonating dimer crystals with no long range entanglement.  Larger cylinders produce topologically ordered spin liquids.  We describe a protocol of creating and studying these systems in Rydberg atom arrays.
\end{abstract}

\maketitle

\section{Introduction}\label{sec:intro}

Dimer models, where the quantum states are labeled by the locations of active bonds, are one of our best settings to explore the impact of constraints on many-body quantum systems \cite{Rokhsar1988,Sachdev1989,Misguich2002,moessner2008quantumdimermodels,Moessner2010}.  On the kagome lattice of corner sharing triangles,   the most natural dimer model gives rise to a topologically ordered $\mathbb{Z}_2$ spin liquid, which can be understood in terms of gauge theory \cite{PhysRevX.11.031005,Glaetzle2014,Samajdar2021,PhysRevX.10.021057,Giudici2022,Verresen2022,Cheng2023,Samajdar2023}.  Rydberg atom experiments have seen signatures of this topological order \cite{Semeghini2021,Satzinger2021}.  Boundary conditions matter here, and it is natural to impose periodic boundary conditions, rolling the lattice into a cylinder or a torus.  
%Motivated by thin torus studies of the quantum hall effect \cite{PhysRevB.28.1142,PhysRevB.50.17199,PhysRevB.76.155101,Bergholtz,PhysRevB.77.155308}, we first study the dimer configurations in the thin cylinder limit.  
We find convenient matrix product state representations of the Rokhsar-Kivelson state consisting of a uniform superposition of all dimer configurations \cite{Rokhsar1988} on cylinders.  For the thinnest cylinders this reduces to a resonating dimer crystal (a plaquette phase) with no long range entanglement.  Slightly larger cylinders produce an entangled  state with 
%symmetry protected topological order 
properties similar to
the Affleck-Kennedy-Lieb-Tasaki (AKLT) state from spin-1 chains \cite{AKLT}.  Larger cylinders correspond to topologically ordered spin liquids.
%  Slightly larger cylinders produce a topologically ordered spin liquid, with properties similar to the AKLT state from spin-1 chains \cite{AKLT}.  
%For larger cylinders 
%We further extend this construction to arbitrary cylinders.  
We develop a protocol for generating these resonating dimer states for arbitrary cylinders, or even tori.
%We explain how to experimentally produce these states in both the thin cylinder limit.  
%Additionally, we present a protocol for generating the Rokhsar Kivelson state for arbitrarily large cylinders.  
Our approach can be implemented using reconfigurable planar arrangements of atoms,
as the cylinder/torus topology need only be imposed in small patches where gates are being applied.  We also discuss implementation on other quantum computing hardware, such as transmon arrays.
  Our main state creation
algorithm takes a time which scales linearly with the length of the cylinder, but is independent of the circumference.    We also give an algorithm which scales with the circumference, but is independent of the length.
%and study their properties. 
%Inspired by the thin-cylinder approach, we develop a method to study cylinders of arbitrary width, namely the "module-connection formalism." Within this framework, we can systematically characterize the structure of dimer coverings and their topological features for cylinders of any width. Furthermore, we demonstrate how to prepare the corresponding Rokhsar-Kivelson state in a Rydberg atom system.

Resonating dimer states arise in a wide range of contexts, from the orbitals in organic molecules %:  Organic molecules, such as benzene are often described as having valence electrons occupying superpositions of different bonding orbitals
\cite{pauling}
%.  
%These states are also important for 
to models of magnets and superconductors \cite{sutherland,anderson}.  Typically there is a constraint that every site in the lattice touches exactly one dimer.  Thus these models can be mapped onto highly constrained spin systems, where there is a two-level system  located at the center of every bond in the lattice.  Exciting this spin corresponds to having a dimer on that bond.  The spin configurations are restricted to those in which one cannot   simultaneously excite more than one bond that
%on bonds which 
touches a given  lattice site.  Rydberg atom experiments explicitly implement this constrained spin system -- using the strong dipole-dipole interactions between the excited atoms to enforce the constraint \cite{Samajdar2021}.

The extensive set of constraints leads to rich physics, including topological order and fractionalized excitations \cite{Sachdev1989,Misguich2002,moessner2008quantumdimermodels,Moessner2010,fradkin}.  
This physics is exemplified by the Rokhsar-Kivelson state, $|\Psi\rangle$, consisting of a uniform quantum superposition of all valid dimer configurations \cite{Rokhsar1988}.  As argued by  Verresen et al.
\cite{PhysRevX.11.031005,Verresen2022}, and described in detail in Sec.~\ref{sec:kagome}, one can define two types of loop operators, and $|\Psi\rangle$ is an eigenstate of all such closed loops.  This property can be interpreted as a manifestation of a gauge symmetry.  Importantly, the gauge structure is a feature of the state itself, and one does not need to refer to a Hamiltonian or energetics in order to study this physics.  
%{\color{red} Should we mention the parent Hamiltonian of RK state? See \cite{Verresen2022}, it is a stabilizer Hamitonian. } 
Thus we are motivated to devise an experimental protocol to produce $|\Psi\rangle$ and measure its properties.  We emphasize that we are not concerned with finding the equilibrium ground state of any particular Hamiltonian, rather we are devising a dynamical process which creates the desired state. %{\color{red} Indeed it's hard to creat the many-body interaction required by that parent Hamitonian. Should we also mention that?} 
This is somewhat analogous to how a sequence of gates can produce interesting states in a quantum computer \cite{Song2019,PhysRevLett.132.130604,Pont2024,Cao2023,jiang2025}.
% constructing a Hamiltonian or concerning ourselves with energetics.  

Numerical calculations often work with a cylindrical geometry, with circumference $L_y$.  This is typically treated as a purely computational tool and it is common to attempt an extrapolation to the large cylinder limit $L_y\to\infty$.  It can also be useful to  take the the opposite tack, and explore the properties of $|\Psi\rangle$ in the limit of small $L$.  For example, studies of thin torus quantum Hall systems have given us enormous insight \cite{PhysRevB.28.1142,PhysRevB.50.17199,PhysRevB.76.155101,Bergholtz,PhysRevB.77.155308}.  
In this paper we consider both the small $L_y$ and large $L_y$ limits.  We gain intuition from studying small $L_y$ cylinders, before considering the arbitrary $L_y$ case.  
%Restricting to small $L$ simplifies the experimental implementation, and helps us build intuition.  For example,  for small $L$ there are planar configurations of atoms which are equivalent to arrangements on a cylinder.  
%Our results are not restricted to thin cylinders.  

We describe the properties of the Rokhsar-Kivelson state for arbitrary cylinders, and give a protocol for experimentally producing them. For small $L_y$ we construct planar arrangements of atoms whose connectivity is equivalent to that of a cylinder. 
Away from this limit, however, producing  cylindrical atomic configurations naively require a three-dimensional arrangement of the atoms.  We show how to circumvent this challenge, and study this physics with a purely planar geometry.  These experiments with different diameter cylinders can probe the connection between topological order in 2D and its 1D antecedents \cite{wenrmp}.    The extension to toroidal geometries is discussed in Appendix~\ref{sec:Torus}.

%{\color{red} Tristan: A question is that there is no 1D TO, so maybe we should refine this claim.} {\color{blue}  Erich:  I think its fine.  The topological order we are talking about in 1D is a SPT.  The symmetry operators are simply the strings that take us between the topological sectors.  I am happy to consi}

%{\color{red} Our second, and more important, innovation, is

%We give a protocol for experimentally studying the %thin cylinder 
%Rokhsar-Kivelson state in Rydberg atom experiments. 
%We are not concerned with energetics, phases, or phase diagrams.  As we argue, t
%This state has a number of exotic properties which we describe.  We explore how it can be used to connect 1D and 2D topological order, as the radius of the cylinder is varied.
%nontrivial properties which are worth exploring.  
%The thin cylinder limit is particularly appealing, as it allows one to connect 1D and 2D  topological order.

In our algorithm we start with a uniform system where all of the atoms are in their ground state.  We then perform a sequence of local gates 
%on one end of the cylinder which form a seed.  Further local gates  
which `grow' the Rokhsar-Kivelson state from one end of the cylinder to the other.  The gates in each annular strip can be performed in parallel, leading to a  state preparation time which scales with  the length of the cylinder but is independent of its width. 
  In Appendix~\ref{sec:Torus} we give an alternative grown algorithm which scales with the width of the cylinder, but is independent of its length.
Regardless, for a $L\times L$ arrangement of $N \sim L^2$ atoms, state preparation takes a time of order $\sqrt{N}$.
This scaling saturates a fundamental bound 
on the rate at 
 which entanglement can be created  through quantum gates \cite{bravyi2006lieb,bravyi2010topological}. 
 The vacuum state and the kagome lattice Rokhsar-Kivelson state can be identified as two different quantum states of matter \cite{zeng2019quantum,chen2010local}, and can only be transformed into one-another by a local circuit whose depth scales as the system's diameter \cite{bravyi2006lieb}.

 A number of works have explored the  idea of preparing states by sequentially applying local gates in ways which are analogous to our protocol. 
 Sch\"on and collaborators presented a generic approach for producing arbitrary matrix product states by using a set of ancilla degrees of freedom which sequentially interact with a single qubit \cite{schon,schonQED}.  
 Other authors generalized these ideas to producing a broader range of tensor network states \cite{wei2022sequential,banuls}.  Other work has characterized the limitations of such approaches \cite{lamata}, and explored their implementation \cite{zhang}.
Liu et al. constructed a protocol to produce string-net states, including the quantum states asociated with the toric code and the double semion model, by applying local unitary operations to rows of plaquettes
 \cite{liu2022methods}.
Kim et al. developed a strategy using quantum channels which only relies upon knowing local properties of the state \cite{kim2024learning}.
%{\color{blue} I have to point out there are key differences in Kim et al.'s work. They do not (only) use unitary gates. They use quantum channel, including non-unitary operations.}
 Chen et al. discussed general principles, and providing a number of additional examples \cite{chen2024sequential}.
Experiments on transmon arrays have used sequential gates to produce the state associated with the toric code \cite{satzinger2021realizing}.

It is also important to note that
there are other approaches to producing the Rokhsar-Kivelson state in  a Rydberg atom array.  Notably,
Giudici et al \cite{giudici2023}, explored a scheme in which one uniformly varies system parameters in a quasi-adiabatic manor.  

{
The remainder of the paper is structured as follows. In Sec.~\ref{sec:kagome} we describe the properties of dimer configurations on the kagome lattice, introducing the string operators and the nomenclature that we use to describe cylindrical arrangements.  Section~\ref{sec:thin} considers the thin-cylinder limit, while Sec.~\ref{sec:Larger_cylinder} constructs matrix product state representations of superpositions of dimer coverings on arbitrary cylinders.  We present our state creation algorithm in Sec.~\ref{sec:StateCreation}.  In Sec.~\ref{sec:probes} we discuss experimental probes, and we summarize in Sec.~\ref{sec:summary}.  Appendix~\ref{sec:proof_of_modules} 
through \ref{seedimp}
%and \ref{sec:Realization} 
give further details of our matrix product state construction, and the physical implementation of our algorithm.  
%Appendix~\ref{seed} describes how one seeds the growth process, and 
Appendix~\ref{sec:Torus} explains how to connect cylinders together.  This latter protocol enables the creation of toroidal geometries and can be used to implement an alternative approach to state preparation.}

{\color{purple}
}

%{\color{red}  Do we want to add something about creation algorithm being motivated by MPS structure?}

%Although our algorithm is novel, there are important connections to the ideas of Sch\"on et al. \cite{schon}, who constructed generic algorithms for sequentially building Matrix Product States, or generalizations  \cite{lamata,wei2022sequential}.  Related ideas have been used for building cluster states \cite{gross,ballester}, and for entangling ions using a moving cavity \cite{zhang}.  
%{\color{red} A recent paper arxiv.2410.23544 has a similar story of preparing states by growing. I met the author Daniel recently and chatted with him. Although we both agree there's significant difference in our papers, I think we should cite this paper and mention the similarity of 'growing'. See \cite{kim2024learning}}

\section{Dimer configurations on the kagome lattice}\label{sec:kagome}

\begin{figure}[tbh]
    %\begin{subfigure}[b]{0.22\textwidth}
     (a)
     \raisebox{5mm}{
     \begin{minipage}[t]{0.22\textwidth}  
     \vspace{0pt}
     \includegraphics[width=\textwidth]{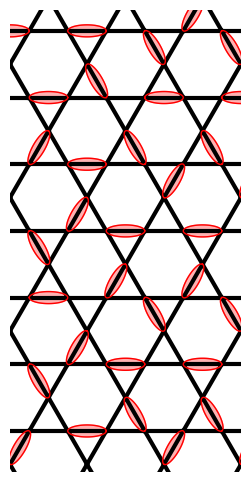}
        %\caption{}
        %\label{fig:Kagome_lattice_a}
    %\end{subfigure}
    \end{minipage}}
    %\hfill
    \begin{minipage}[t]{0.22\textwidth}
    \raggedright 
    \vspace{0pt}
%    \begin{subfigure}[b]{0.22\textwidth} 
%        \begin{minipage}[b]{\textwidth} 
 %           \begin{subfigure}[b]{\textwidth} 
 (b)\\
 \includegraphics[width=\textwidth]{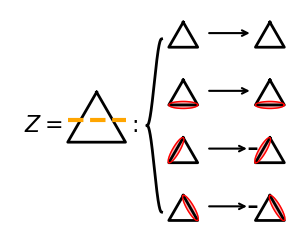}
 %               \caption{}
 %               \label{fig:Zstring_b}
 %           \end{subfigure}
%\vspace{-2cm} 
            (c)\\
 %           \begin{subfigure}[b]{\textwidth} 
                \includegraphics[width=\textwidth]{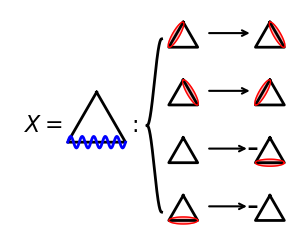}
 %               \caption{}
 %               \label{fig:Xstring_c}
 %           \end{subfigure}
        \end{minipage}
  %  \end{subfigure}
\caption{Kagome dimer covering and string operators. (a) A sample dimer covering on a kagome lattice. The kagome lattice consists of a honeycomb of corner-sharing triangles. Each red bond corresponds to a dimer, and each vertex is touched by exactly one dimer. (b) Illustration of the Z-string operator. If the Z-string passes through a dimer, it acts on the state by multiplying it by -1. (c) Illustration of the X-string operator. It shuffles dimer configurations.
%{\color{red} Z-Gate has a typo -- it should not move the bonds around.}
}\label{fig:kagome_lattice}
\end{figure}

As shown in Fig.~\ref{fig:kagome_lattice}(a), the kagome lattice consists of a honeycomb network of corner-sharing triangles. 
Dimers sit on the bonds, forming a ruby-lattice structure \cite{PhysRevX.11.031005}.  They obey the constraint that exactly one dimer is in contact with each site.  We take the dimer coverings to form an orthornormal basis.  The Rokhsar-Kivelson state consists of a linear superposition of all valid dimer coverings, possibly obeying some non-local constraints which define distinct topological sectors.

%This lattice exhibits a vast number of possible dimer coverings, where each kagome vertex touches exactly one dimer. Given a specific geometry for the space hosting the kagome lattice, a key focus is to systematically explore all possible dimer covering configurations, their interconnections, and their classification into distinct topological sectors.

%{\color{purple} Motivated by the $\mathbb{Z}_2$
% lattice gauge theory that describes the corresponding spin liquid, we introduce string operators to study dimer coverings.}  
 As argued in \cite{PhysRevX.11.031005, Semeghini2021,Samajdar2023,Tarabunga2022}, 
 This superposition of coverings has the structure of a $\mathbb{Z}_2$
 lattice gauge theory.  This property is best elucidated by considering the string operators discussed in those works and illustrated in Fig.~\ref{fig:kagome_lattice}(b) and  \ref{fig:kagome_lattice}(c).  
 These operators are both Unitary and Hermitian -- and hence can be viewed as ``gates" which act on states, or ``observables" which can be measured.
 A $Z$-string segment is   drawn as a dashed line which extends through the apex of a triangle (Fig.~\ref{fig:kagome_lattice}(b)).  An individual dimer covering is an eigenstate of this operator, with eigenvalue $(-1)^{s}$, where $s$ is the number of dimers it passes through.  A $X$-string segment is drawn as a squiggly line that extends between two neighboring sites on the lattice (Fig.~\ref{fig:kagome_lattice}(b)).  As illustrated, it rearranges dimers which touch those  two sites.

 %one can move between dimer configurations using the $X$-loop operator, as illustrated in Fig.~\ref{fig:Xstring_c}, and can count the parity of dimers using the $Z$-loop, as illustrated in Fig.~\ref{fig:Zstring_b}. We adopt the convention used in \cite{Semeghini2021}. 
 %{\color{blue}  Summarize the construction in Eqs 3 and 7 of \cite{PhysRevX.11.031005}.}

One can make closed loops out of $Z$-string  or $X$-string segments.  The smallest $Z$-loop encloses a single vertex, and any valid covering will be an eigenstate of such operators with eigenvalue $-1$.  In general a contractible $Z$-loop will enclose $n$ vertices, and the eigenvalue is $(-1)^n$.  
In the framework of \(\mathbb{Z}_2\) lattice gauge theory, the $Z$-loop is interpreted as measuring the electric flux through its surface -- yielding a value which only depends on the number of charges (vertices) that it encloses.
A closed $X$-loop converts one valid dimer configuration into another.  The Rokhsar-Kivelson state, which is a uniform superposition of all possible dimer coverings, is an eigenstate of all contractable loop operators.

If one wraps the kagome lattice onto a cylinder or torus {(see Appendix~\ref{sec:Torus})}, there will be non-contractable $Z$-loops and $X$-loops.  One can break the dimer configurations into different topological sectors, based upon if they are $+1$ or $-1$ eigenstates of the non-contractable $Z$-loops.  Perpendicular $X$-loops move one between these sectors.  This structure is elucidated by the examples in Sec.~\ref{sec:thin}. The Rokhsar-Kivelson state in a fixed topological sector is an eigenstate of contractable loop operators, but not necessarily the non-contractable loops.

%We will construct superpositions of dimer coverings, typically restricted to one of these topological sectors.

%the \(Z\) string is associated with the electric field \(E\) and satisfies the corresponding Gauss law. For any contractible closed \(Z\) loop, the measurement for any given dimer covering yields \( (-1)^{\# \text{vertices inside}} \). In contrast, non-contractible \(Z\) loops serve to distinguish different topological sectors. The \(X\) string corresponds to the gauge field, where each given \(X\)-loop pairs up dimer configurations according to a specific rule, allowing for transformations between them. A contractible \(X\)-loop induces transformations within the same topological sector, whereas a non-contractible \(X\)-loop corresponds to transformations between distinct topological sectors.
%Rokhsar-Kivelson state, which is a uniform superposition of all possible dimer covering states, is an eigenstate of all the loop operators.

\begin{figure}[tbh]

\begin{minipage}{0.5\columnwidth}
\includegraphics[width=\textwidth]{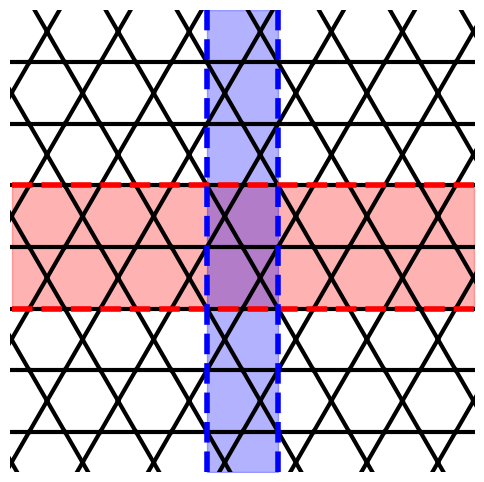}\\
{\centerline{(a)}}
\end{minipage}
\begin{minipage}{0.47\columnwidth}
\includegraphics[width=\textwidth]
{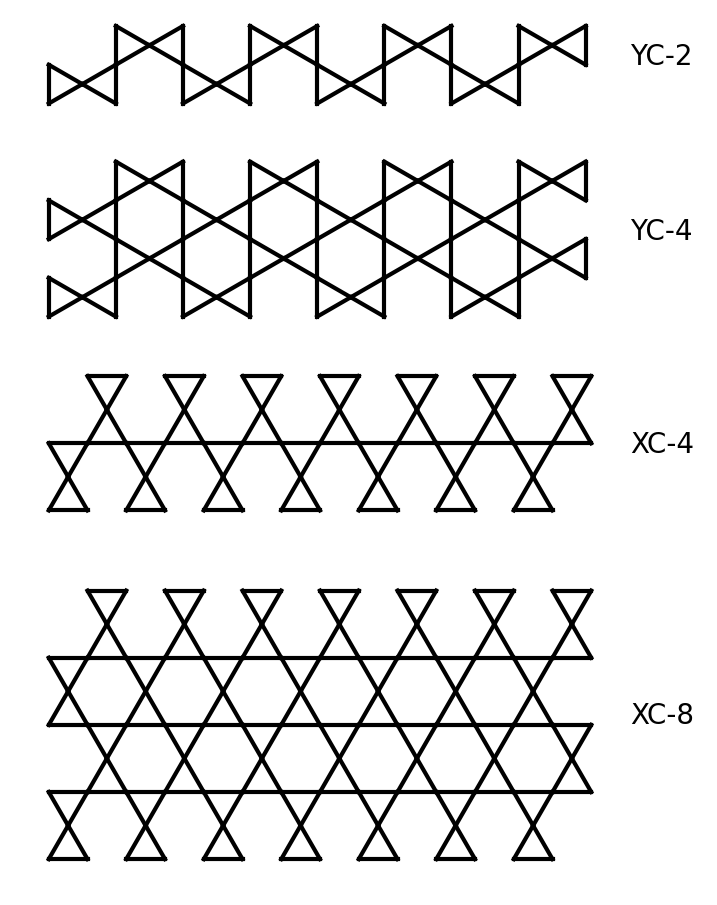}
{\centerline{(b)}}
\end{minipage}
    \caption{
     Constructing kagome lattice  cylinders. (a)  The blue vertical and red horizontal strips can be rolled into  YC-2 and XC-4 cylinders  \cite{yan2011spin}. (b)  Further examples:  the labeling XC-$2N$ or YC-$2N$,  specifies the orientation and the number of triangular rows along the circumference.  In each of these, the strip has been oriented so that periodic boundary conditions are applied in the vertical direction.}
    \label{fig:cylinder_structure}
\end{figure}

We will predominantly consider cylindrical geometries, where the lattice is infinite in one direction, and periodic in the other.
%As mentioned in Sec.~\ref{sec:intro}, it's common to embed the kagome lattice into a cylinder geometry, adding a periodic boundary condition in the horizontal or vertical direction.
Figure~\ref{fig:cylinder_structure} shows strips along high symmetry directions, which can be wrapped into cylinders by applying periodic boundary conditions in either the  $x$ or $y$ directions.  Follow the nomenclature from \cite{yan2011spin},
we denote the two configurations shown there as YC-$2N$ or XC-$2N$, where $2N$ counts the number of rows of triangles which appear in the periodic direction.

%to refer to different cylinder geometries. For example, in a YC-4 cylinder, the "Y" indicates that the cylinder is oriented along the Y-direction (vertical direction) in Fig.~\ref{fig:kagome_lattice_a}, while the periodic boundary condition is imposed in the horizontal direction. The number "4" denotes a circumference of 4 lattice spacings, and "C" simply stands for "cylinder."
%See Fig~\ref{fig:cylinder_structure} for further examples.

% {\color{red} 
% \begin{enumerate}
% \item Introduce X and Z loop operators
% \begin{itemize}
% \item Rokhsar-Kivelson state is eigenstate
% \item Topological sectors from non-contractable loops
% \end{itemize}
% \item Nomenclature for making finite cylinders
% \end{enumerate}}

\section{Thin Cylinder Limit}\label{sec:thin}

Here we analyze the limit of thin cylinders, which are particularly amenable to experimental study and provide key physical intuitions.  More general cases will be discussed in Sec.~\ref{sec:Larger_cylinder}.

\subsection{Eye Model}\label{sec:EyeModel}
\def\eye{\tikz[baseline=-0.6ex]{
  \coordinate (l) at (-1ex,0);
  \coordinate (t) at (0,1ex);
  \coordinate (b) at (0,-1ex);
  \coordinate (r) at (1ex,0);
  \draw (l) -- (t);
  \draw (t) -- (r);
  \draw (l) -- (b);
  \draw (b) -- (r);
  \draw (t) to[out=-45,in=45] (b);
  \draw (b) to[out=135,in=-135] (t);
}}

\begin{figure}[tbh]
 %   \centering
 %   \begin{subfigure}[b]{0.3\textwidth} % 子图 (a)
%\includegraphics[width=0.4\textwidth]{cylinder_construction.png}\\
%{\centerline{(a)}}
  %      \caption{}      %\label{fig:cylinder_construction}
    %\end{subfigure}
 %   \vspace{0.5cm} % 子图 (a) 和 (b) 之间的间距
%    \begin{subfigure}[b]{0.45\textwidth} % 子图 (b)
\includegraphics[width=0.5\textwidth]{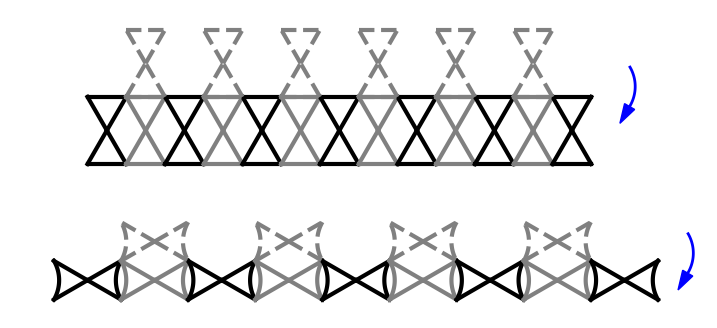}
%{\centerline{(b)}}
 %       \caption{}
  %      \label{fig:flipping}
  %  \end{subfigure}
    \caption{Utilizing periodic boundary conditions, the XC-4 cylinder (top) and YC-2 cylinder (bottom) can be transformed into planar structures, referred to in the text as the {\em hourglass model}  and {\em eye model}, due to the shapes of the unit cells.}
    \label{fig:Eye_Hourglass}
\end{figure}

The simplest case we can consider is the YC-2 cylinder, corresponding to the blue shaded area in Fig.~\ref{fig:cylinder_structure}.  Due to the periodic boundary conditions, the unit cell, consisting of 6 bonds, can be compactly expressed as planar eye-shaped symbol, \eye, as shown in Fig.~\ref{fig:Eye_Hourglass}.  In an experiment one would arrange the atoms in this planar shape to effectively realize a cylinder geometry.

%If one constructs dimer configurations, one quickly discovers that the allowed patterns are highly constrained.

\def\eyeA{\tikz[baseline=-0.6ex]{
  \coordinate (l) at (-1ex,0);
  \coordinate (t) at (0,1ex);
  \coordinate (b) at (0,-1ex);
  \coordinate (r) at (1ex,0);
  \draw (l) -- (t);
  \draw (t) -- (r);
  \draw (l) -- (b);
  \draw (b) -- (r);
  \draw (t) to[out=-45,in=45] (b);
  \draw (b) to[out=135,in=-135] (t);
  \draw [fill=red] (t) to[out=-45,in=45] (b) to[out=135,in=-135] (t);
}}

\def\eyeB{\tikz[baseline=-0.6ex]{
  \coordinate (l) at (-1ex,0);
  \coordinate (t) at (0,1ex);
  \coordinate (b) at (0,-1ex);
  \coordinate (r) at (1ex,0);
  \draw (l) -- (t);
  \draw (t) -- (r);
  \draw (l) -- (b);
  \draw (b) -- (r);
  \draw (t) to[out=-45,in=45] (b);
  \draw (b) to[out=135,in=-135] (t);
  \draw [fill=blue] (l)--(t) to[out=-135,in=135] (b) --(l);
  \draw [fill=blue] (r)--(t) to[out=-45,in=45] (b) --(r);
}}

\def\eyeAa{\tikz[baseline=-0.6ex]{
  \coordinate (l) at (-1ex,0);
  \coordinate (t) at (0,1ex);
  \coordinate (b) at (0,-1ex);
  \coordinate (r) at (1ex,0);
  \draw (l) -- (t);
  \draw (t) -- (r);
  \draw (l) -- (b);
  \draw (b) -- (r);
  \draw (t) to[out=-45,in=45] (b);
  \draw (b) to[out=135,in=-135] (t);
 \draw [line width=1.5] (t) to[out=-45,in=45] (b);
}}

\def\eyeAb{\tikz[baseline=-0.6ex]{
  \coordinate (l) at (-1ex,0);
  \coordinate (t) at (0,1ex);
  \coordinate (b) at (0,-1ex);
  \coordinate (r) at (1ex,0);
  \draw (l) -- (t);
  \draw (t) -- (r);
  \draw (l) -- (b);
  \draw (b) -- (r);
  \draw (t) to[out=-45,in=45] (b);
  \draw (b) to[out=135,in=-135] (t);
 \draw [line width=1.5] (b) to[out=135,in=-135] (t);
}}

\def\eyeBa{\tikz[baseline=-0.6ex]{
  \coordinate (l) at (-1ex,0);
  \coordinate (t) at (0,1ex);
  \coordinate (b) at (0,-1ex);
  \coordinate (r) at (1ex,0);
  \draw (l) -- (t);
  \draw (t) -- (r);
  \draw (l) -- (b);
  \draw (b) -- (r);
  \draw (t) to[out=-45,in=45] (b);
  \draw (b) to[out=135,in=-135] (t);
 \draw [line width=1.5] (l) -- (t);
 \draw [line width=1.5] (b) -- (r);
}}

\def\eyeBb{\tikz[baseline=-0.6ex]{
  \coordinate (l) at (-1ex,0);
  \coordinate (t) at (0,1ex);
  \coordinate (b) at (0,-1ex);
  \coordinate (r) at (1ex,0);
  \draw (l) -- (t);
  \draw (t) -- (r);
  \draw (l) -- (b);
  \draw (b) -- (r);
  \draw (t) to[out=-45,in=45] (b);
  \draw (b) to[out=135,in=-135] (t);
 \draw [line width=1.5] (l) -- (b);
 \draw [line width=1.5] (t) -- (r);
}}

Here there are two topologically inequivalent Rokhsar-Kivelson states, related by translation.  We
%ne of which can be expressed 
express one of these as
as a product state over the unit cells as
\begin{equation}\label{eyepsi}
|\psi\rangle=\cdots\eyeA\eyeB\eyeA\eyeB\cdots
\end{equation}
 The shaded symbols represent local resonating bonds:
\begin{align}\label{eyeplaquette}
\eyeA &=\frac{\eyeAa+\eyeAb}{\sqrt{2}}&
\eyeB =\frac{\eyeBa+\eyeBb}{\sqrt{2}}.
\end{align}
The dark lines represent dimers.
States of this form, with local resonating bonds, are often referred to as {\em plaquette} states.  By simply drawing out all possible dimer patterns, one can readily convince oneself that the patterns in Eq.~(\ref{eyeplaquette}) exhaust the possibilities on a single unit cell, given the constraint that every vertex is touched by exactly one dimer.

\begin{figure}[tbh]
    \centering
    \begin{minipage}{0.45\textwidth}
        \includegraphics[width=\textwidth]{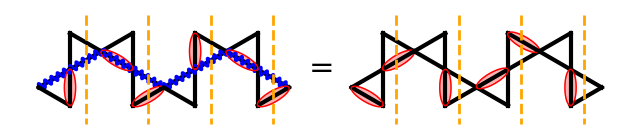}
        \centering
        (a)
        \label{fig:topoYC}
    \end{minipage}
    \hspace{0.05\textwidth} % 调整两张图片之间的水平间距
    \begin{minipage}{0.45\textwidth}
        \includegraphics[width=\textwidth]{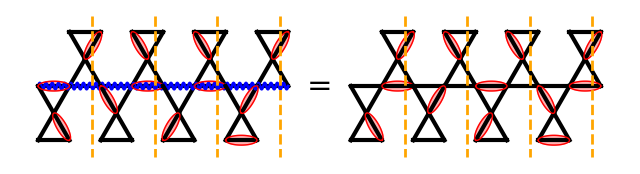}
        \centering
        (b)
        \label{fig:topoXC}
    \end{minipage}
    \caption{A schematic illustration of transitions between topological sectors in YC-2 and XC-4 geometries using a horizontal $X$-string.  
    (a) Sequential $Z$-loops acting on YC-2 geometry gives a $\{1,-1,1,-1...\}$ pattern on the left-hand side, and a $\{-1,1,-1,1...\}$ pattern on the right-hand side.  
    (b) Sequential $Z$-loops acting on XC-4 geometry gives a $\{-1,-1,-1,-1...\}$ pattern on the left-hand side, and a $\{1,1,1,1...\}$ pattern on the right-hand side.}
    \label{fig:topoXCYC}
\end{figure}

The configuration in Eq.~(\ref{eyepsi}) breaks translational symmetry, because the two cells are inequivalent.  
The second Rokhsar-Kivelson state is constructed by shifting the pattern by one unit cell.  These patterns are connected by the string operator shown in Fig.~\ref{fig:topoXCYC}. 
%More specifically, 
%a topologically nontrivial 
An
$X$-string segment oriented along the horizontal direction interchanges  \eyeA~and \eyeB~configurations, and hence an infinitely long horizontal $X$-string connects the patterns in the two distinct topological sectors.

As anticipated in Sec.~\ref{sec:kagome}, the different topological sectors can be distinguished by the properties of the non-contractible vertical $Z$-loops, in the circumferential direction.  Fig.~\ref{fig:topoXCYC} (a) shows that the eigenvalues of sequential $Z$ loops follow a pattern $\{1,-1,1,-1\ldots\}$.  Switching between these sectors shifts this to $\{-1,1-1,1\ldots\}$.  
%A more detailed discussion on the effect of the string operator in dimer coverings is deferred to Sec.~\ref{sec:Larger_cylinder}.

The symmetry breaking found here is somewhat reminiscent of the thin torus limit of the quantum Hall effect \cite{Rokhsar1988,Sachdev1989,Misguich2002,moessner2008quantumdimermodels,Moessner2010}.  There the topologically ordered two-dimensional state evolves into a charge density wave as the boundaries are squeezed together.  The wavefunction in Eq.~(\ref{eyepsi}) is analogous to that density wave.

%{\color{red}
For finite length cylinders we should also  consider how these structures can terminate.  %The pattern in Eq.~(\ref{eyepsi}) is consistent with 
We first consider
finite size systems which contain an integer number of eye-shaped unit cells.  Terminated in this way, the quantum states span a two-dimensional space, corresponding to the two topological sectors.  There are no other degrees of freedom

If we terminate the system in the middle of a unit cell, however, then we we must specify the configuration of the partial  unit cell.  The available Hilbert space will typically be spanned by two different 
%static (non-resonating) 
dimer configurations on that last partial cell, giving an extra spin-1/2 degree of freedom.
%}

\subsection{Hourglass Model}
\def\X{\tikz{
  \coordinate (lb) at (-1ex,-1ex);
  \coordinate (rb) at (1ex,-1ex);
  \coordinate (c) at (0,0);
  \coordinate (lt) at (-1ex,1ex);
  \coordinate (rt) at (1ex,1ex);
  \draw (lb) -- (rb);
  \draw (lb) -- (rt);
  \draw (lt) -- (rb);
  \draw (lt) -- (rt);
}}

\def\XX{\tikz{
  % 第一个结构
  \coordinate (lb1) at (-1ex,-1ex);
  \coordinate (rb1) at (1ex,-1ex);
  \coordinate (c1) at (0,0);
  \coordinate (lt1) at (-1ex,1ex);
  \coordinate (rt1) at (1ex,1ex);
  \draw (lb1) -- (rb1);
  \draw (lb1) -- (rt1);
  \draw (lt1) -- (rb1);
  \draw (lt1) -- (rt1);

  % 第二个结构，向右平移 2ex
  \coordinate (lb2) at ([xshift=2ex] -1ex,-1ex);
  \coordinate (rb2) at ([xshift=2ex] 1ex,-1ex);
  \coordinate (c2) at ([xshift=2ex] 0,0);
  \coordinate (lt2) at ([xshift=2ex] -1ex,1ex);
  \coordinate (rt2) at ([xshift=2ex] 1ex,1ex);
  \draw (lb2) -- (rb2);
  \draw (lb2) -- (rt2);
  \draw (lt2) -- (rb2);
  \draw (lt2) -- (rt2);
}}

\def\XABBB{\tikz{
  \coordinate (lb) at (-1ex,-1ex);
  \coordinate (rb) at (1ex,-1ex);
  \coordinate (c) at (0,0);
  \coordinate (lt) at (-1ex,1ex);
  \coordinate (rt) at (1ex,1ex);
  \draw (lb) -- (rb);
  \draw (lb) -- (rt);
  \draw (lt) -- (rb);
  \draw (lt) -- (rt);
  \draw [line width=2] (lt) -- (c);
}}

\def\XBBAB{\tikz{
  \coordinate (lb) at (-1ex,-1ex);
  \coordinate (rb) at (1ex,-1ex);
  \coordinate (c) at (0,0);
  \coordinate (lt) at (-1ex,1ex);
  \coordinate (rt) at (1ex,1ex);
  \draw (lb) -- (rb);
  \draw (lb) -- (rt);
  \draw (lt) -- (rb);
  \draw (lt) -- (rt);
  \draw [line width=2] (rt) -- (c);
}}

\def\XBBBA{\tikz{
  \coordinate (lb) at (-1ex,-1ex);
  \coordinate (rb) at (1ex,-1ex);
  \coordinate (c) at (0,0);
  \coordinate (lt) at (-1ex,1ex);
  \coordinate (rt) at (1ex,1ex);
  \draw (lb) -- (rb);
  \draw (lb) -- (rt);
  \draw (lt) -- (rb);
  \draw (lt) -- (rt);
  \draw [line width=2] (rb) -- (c);
}}

\def\XBABB{\tikz{
  \coordinate (lb) at (-1ex,-1ex);
  \coordinate (rb) at (1ex,-1ex);
  \coordinate (c) at (0,0);
  \coordinate (lt) at (-1ex,1ex);
  \coordinate (rt) at (1ex,1ex);
  \draw (lb) -- (rb);
  \draw (lb) -- (rt);
  \draw (lt) -- (rb);
  \draw (lt) -- (rt);
  \draw [line width=2] (lb) -- (c);
}}

\def\XABAA{\tikz{
  \coordinate (lb) at (-1ex,-1ex);
  \coordinate (rb) at (1ex,-1ex);
  \coordinate (c) at (0,0);
  \coordinate (lt) at (-1ex,1ex);
  \coordinate (rt) at (1ex,1ex);
  \draw (lb) -- (rb);
  \draw (lb) -- (rt);
  \draw (lt) -- (rb);
  \draw (lt) -- (rt);
  \draw [line width=2] (lt) -- (rt);
  \draw [line width=2] (rb) -- (c);
}}

\def\XAAAB{\tikz{
  \coordinate (lb) at (-1ex,-1ex);
  \coordinate (rb) at (1ex,-1ex);
  \coordinate (c) at (0,0);
  \coordinate (lt) at (-1ex,1ex);
  \coordinate (rt) at (1ex,1ex);
  \draw (lb) -- (rb);
  \draw (lb) -- (rt);
  \draw (lt) -- (rb);
  \draw (lt) -- (rt);
  \draw [line width=2] (lt) -- (rt);
  \draw [line width=2] (lb) -- (c);
}}

\def\XAABA{\tikz{
  \coordinate (lb) at (-1ex,-1ex);
  \coordinate (rb) at (1ex,-1ex);
  \coordinate (c) at (0,0);
  \coordinate (lt) at (-1ex,1ex);
  \coordinate (rt) at (1ex,1ex);
  \draw (lb) -- (rb);
  \draw (lb) -- (rt);
  \draw (lt) -- (rb);
  \draw (lt) -- (rt);
  \draw [line width=2] (lb) -- (rb);
  \draw [line width=2] (lt) -- (c);
}}

\def\XBAAA{\tikz{
  \coordinate (lb) at (-1ex,-1ex);
  \coordinate (rb) at (1ex,-1ex);
  \coordinate (c) at (0,0);
  \coordinate (lt) at (-1ex,1ex);
  \coordinate (rt) at (1ex,1ex);
  \draw (lb) -- (rb);
  \draw (lb) -- (rt);
  \draw (lt) -- (rb);
  \draw (lt) -- (rt);
  \draw [line width=2] (lb) -- (rb);
  \draw [line width=2] (rt) -- (c);
}}

\def\XUP{\tikz{
  \coordinate (lb) at (-1ex,-1ex);
  \coordinate (rb) at (1ex,-1ex);
  \coordinate (c) at (0,0);
  \coordinate (lt) at (-1ex,1ex);
  \coordinate (rt) at (1ex,1ex);
  \draw (lb) -- (rb);
  \draw (lb) -- (rt);
  \draw (lt) -- (rb);
  \draw (lt) -- (rt);
  \draw [line width=2] (lt) -- (rt);
}}

\def\XDO{\tikz{
  \coordinate (lb) at (-1ex,-1ex);
  \coordinate (rb) at (1ex,-1ex);
  \coordinate (c) at (0,0);
  \coordinate (lt) at (-1ex,1ex);
  \coordinate (rt) at (1ex,1ex);
  \draw (lb) -- (rb);
  \draw (lb) -- (rt);
  \draw (lt) -- (rb);
  \draw (lt) -- (rt);
  \draw [line width=2] (lb) -- (rb);
}}

\def\ISigA{\tikz{
  \coordinate (lb) at (-1ex,-1ex);
  \coordinate (rb) at (1ex,-1ex);
  \coordinate (c) at (0,0);
  \coordinate (lt) at (-1ex,1ex);
  \coordinate (rt) at (1ex,1ex);
  \draw (lb) -- (rb);
  \draw (rb) -- (c);
  \draw (rt) -- (c);
  \draw (lt) -- (rt);
  \draw [line width=2] (lt) -- (rt);
}}

\def\ISigB{\tikz{
  \coordinate (lb) at (-1ex,-1ex);
  \coordinate (rb) at (1ex,-1ex);
  \coordinate (c) at (0,0);
  \coordinate (lt) at (-1ex,1ex);
  \coordinate (rt) at (1ex,1ex);
  \draw (lb) -- (rb);
  \draw (rb) -- (c);
  \draw (rt) -- (c);
  \draw (lt) -- (rt);
  \draw [line width=2] (rt) -- (c);
}}

\def\ISigC{\tikz{
  \coordinate (lb) at (-1ex,-1ex);
  \coordinate (rb) at (1ex,-1ex);
  \coordinate (c) at (0,0);
  \coordinate (lt) at (-1ex,1ex);
  \coordinate (rt) at (1ex,1ex);
  \draw (lb) -- (rb);
  \draw (rb) -- (c);
  \draw (rt) -- (c);
  \draw (lt) -- (rt);
  \draw [line width=2] (rb) -- (c);
}}

\def\ISigD{\tikz{
  \coordinate (lb) at (-1ex,-1ex);
  \coordinate (rb) at (1ex,-1ex);
  \coordinate (c) at (0,0);
  \coordinate (lt) at (-1ex,1ex);
  \coordinate (rt) at (1ex,1ex);
  \draw (lb) -- (rb);
  \draw (rb) -- (c);
  \draw (rt) -- (c);
  \draw (lt) -- (rt);
  \draw [line width=2] (lb) -- (rb);
}}

\def\ISigE{\tikz{
  \coordinate (lb) at (-1ex,-1ex);
  \coordinate (rb) at (1ex,-1ex);
  \coordinate (c) at (0,0);
  \coordinate (lt) at (-1ex,1ex);
  \coordinate (rt) at (1ex,1ex);
  \draw (lb) -- (rb);
  \draw (rb) -- (c);
  \draw (rt) -- (c);
  \draw (lt) -- (rt);
  %\draw [line width=2] (lb) -- (rb);
}}

\def\ISigF{\tikz{
  \coordinate (lb) at (-1ex,-1ex);
  \coordinate (rb) at (1ex,-1ex);
  \coordinate (c) at (0,0);
  \coordinate (lt) at (-1ex,1ex);
  \coordinate (rt) at (1ex,1ex);
  \draw (lb) -- (rb);
  \draw (rb) -- (c);
  \draw (rt) -- (c);
  \draw (lt) -- (rt);
  \draw [line width=2] (rt) -- (c);
  \draw [line width=2] (lb) -- (rb);
}}

\def\ISigG{\tikz{
  \coordinate (lb) at (-1ex,-1ex);
  \coordinate (rb) at (1ex,-1ex);
  \coordinate (c) at (0,0);
  \coordinate (lt) at (-1ex,1ex);
  \coordinate (rt) at (1ex,1ex);
  \draw (lb) -- (rb);
  \draw (rb) -- (c);
  \draw (rt) -- (c);
  \draw (lt) -- (rt);
  \draw [line width=2] (lt) -- (rt);
  \draw [line width=2] (rb) -- (c);
}}

\def\SigA{\tikz{
  \coordinate (lb) at (-1ex,-1ex);
  \coordinate (rb) at (1ex,-1ex);
  \coordinate (c) at (0,0);
  \coordinate (lt) at (-1ex,1ex);
  \coordinate (rt) at (1ex,1ex);
  \draw (lb) -- (rb);
  \draw (lb) -- (c);
  \draw (lt) -- (c);
  \draw (lt) -- (rt);
}}

\def\SigB{\tikz{
  \coordinate (lb) at (-1ex,-1ex);
  \coordinate (rb) at (1ex,-1ex);
  \coordinate (c) at (0,0);
  \coordinate (lt) at (-1ex,1ex);
  \coordinate (rt) at (1ex,1ex);
  \draw (lb) -- (rb);
  \draw (lb) -- (c);
  \draw (lt) -- (c);
  \draw (lt) -- (rt);
  \draw [line width=2] (lt) -- (rt);
}}

\def\SigC{\tikz{
  \coordinate (lb) at (-1ex,-1ex);
  \coordinate (rb) at (1ex,-1ex);
  \coordinate (c) at (0,0);
  \coordinate (lt) at (-1ex,1ex);
  \coordinate (rt) at (1ex,1ex);
  \draw (lb) -- (rb);
  \draw (lb) -- (c);
  \draw (lt) -- (c);
  \draw (lt) -- (rt);
  \draw [line width=2] (lt) -- (c);
}}

\def\SigD{\tikz{
  \coordinate (lb) at (-1ex,-1ex);
  \coordinate (rb) at (1ex,-1ex);
  \coordinate (c) at (0,0);
  \coordinate (lt) at (-1ex,1ex);
  \coordinate (rt) at (1ex,1ex);
  \draw (lb) -- (rb);
  \draw (lb) -- (c);
  \draw (lt) -- (c);
  \draw (lt) -- (rt);
  \draw [line width=2] (lb) -- (c);
}}

\def\SigE{\tikz{
  \coordinate (lb) at (-1ex,-1ex);
  \coordinate (rb) at (1ex,-1ex);
  \coordinate (c) at (0,0);
  \coordinate (lt) at (-1ex,1ex);
  \coordinate (rt) at (1ex,1ex);
  \draw (lb) -- (rb);
  \draw (lb) -- (c);
  \draw (lt) -- (c);
  \draw (lt) -- (rt);
  \draw [line width=2] (lb) -- (rb);
}}

\def\SigF{\tikz{
  \coordinate (lb) at (-1ex,-1ex);
  \coordinate (rb) at (1ex,-1ex);
  \coordinate (c) at (0,0);
  \coordinate (lt) at (-1ex,1ex);
  \coordinate (rt) at (1ex,1ex);
  \draw (lb) -- (rb);
  \draw (lb) -- (c);
  \draw (lt) -- (c);
  \draw (lt) -- (rt);
  \draw [line width=2] (lt) -- (rt);
  \draw [line width=2] (lb) -- (c);
}}

\def\SigG{\tikz{
  \coordinate (lb) at (-1ex,-1ex);
  \coordinate (rb) at (1ex,-1ex);
  \coordinate (c) at (0,0);
  \coordinate (lt) at (-1ex,1ex);
  \coordinate (rt) at (1ex,1ex);
  \draw (lb) -- (rb);
  \draw (lb) -- (c);
  \draw (lt) -- (c);
  \draw (lt) -- (rt);
  \draw [line width=2] (lb) -- (rb);
  \draw [line width=2] (lt) -- (c);
}}

\def\cupA{\tikz{
  \coordinate (lb) at (-1ex,-1ex);
  \coordinate (rb) at (1ex,-1ex);
  \coordinate (c) at (0,0);
  \coordinate (lt) at (-1ex,1ex);
  \coordinate (rt) at (1ex,1ex);
  \draw (rt) -- (c);
  \draw (rb) -- (c);
}}

\def\bcupA{\tikz{
  \coordinate (lb) at (-1ex,-1ex);
  \coordinate (rb) at (1ex,-1ex);
  \coordinate (c) at (0,0);
  \coordinate (lt) at (-1ex,1ex);
  \coordinate (rt) at (1ex,1ex);
  \draw (lt) -- (c);
  \draw (lb) -- (c);
}}

\def\cupB{\tikz{
  \coordinate (lb) at (-1ex,-1ex);
  \coordinate (rb) at (1ex,-1ex);
  \coordinate (c) at (0,0);
  \coordinate (lt) at (-1ex,1ex);
  \coordinate (rt) at (1ex,1ex);
  \draw (rt) -- (c);
  \draw (rb) -- (c);
  \draw [line width=2] (rt) -- (c);
}}

\def\bcupB{\tikz{
  \coordinate (lb) at (-1ex,-1ex);
  \coordinate (rb) at (1ex,-1ex);
  \coordinate (c) at (0,0);
  \coordinate (lt) at (-1ex,1ex);
  \coordinate (rt) at (1ex,1ex);
  \draw (lt) -- (c);
  \draw (lb) -- (c);
  \draw [line width=2] (lt) -- (c);
}}

\def\cupC{\tikz{
  \coordinate (lb) at (-1ex,-1ex);
  \coordinate (rb) at (1ex,-1ex);
  \coordinate (c) at (0,0);
  \coordinate (lt) at (-1ex,1ex);
  \coordinate (rt) at (1ex,1ex);
  \draw (rt) -- (c);
  \draw (rb) -- (c);
  \draw [line width=2] (rb) -- (c);
}}

\def\bcupC{\tikz{
  \coordinate (lb) at (-1ex,-1ex);
  \coordinate (rb) at (1ex,-1ex);
  \coordinate (c) at (0,0);
  \coordinate (lt) at (-1ex,1ex);
  \coordinate (rt) at (1ex,1ex);
  \draw (lt) -- (c);
  \draw (lb) -- (c);
  \draw [line width=2] (lb) -- (c);
}}

% 修正后的 \XAAABX = \XAAAB (左) + \X (右)
\def\XAAABX{\tikz{
  % 左侧的 \XAAAB（原点位置，不平移）
  \begin{scope}
    \coordinate (lb) at (-1ex,-1ex);
    \coordinate (rb) at (1ex,-1ex);
    \coordinate (c) at (0,0);
    \coordinate (lt) at (-1ex,1ex);
    \coordinate (rt) at (1ex,1ex);
    \draw (lb) -- (rb);
    \draw (lb) -- (rt);
    \draw (lt) -- (rb);
    \draw (lt) -- (rt);
    \draw [line width=2] (lt) -- (rt);  % 顶部横线加粗
    \draw [line width=2] (lb) -- (c);   % 左下到中心加粗
  \end{scope}
  
  % 右侧的 \X（向右平移 2ex，左边界对齐原点的右边界）
  \begin{scope}[shift={(2ex, 0)}]
    \coordinate (lb) at (-1ex,-1ex);
    \coordinate (rb) at (1ex,-1ex);
    \coordinate (c) at (0,0);
    \coordinate (lt) at (-1ex,1ex);
    \coordinate (rt) at (1ex,1ex);
    \draw (lb) -- (rb);
    \draw (lb) -- (rt);
    \draw (lt) -- (rb);
    \draw (lt) -- (rt);
  \end{scope}
}}

% 修正后的 \XAABAX = \XAABA (左) + \X (右)
\def\XAABAX{\tikz{
  % 左侧的 \XAABA
  \begin{scope}
    \coordinate (lb) at (-1ex,-1ex);
    \coordinate (rb) at (1ex,-1ex);
    \coordinate (c) at (0,0);
    \coordinate (lt) at (-1ex,1ex);
    \coordinate (rt) at (1ex,1ex);
    \draw (lb) -- (rb);
    \draw (lb) -- (rt);
    \draw (lt) -- (rb);
    \draw (lt) -- (rt);
    \draw [line width=2] (lb) -- (rb);  % 底部横线加粗
    \draw [line width=2] (lt) -- (c);   % 左上到中心加粗
  \end{scope}
  
  % 右侧的 \X
  \begin{scope}[shift={(2ex, 0)}]
    \coordinate (lb) at (-1ex,-1ex);
    \coordinate (rb) at (1ex,-1ex);
    \coordinate (c) at (0,0);
    \coordinate (lt) at (-1ex,1ex);
    \coordinate (rt) at (1ex,1ex);
    \draw (lb) -- (rb);
    \draw (lb) -- (rt);
    \draw (lt) -- (rb);
    \draw (lt) -- (rt);
  \end{scope}
}}

% 修正后的 \XBBABX = \XBBAB (左) + \X (右)
\def\XBBABX{\tikz{
  % 左侧的 \XBBAB
  \begin{scope}
    \coordinate (lb) at (-1ex,-1ex);
    \coordinate (rb) at (1ex,-1ex);
    \coordinate (c) at (0,0);
    \coordinate (lt) at (-1ex,1ex);
    \coordinate (rt) at (1ex,1ex);
    \draw (lb) -- (rb);
    \draw (lb) -- (rt);
    \draw (lt) -- (rb);
    \draw (lt) -- (rt);
    \draw [line width=2] (rt) -- (c);   % 右上到中心加粗
  \end{scope}
  
  % 右侧的 \X
  \begin{scope}[shift={(2ex, 0)}]
    \coordinate (lb) at (-1ex,-1ex);
    \coordinate (rb) at (1ex,-1ex);
    \coordinate (c) at (0,0);
    \coordinate (lt) at (-1ex,1ex);
    \coordinate (rt) at (1ex,1ex);
    \draw (lb) -- (rb);
    \draw (lb) -- (rt);
    \draw (lt) -- (rb);
    \draw (lt) -- (rt);
  \end{scope}
}}

% 修正后的 \XBBBAX = \XBBBA (左) + \X (右)
\def\XBBBAX{\tikz{
  % 左侧的 \XBBBA
  \begin{scope}
    \coordinate (lb) at (-1ex,-1ex);
    \coordinate (rb) at (1ex,-1ex);
    \coordinate (c) at (0,0);
    \coordinate (lt) at (-1ex,1ex);
    \coordinate (rt) at (1ex,1ex);
    \draw (lb) -- (rb);
    \draw (lb) -- (rt);
    \draw (lt) -- (rb);
    \draw (lt) -- (rt);
    \draw [line width=2] (rb) -- (c);   % 右下到中心加粗
  \end{scope}
  
  % 右侧的 \X
  \begin{scope}[shift={(2ex, 0)}]
    \coordinate (lb) at (-1ex,-1ex);
    \coordinate (rb) at (1ex,-1ex);
    \coordinate (c) at (0,0);
    \coordinate (lt) at (-1ex,1ex);
    \coordinate (rt) at (1ex,1ex);
    \draw (lb) -- (rb);
    \draw (lb) -- (rt);
    \draw (lt) -- (rb);
    \draw (lt) -- (rt);
  \end{scope}
}}

% 定义 \XAAABXBABB = \XAAAB + \XBABB 并排
\def\XAAABXBABB{\tikz{
  % 左侧的 \XAAAB
  \begin{scope}
    \coordinate (lb) at (-1ex,-1ex);
    \coordinate (rb) at (1ex,-1ex);
    \coordinate (c) at (0,0);
    \coordinate (lt) at (-1ex,1ex);
    \coordinate (rt) at (1ex,1ex);
    \draw (lb) -- (rb);
    \draw (lb) -- (rt);
    \draw (lt) -- (rb);
    \draw (lt) -- (rt);
    \draw [line width=2] (lt) -- (rt);  % 顶部横线加粗
    \draw [line width=2] (lb) -- (c);    % 左下到中心加粗
  \end{scope}
  
  % 右侧的 \XBABB (向右平移 2ex)
  \begin{scope}[shift={(2ex, 0)}]
    \coordinate (lb) at (-1ex,-1ex);
    \coordinate (rb) at (1ex,-1ex);
    \coordinate (c) at (0,0);
    \coordinate (lt) at (-1ex,1ex);
    \coordinate (rt) at (1ex,1ex);
    \draw (lb) -- (rb);
    \draw (lb) -- (rt);
    \draw (lt) -- (rb);
    \draw (lt) -- (rt);
    \draw [line width=2] (lb) -- (c);    % 左下到中心加粗
  \end{scope}
}}

% 定义 \XAAABXDO = \XAAAB + \XDO 并排
\def\XAAABXDO{\tikz{
  % 左侧的 \XAAAB
  \begin{scope}
    \coordinate (lb) at (-1ex,-1ex);
    \coordinate (rb) at (1ex,-1ex);
    \coordinate (c) at (0,0);
    \coordinate (lt) at (-1ex,1ex);
    \coordinate (rt) at (1ex,1ex);
    \draw (lb) -- (rb);
    \draw (lb) -- (rt);
    \draw (lt) -- (rb);
    \draw (lt) -- (rt);
    \draw [line width=2] (lt) -- (rt);
    \draw [line width=2] (lb) -- (c);
  \end{scope}
  
  % 右侧的 \XDO (向右平移 2ex)
  \begin{scope}[shift={(2ex, 0)}]
    \coordinate (lb) at (-1ex,-1ex);
    \coordinate (rb) at (1ex,-1ex);
    \coordinate (c) at (0,0);
    \coordinate (lt) at (-1ex,1ex);
    \coordinate (rt) at (1ex,1ex);
    \draw (lb) -- (rb);
    \draw (lb) -- (rt);
    \draw (lt) -- (rb);
    \draw (lt) -- (rt);
    \draw [line width=2] (lb) -- (rb);  % 底部横线加粗
  \end{scope}
}}

% 定义 \XBBABXBABB = \XBBAB + \XBABB 并排
\def\XBBABXBABB{\tikz{
  % 左侧的 \XBBAB
  \begin{scope}
    \coordinate (lb) at (-1ex,-1ex);
    \coordinate (rb) at (1ex,-1ex);
    \coordinate (c) at (0,0);
    \coordinate (lt) at (-1ex,1ex);
    \coordinate (rt) at (1ex,1ex);
    \draw (lb) -- (rb);
    \draw (lb) -- (rt);
    \draw (lt) -- (rb);
    \draw (lt) -- (rt);
    \draw [line width=2] (rt) -- (c);   % 右上到中心加粗
  \end{scope}
  
  % 右侧的 \XBABB (向右平移 2ex)
  \begin{scope}[shift={(2ex, 0)}]
    \coordinate (lb) at (-1ex,-1ex);
    \coordinate (rb) at (1ex,-1ex);
    \coordinate (c) at (0,0);
    \coordinate (lt) at (-1ex,1ex);
    \coordinate (rt) at (1ex,1ex);
    \draw (lb) -- (rb);
    \draw (lb) -- (rt);
    \draw (lt) -- (rb);
    \draw (lt) -- (rt);
    \draw [line width=2] (lb) -- (c);    % 左下到中心加粗
  \end{scope}
}}

% 定义 \XBBABXDO = \XBBAB + \XDO 并排
\def\XBBABXDO{\tikz{
  % 左侧的 \XBBAB
  \begin{scope}
    \coordinate (lb) at (-1ex,-1ex);
    \coordinate (rb) at (1ex,-1ex);
    \coordinate (c) at (0,0);
    \coordinate (lt) at (-1ex,1ex);
    \coordinate (rt) at (1ex,1ex);
    \draw (lb) -- (rb);
    \draw (lb) -- (rt);
    \draw (lt) -- (rb);
    \draw (lt) -- (rt);
    \draw [line width=2] (rt) -- (c);   % 右上到中心加粗
  \end{scope}
  
  % 右侧的 \XDO (向右平移 2ex)
  \begin{scope}[shift={(2ex, 0)}]
    \coordinate (lb) at (-1ex,-1ex);
    \coordinate (rb) at (1ex,-1ex);
    \coordinate (c) at (0,0);
    \coordinate (lt) at (-1ex,1ex);
    \coordinate (rt) at (1ex,1ex);
    \draw (lb) -- (rb);
    \draw (lb) -- (rt);
    \draw (lt) -- (rb);
    \draw (lt) -- (rt);
    \draw [line width=2] (lb) -- (rb);  % 底部横线加粗
  \end{scope}
}}

% 定义 \XAABAXABBB = \XAABA (左) + \XABBB (右)
\def\XAABAXABBB{\tikz{
  % 左侧的 \XAABA
  \begin{scope}
    \coordinate (lb) at (-1ex,-1ex);
    \coordinate (rb) at (1ex,-1ex);
    \coordinate (c) at (0,0);
    \coordinate (lt) at (-1ex,1ex);
    \coordinate (rt) at (1ex,1ex);
    \draw (lb) -- (rb);
    \draw (lb) -- (rt);
    \draw (lt) -- (rb);
    \draw (lt) -- (rt);
    \draw [line width=2] (lb) -- (rb);  % 底部横线加粗
    \draw [line width=2] (lt) -- (c);   % 左上到中心加粗
  \end{scope}
  
  % 右侧的 \XABBB (向右平移 2ex)
  \begin{scope}[shift={(2ex, 0)}]
    \coordinate (lb) at (-1ex,-1ex);
    \coordinate (rb) at (1ex,-1ex);
    \coordinate (c) at (0,0);
    \coordinate (lt) at (-1ex,1ex);
    \coordinate (rt) at (1ex,1ex);
    \draw (lb) -- (rb);
    \draw (lb) -- (rt);
    \draw (lt) -- (rb);
    \draw (lt) -- (rt);
    \draw [line width=2] (lt) -- (c);    % 左下到中心加粗
  \end{scope}
}}

% 定义 \XAABAXUP = \XAABA (左) + \XUP (右)
\def\XAABAXUP{\tikz{
  % 左侧的 \XAABA
  \begin{scope}
    \coordinate (lb) at (-1ex,-1ex);
    \coordinate (rb) at (1ex,-1ex);
    \coordinate (c) at (0,0);
    \coordinate (lt) at (-1ex,1ex);
    \coordinate (rt) at (1ex,1ex);
    \draw (lb) -- (rb);
    \draw (lb) -- (rt);
    \draw (lt) -- (rb);
    \draw (lt) -- (rt);
    \draw [line width=2] (lb) -- (rb);  % 底部横线加粗
    \draw [line width=2] (lt) -- (c);   % 左上到中心加粗
  \end{scope}
  
  % 右侧的 \XUP (向右平移 2ex)
  \begin{scope}[shift={(2ex, 0)}]
    \coordinate (lb) at (-1ex,-1ex);
    \coordinate (rb) at (1ex,-1ex);
    \coordinate (c) at (0,0);
    \coordinate (lt) at (-1ex,1ex);
    \coordinate (rt) at (1ex,1ex);
    \draw (lb) -- (rb);
    \draw (lb) -- (rt);
    \draw (lt) -- (rb);
    \draw (lt) -- (rt);
    \draw [line width=2] (lt) -- (rt);  % 顶部横线加粗
  \end{scope}
}}

% 定义 \XBBBAXABBB = \XBBBA (左) + \XABBB (右)
\def\XBBBAXABBB{\tikz{
  % 左侧的 \XBBBA
  \begin{scope}
    \coordinate (lb) at (-1ex,-1ex);
    \coordinate (rb) at (1ex,-1ex);
    \coordinate (c) at (0,0);
    \coordinate (lt) at (-1ex,1ex);
    \coordinate (rt) at (1ex,1ex);
    \draw (lb) -- (rb);
    \draw (lb) -- (rt);
    \draw (lt) -- (rb);
    \draw (lt) -- (rt);
    \draw [line width=2] (rb) -- (c);   % 右下到中心加粗
  \end{scope}
  
  % 右侧的 \XABBB (向右平移 2ex)
  \begin{scope}[shift={(2ex, 0)}]
    \coordinate (lb) at (-1ex,-1ex);
    \coordinate (rb) at (1ex,-1ex);
    \coordinate (c) at (0,0);
    \coordinate (lt) at (-1ex,1ex);
    \coordinate (rt) at (1ex,1ex);
    \draw (lb) -- (rb);
    \draw (lb) -- (rt);
    \draw (lt) -- (rb);
    \draw (lt) -- (rt);
    \draw [line width=2] (lt) -- (c);    % 左下到中心加粗
  \end{scope}
}}

% 定义 \XBBBAXUP = \XBBBA (左) + \XUP (右)
\def\XBBBAXUP{\tikz{
  % 左侧的 \XBBBA
  \begin{scope}
    \coordinate (lb) at (-1ex,-1ex);
    \coordinate (rb) at (1ex,-1ex);
    \coordinate (c) at (0,0);
    \coordinate (lt) at (-1ex,1ex);
    \coordinate (rt) at (1ex,1ex);
    \draw (lb) -- (rb);
    \draw (lb) -- (rt);
    \draw (lt) -- (rb);
    \draw (lt) -- (rt);
    \draw [line width=2] (rb) -- (c);   % 右下到中心加粗
  \end{scope}
  
  % 右侧的 \XUP (向右平移 2ex)
  \begin{scope}[shift={(2ex, 0)}]
    \coordinate (lb) at (-1ex,-1ex);
    \coordinate (rb) at (1ex,-1ex);
    \coordinate (c) at (0,0);
    \coordinate (lt) at (-1ex,1ex);
    \coordinate (rt) at (1ex,1ex);
    \draw (lb) -- (rb);
    \draw (lb) -- (rt);
    \draw (lt) -- (rb);
    \draw (lt) -- (rt);
    \draw [line width=2] (lt) -- (rt);  % 顶部横线加粗
  \end{scope}
}}

The next simple cylinder is the XC-4 cylinder, corresponding to the red shaded area in Fig.~\ref{fig:cylinder_structure}, along with its  planar representation in Fig.~\ref{fig:Eye_Hourglass}, consisting of a repeating hourglass pattern of 6 bonds, \X.  Again, the different topological sectors correspond to period 2 symmetry broken states, distinguished by the parity of $Z$-strings that wrap along the short axis of the cylinder. As illustrated in Fig.~\ref{fig:topoXCYC}(b), these two topological sectors are connected by a horizontal $X$-string.

Unlike the eye model, however, $|\Psi\rangle$ is not a product of resonating plaquettes.
Instead the Rokhsar-Kivelson state in the hourglass model is represented as a bond-dimension 2 matrix product state (MPS),
\begin{equation}\label{hourmps}
|\Psi\rangle=\cdots
\left(
\begin{array}{cc}
\XABAA&\XABBB\\
\XBAAA&\XBABB
\end{array}
\right)
\left(
\begin{array}{cc}
\XBBBA&\XBBAB\\
\XAABA&\XAAAB
\end{array}
\right)\cdots
\end{equation}
or its translation by one unit cell.  Multiplying out the matrices gives a sum of dimer configurations.  All possible configurations appear in this sum: We have exhausted the valid arrangements in each unit cell, and all allowed connections between them.  When discussing wider cylinders we will find it convenient to double the unit cell.

To make a connection to
 the AKLT state we 
 map the plaquette configurations onto pairs of spins -- using a sublattice dependent mapping.
% can show that $|\Psi\rangle$  can be considered to have a `symmetry protected topologically order', where the symmetry is related to the action of $X$ loops on the diamond shaped plaquettes between two hourglasses.   
 On a given sublattice one only encounters four configurations.  On the first sublattice we define
\begin{align}\nonumber
\uparrow\uparrow&=\XABAA&
\uparrow\downarrow&=-\XABBB\\
\downarrow\uparrow&=\XBAAA&
\downarrow\downarrow&=-\XBABB.
\end{align}
On the second sublattice we instead define
\begin{align}
\uparrow\uparrow&=\XAAAB&
\uparrow\downarrow&=-\XAABA\\
\downarrow\uparrow&=\XBBAB&
\downarrow\downarrow&=-\XBBBA.
\end{align}
The state in Eq~(\ref{hourmps}) can then be represented as a product state, where the second spin in each pair forms a singlet with the first spin in the next pair.  For example, a chain of unit cells might be represented as
$|\Psi\rangle=\uparrow(\uparrow\downarrow-\downarrow\uparrow)(\uparrow\downarrow-\downarrow\uparrow)\uparrow$.  The entanglement is hidden by the fact that the transformation from spins to bonds is non-local.

%$$|\Psi\rangle={\color{red}\uparrow}(
%{\color{red}\uparrow}{\color{blue}\downarrow}-{\color{red}\downarrow}{\color{blue}\uparrow})({\color{blue}\uparrow}\downarrow-{\color{blue}\downarrow}\uparrow)\uparrow$$

This mapping  illustrates two other important features of $|\Psi\rangle$.  First, when it is cut in two, between two unit cells, it has an entanglement entropy of $\ln 2$.  Second, a finite length chain will naturally possess effectively spin-1/2 edge modes -- corresponding to the fact that there are two natural terminations for any dimer covering on a finite length chain.  These edge modes are in addition to the global degrees of freedom corresponding to the topological sectors.

%{\color{red} TO DO: Add figure which shows tree structure of MPS to this section, and explain how it works.}

%For our state creation algorithm, it will also be useful to think about partial unit cells, decomposing the dimer patterns as
%\begin{equation}
%\ket{\psi} = 
%\left(
%\begin{array}{cc}
%\SigF&\SigG
%\end{array}
%\right)
%\left(
% \begin{array}{cc}
% \cupA&\\
% &\cupA
% \end{array}
% \right)
% \left(
% \begin{array}{ccc}
% \SigD&\SigE&\\
% \SigC&&\SigB
% \end{array}
% \right)
%\end{equation}

%{\color{red}  Breaking up the unit cell -- used for the state creation stuff}
% \begin{align}
% \left(
% \begin{array}{cc}
% \XABAA&\XABBB\\
% \XBAAA&\XBABB
% \end{array}
% \right)&=
% \left(
% \begin{array}{ccc}
% \SigB&&\SigC\\
% &\SigE&\SigD
% \end{array}
% \right)
% \left(
% \begin{array}{cc}
% \cupC\\
% \cupB\\
% &\cupA
% \end{array}
% \right)\\
% \left(
% \begin{array}{cc}
% \XBBBA&\XBBAB\\
% \XAABA&\XAAAB
% \end{array}
% \right)&=
% \left(
% \begin{array}{ccc}
% \SigA\\
% &\SigF&\SigG\\
% \end{array}
% \right)
% \left(
% \begin{array}{cc}
% \cupC&\cupB\\
% \cupA\\
% &\cupA\\
% \end{array}
% \right)
% \end{align}

\section{Larger Cylinders}\label{sec:Larger_cylinder}

In contrast to the thin-cylinder limit, dimer coverings on larger cylinders cannot be embedded in a plane, naively necessitating a three-dimensional arrangement in experiments. Nonetheless, as we argue in Sec.~\ref{sec:StateCreation}, if we can move the sites around during the state creation process (or perform gates on qubits which are sufficiently far apart), we can construct these resonating dimer states through dynamical planar geometries.

Here we generalize the constructions of Sec.~\ref{sec:thin} by breaking our cylinder into annular strips, which are analogous to the unit cells of the eye or hourglass models. For a cylinder of arbitrary width, we find that there exists a systematic method to describe the dimer coverings, and show that the Rokhsar-Kivelson state can accordingly be written as a matrix product state. Some details are relegated to Appendix~\ref{sec:proof_of_modules}.

\begin{figure}[tbh]
\includegraphics[width=0.45\textwidth]{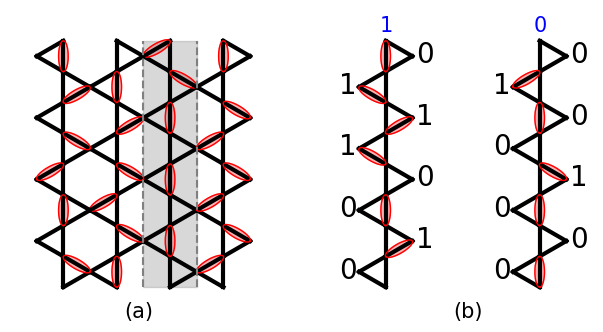}
    \caption{Dimer coverings on the YC-8 cylinder.
    %The module and its naming convention, taking  YC-8 geometry as an example: 
(a) One annular strip is highlighted. 
%The diagram shows how the cylinder can be decomposed into connections of modules.
(b) Each strip is labeled  by 9 numbers, corresponding to which external vertices have dimers touching them.  The two diagrams correspond to $(L,R,u)=$
%The diagram provides an illustration of the naming method. The left and right modules are labeled as 
(1100,0101,1) and (1000,0010,0), respectively.}
    \label{fig:module_conn_1}
\end{figure}

We begin our discussion with the YC-2$N$ case (see Fig.~\ref{fig:module_conn_1}).
%for the YC-8 example, illustrating the general construction.
%To characterize dimer covering configurations on a cylinder, we divide the problem into two parts:
%
%\vspace{1em}  
%\noindent
%{\bf (1) Valid configurations on a single annular strip.}
%\noindent
%
  Each annular strip contains $2N$ external vertices.
   Alternate strips have the first external vertex pointing to the left or to the right, related by a mirror symmetry.
   Each strip has
   $2^{2N}$ allowed dimer configurations, where each internal vertex is touched by a single dimer, and where no vertex is touched by more than one dimer. These configurations can be conveniently labeled by assigning a binary digit to each leftward- and rightward-facing triangle: $1$ if the external vertex is touched by a dimer, and $0$ otherwise. Equivalently, these binary digits correspond to the eigenvalues of $Z$-strings passing through the noses of the triangles.   We denote the resulting binary %digits on the left and right sides of the strip as $L_j$ and $R_j$, with $j=1,2,\cdots N$ indexed from the top to bottom.  The string of digits are 
  strings as
  $L$ and $R$. We also label the top-most vertex with $u=1$ if it is touched by a dimer from below, and $u=0$ if it instead is touched by a dimer connecting through the opposite end of the strip due to periodic boundary conditions. The dimer configuration is uniquely specified by $(L,R,u)$. As shown in Appendix~\ref{sec:proof_of_modules}, the parity of $L$ (\emph{i.e.} sum of the digits modulo 2) must equal the parity of $R$, and we refer to this constraint as the \emph{parity condition}. 

  We denote the quantum state of a strip as $A_{L_jR_j}^{u_j}$, and set it to zero if the parity of $L_j$ and $R_j$ does not match.
%\vspace{1em}  
%\noindent
%{\bf(2) Connecting neighboring strips into a cylinder.}
%\noindent
  The dimer configurations on neighboring strips are constrained by the requirement that  exactly one dimer touches every site of the lattice. We write this condition as $\bar{L}_{j+1} = R_j$, which defines 
   $\bar{L}$ as the bitwise complement of $L$ ({\em i.e.} 1's and 0's are exchanged). This constraint is referred to as the \emph{connection condition}.
  %The \emph{connection condition} requires that t
  %For any valid dimer configuration, 
  %the left boundary of one strip is the complement of the right boundary of the preceding strip: $\bar{L}_{j+1} = R_j$.

%Putting everything together, 
Given these constraints, one can write 
the equal-weight superposition of all valid dimer configurations
%—the RK state—can be written 
as a matrix product state:
\begin{equation}\label{mpsY}
|\Psi\rangle = \sum_{\{S_j\}, u_j} A_{\bar S_1 S_2}^{u_1} A_{\bar S_2 S_3}^{u_2} A_{\bar S_3 S_4}^{u_3} \cdots,
\end{equation}
where $S_j = (\sigma_j^1, \sigma_j^2, \dots, \sigma_j^N)$ is a binary string of length $N$, and $u_j \in \{0, 1\}$.  

\begin{figure}[tbh]
    \centering
    \begin{minipage}{0.45\textwidth}
        \includegraphics[width=\textwidth]{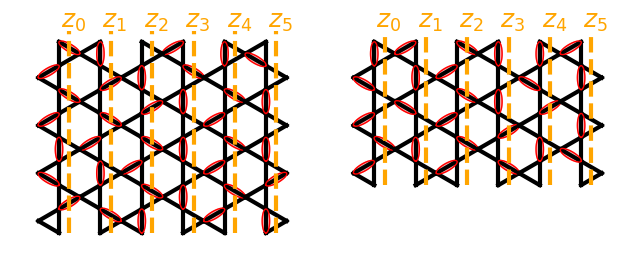}
        \centering 
        (a)
        \label{fig:toposector}
    \end{minipage}
    \hspace{0.05\textwidth}
    \begin{minipage}{0.45\textwidth}
        \includegraphics[width=\textwidth]{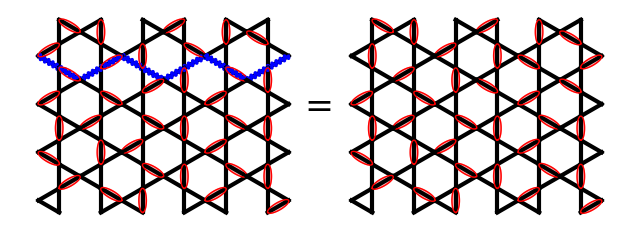} 
        \centering 
        (b)
        \label{fig:topotrans}
    \end{minipage}
    \caption{Topological Sectors and String Operators on YC Geometries.  
    (a) Schematic representation of the topological sectors for YC-8 (left) and YC-6 (right) cylinders. The dimer configurations are eigenstates of the non-contractable Z-loops, with eigenvalues $\mathcal{Z} = \{z_0, z_1, z_2, z_3, z_4, z_5\}$. 
    On the left, $\mathcal{Z}= \{-1,-1,-1,-1,-1\}$, while on the right, $\mathcal{Z} = \{-1,1,-1,1,-1,1\}$ .   
    (b) As illustrated, a horizontal $X$-string  connects different topological sectors, reversing the parity of each Z-string.
    }
    \label{fig:toposec}
\end{figure}

This Rokhsar-Kivelson state exhibits two topological sectors, distinguished by the parity of the leftmost binary string, denoted by $\pi(S_1)$. This initial parity determines the parity of all subsequent strings via the recurrence relation
%\begin{equation}
 $   \pi(S_j) = (-1)^N \pi(S_{j-1}),$
%end{equation}
where $N$ is the width of the YC-$2N$ cylinder.
%To see this, we can fix the parity of the first binary array $X_1$, thereby restricting the system to a specific topological sector. Under this constraint, all nonzero terms in the sum correspond to configurations where each $X_j$ has a fixed parity. If $N$ is even, then all $X_j$ share the same parity as $X_1$; if $N$ is odd, the parity of $X_j$ alternates periodically: for odd $j$, $X_j$ has the same parity as $X_1$, while for even $j$, it has the opposite parity. This structure follows directly from the {parity condition} and {connection condition}.
As illustrated in Fig.~\ref{fig:toposec}, %this parity 
$\pi(S_j)$
corresponds to the eigenvalue of a $Z$-string operator, and it can be flipped by acting with a horizontal $X$-string. When restricted to a single topological sector, the wavefunction in Eq.~(\ref{mpsY}) has bond dimension $2^{N-1}$ and exhibits an entanglement entropy of $S = (N-1)\ln 2$ when bipartitioned between any two annular strips.  At left edge there are $N$ spin-1/2 degrees for freedom, corresponding to the choice of $\bar S_1$.  At the right edge there are nominally another $N$ degrees of freedom, however, due to the constraints on the parity, one of these degrees of freedom is redundant.

%We can restrict ourselves to a particular topological sector by  specifying the parity of the first string, $X_1$ -- in which case $X_j$ has a fixed parity for all the nonzero terms in the sum. As illustrated in Fig.~\ref{fig:toposec}, this parity corresponds to the eigenvalue of a Z-string operator, and it can be switched by acting with a horizontal X-string.   We can also consider other terminations, where $X_1$ is fully specified, which gives a $N-1$ dimensional boundary Hilbert space, in each topological sector.  When restricted to a single sector, the wavefunction in Eq.~(\ref{mpsY}) has bond dimension $N-1$ and, when cut between any two annular strips, has entanglement entropy of $S=(N-1)\ln 2$.

\begin{figure}[tbh]
\includegraphics[width=0.45\textwidth]{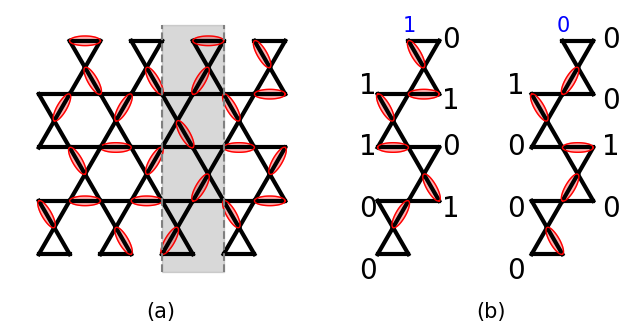}
    \caption{Dimer coverings on the XC-8 cylinder.
    %The module and its naming convention, taking  YC-8 geometry as an example: 
(a) One annular strip is highlighted. (b )Each strip is labeled by 9 bits, indicating which external vertices are touched by dimers. The two diagrams correspond to $(L, R, u) = (1100, 1010, 1)$ and $(1000, 0100, 0)$, respectively. 
Note that when reading the binary array $R$, we start from the second vertex at the top and proceed downward, appending the topmost bit at the end due to periodic boundary conditions.
 }
    \label{fig:module_conn_3}
\end{figure}

We can produce a similar construction with XC-2$N$ structures, making a small change in the naming convention. 
%The naming convention is slightly different from that of the YC-2$N$ case. 
%For YC-2$N$, both the left and right binary arrays, denoted as $L$ and $R$, are read from top to bottom. In contrast, f
For XC-2$N$, as with YC-2$N$, the left binary array $L$ is read from top to bottom, but we define the right array $R$ so that it begins with the second element from the top and proceeds downward, with the topmost binary digit appended at the end due to the periodic boundary conditions (see Fig.~\ref{fig:module_conn_3}).
This naming convention allows us to again have a simple  \emph{connection condition} for neighboring strips, namely $\bar{L}_{j+1} = R_j$, as before. As with the YC-$2N$ structures, $L$ and $R$ must have the same parity, and hence the wavefunction can also be written as Eq.~\eqref{mpsY}.

\section{State Creation}\label{sec:StateCreation}

 Here we describe our central result, namely an approach to creating resonating dimer states in an system of Rydberg atoms trapped in an array of microtraps, or in other quantum computing platforms.  We formulate our procedure in terms of a sequence of {\em gates} which are implemented by moving the microtraps around, and sweeping various fields which can be controlled in the experiment.

 %{\color{red}  This approach is motivated by the structure illustrated in Fig...  }

 Each atom can be in one of two energy levels: $|0\rangle$ and $|1\rangle$.  There are strong dipole-dipole interactions between the atoms in the $|1\rangle$ states, which represent the excited Rydberg atom.  Atoms in their ground state, $|0\rangle$, do not have an appreciable interaction.  The atoms can be driven by a spatially dependent laser which couples the two states.  Up to irrelevant additive constants, the system can be described by a Hamiltonian \cite{Samajdar2021,Semeghini2021},
% {\color{red} Do we want to change any of the factors of 2?}
\begin{equation}\label{ham}
    H = \sum_\alpha \frac{\Omega_\alpha(t)}{2}\sigma^x_\alpha -\sum_\alpha \Delta_\alpha(t) n_\alpha + \sum_{\langle \alpha,\beta \rangle} V_{\alpha\beta}\,n_\alpha n_\beta.
\end{equation}
Here $\Delta_\alpha$ is the detuning of the atom labeled by $\alpha$.  It can be controlled via a spatially dependent magnetic field.  The coupling $\Omega_\alpha$ is proportional to the square of the laser intensity at the atom's location, and $V_{\alpha\beta}$ encodes the interaction between atoms in the excited states.  We have introduced operators $\sigma^x=|1\rangle\langle 0| + |0 \rangle\langle 1|$ and $n=|1\rangle\langle 1|$.

The dipole matrix elements, $V_{\alpha\beta}$, strongly depend on the distance between the atoms.  It is straightforward to engineer a situation where, for any pair of sites, $V_{\alpha\beta}$ is either negligibly small (compared to $\Omega$) or very large.  Thus we treat $V_{\alpha\beta}$ as taking on only the values $0$ or $\infty$.  The latter corresponds to a constraint that the two atoms cannot be simultaneously excited.  We say that they are within the blockade radius.

To realize a dimer model with this array of Rydberg atoms, we follow the procedure in \cite{Semeghini2021}, and envision placing an atom at the center of each bond.  The $|1\rangle$ state is identified as the presence of a dimer, while the $|0\rangle$ state corresponds to the absence.  The available Hilbert space is larger than that of a traditional dimer model, as one is not restricted to dimer coverings, but can also have defects where there are missing dimers.  We will, however, engineer our protocol so that the final state will correspond to a superposition of dimer coverings.

\begin{figure}[tbh]
    %\centering
%\includegraphics[width=0.4\textwidth]{ATOM.png} 
\includegraphics[width=0.3\textwidth]{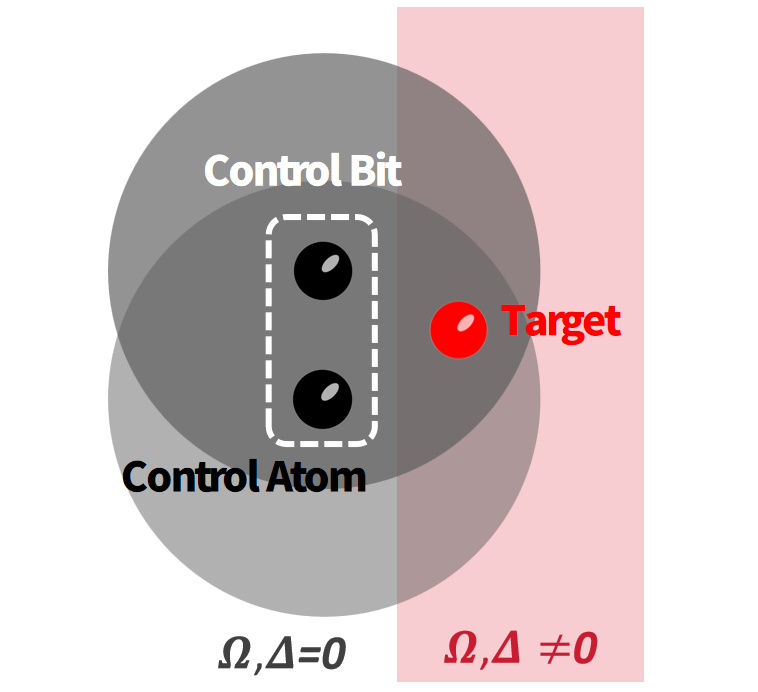}
    \caption{Schematic of key components in our quantum gates.  Two {\em control atoms} are shown, along with grey shaded circles corresponding to their blockade radius. Together these two control atoms  compose a \emph{control bit}. A target atom inside these circles cannot be excited unless all of  the control atoms are in their ground states.  Time dependent control fields $\Omega,\Delta$, as depicted by the shaded red area, are applied to the target atom, but not the control atoms. %{\color{red} Edits to figure:  Increase dpi.  Color code labels, add box for Control Atoms}
    }
    \label{fig:ATOM}
\end{figure}

All of the coupling constants in Eq.~(\ref{ham}) can be made time dependent by moving the microtraps, modulating the magnetic field, or modulating the laser intensity/profile.  We will start with a trivial configuration, where all of the atoms are in the $|0\rangle$ state.  We will then use a sequence of local gates to `grow' the resonating dimer state.  
%As we have already emphasized, this is a very different approach from the one used in \cite{giudici2023} to generate equilibrium spin liquid states.  In that work the parameters of the Hamiltonian were kept {\em spatially uniform} as they were changed.  
%By construction o
Our protocol will take a time which is proportional to the length of the cylinder, but is independent of its width.  {  In Appendix~\ref{sec:Torus} we give an alternative protocol in which the time is proportional to the width, but independent of the length.}
%{\color{blue}  Something about matrix product states???}
%{\color{blue}  Possibly say something about differences between the states in the experiments and our states?  Maybe not?}

\subsection{Gates}\label{sec:Gates}
We begin by introducing the quantum gates employed in our state preparation protocol. 
%To maintain consistency with the standard language of quantum computation, we adopt the convention that $\ket{g}$ corresponds to $\ket{0}$ and $\ket{e}$ to $\ket{1}$.
%{\color{red} Do we want to flip that around?  Calling the excited state $|0\rangle$ and the ground state $|1\rangle$?  Why do we need to introduce $|0\rangle$ and $|1\rangle$.  What's wrong with $\ket{g}$ and $\ket{e}$? } 
Each %quantum gate in our protocol i
of these
involves a combination of \textit{control atoms} and \textit{target atoms}. The control atoms impose constraints on the target atoms via Rydberg blockade, while the quantum state of the target atoms is actively manipulated.  

Practically, we implement these gates by tuning magnetic fields and laser parameters to control $\Delta_\alpha(t)$ and $\Omega_\alpha(t)$ of the target atoms. The relative positions between control and target atoms are adjusted using microtraps to ensure the desired %blockade-induced 
interactions. For control atoms and other uninvolved atoms, we set $\Delta_\alpha = 0$ and $\Omega_\alpha = 0$ throughout the operation. After each gate, we also immediately turn off $\Delta_\alpha$ and $\Omega_\alpha$ for the target atoms to suppress unwanted transitions and accumulated phases.

\begin{figure}
\begin{equation*}
\begin{array}{lrl}
%
% A
(a)\,\, { U^H_{1c1t}}
    &\raisebox{-0.3\height}{\includegraphics{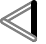}}
    &\to\frac{1}{\sqrt{2}}\left(\raisebox{-0.3\height}{\includegraphics{minifigs/t00g.pdf}}+
    \raisebox{-0.3\height}{\includegraphics{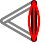}}\right)\\[3mm]
(b)\,\, { U^X_{1c1t}}
    &\raisebox{-0.4\height}{\includegraphics{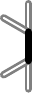}}
    &\to\raisebox{-0.4\height}{\includegraphics{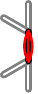}}\\[6mm]
(c)\,\, { U_{1c2t}}
    &\raisebox{-0.3\height}{\includegraphics{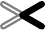}}
    &\to\frac{1}{\sqrt{2}}\left(\raisebox{-0.3\height}{\includegraphics{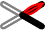}}+
    \raisebox{-0.3\height}{\includegraphics{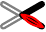}}\right)\\[3mm]
(d)\,\,  U^X_{2c2t}
&\quad\raisebox{-0.45\height}{\includegraphics{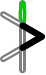}}
    &\to\!
    \raisebox{-0.45\height}{\includegraphics{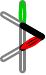}}
    \quad\quad\quad
\raisebox{-0.45\height}{\includegraphics{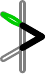}}
    \to\!
    \raisebox{-0.45\height}{\includegraphics{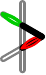}}    
\\
&\raisebox{-0.45\height}{\includegraphics{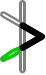}}
    &\to\!
    \raisebox{-0.45\height}{\includegraphics{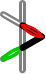}}
    \quad\quad\quad
\raisebox{-0.45\height}{\includegraphics{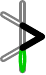}}
    \to\!
    \raisebox{-0.45\height}{\includegraphics{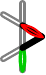}}    
\end{array}
\end{equation*}
\caption{Gates for YC-$2N$ state preparation. Light and dark bonds represent control and target atoms.  Excited atoms are highlighted in green (control) or red (target). Only gate actions which occur during the preparation,
%\footnote{For example, in the case of the $U_{2c2t}^X$ gate, the control qubit state $\ket{0_c 0_c}$—where all control atoms are in the ground state—does not appear in our state preparation protocol, as can be verified by following the procedure.
%{\color{red}  It is better to make clarifying remarks in the text.}
%}
and which change dimer configurations are shown. }\label{fig:flipy}
\end{figure}

\begin{figure}
\begin{equation*}
\begin{array}{ll}
(a)\,\,{ U^H_{1c1t}}&
\raisebox{-0.3\height}{\includegraphics{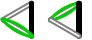}}\\[2mm]
(b)\,\, { U^X_{1c1t}}&
\raisebox{-0.4\height}{\includegraphics{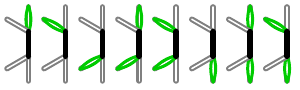}}\\[2mm]
(c)\,\, { U_{1c2t}}&
\raisebox{-0.3\height}{\includegraphics{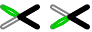}}\\[2mm]
(d)\,\,  U^X_{2c2t}
&
\raisebox{-0.3\height}{\includegraphics{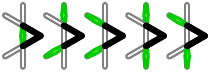}}
\end{array}
\end{equation*}
\caption{Blockaded configurations during YC-$2N$ state preparation, corresponding to the case where the control qubits are in the $|1_c\rangle$ state.  Each target bond is touched by at least one control-bond dimer, 
and the gates leave these spin configurations unchanged.
}\label{fig:noflipy}
\end{figure}

\begin{figure*}
\begin{equation*}
\begin{array}{rlrlrl}
(a)\,\,U^H_{2c2t}
&&
\quad(b)\,\,U_{2c4t}\\
\raisebox{-0.45\height}{\includegraphics{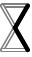}}
    &\to
\frac{1}{\sqrt{2}}\left(
\raisebox{-0.45\height}{\includegraphics{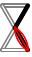}}
+\raisebox{-0.45\height}{\includegraphics{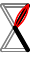}}
\right)
&
\raisebox{-0.45\height}{\includegraphics{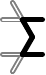}}
    &\to
\frac{1}{\sqrt{2}}\left(
\raisebox{-0.45\height}{\includegraphics{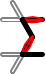}}
+
\raisebox{-0.45\height}{\includegraphics{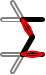}}
\right)\\
\raisebox{-0.45\height}{\includegraphics{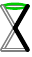}}
    &\to
\raisebox{-0.45\height}{\includegraphics{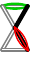}}
%\quad\quad\quad
&
\raisebox{-0.45\height}{\includegraphics{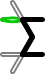}}
    &\to
\frac{1}{\sqrt{2}}\left(
\raisebox{-0.45\height}{\includegraphics{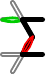}}
+
\raisebox{-0.45\height}{\includegraphics{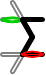}}
\right)
&
\qquad\raisebox{-0.45\height}{\includegraphics{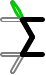}}
    &\to
\frac{1}{\sqrt{2}}\left(
\raisebox{-0.45\height}{\includegraphics{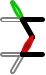}}
+
\raisebox{-0.45\height}{\includegraphics{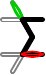}}
\right)\\
\raisebox{-0.45\height}{\includegraphics{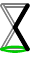}}
    &\to
\raisebox{-0.45\height}{\includegraphics{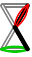}}
&
\raisebox{-0.45\height}{\includegraphics{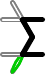}}
    &\to
\frac{1}{\sqrt{2}}\left(
\raisebox{-0.45\height}{\includegraphics{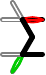}}
+
\raisebox{-0.45\height}{\includegraphics{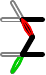}}
\right)
&
\raisebox{-0.45\height}{\includegraphics{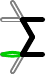}}
    &\to
\frac{1}{\sqrt{2}}\left(
\raisebox{-0.45\height}{\includegraphics{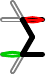}}
+
\raisebox{-0.45\height}{\includegraphics{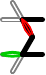}}
\right)
\end{array}
\end{equation*}
\caption{Gates for XC-$2N$ state preparation. Only gate actions which change dimer configurations are shown.}\label{fig:flipx}
\end{figure*}

%In each of our gates there will be a set of control atoms and a set of target atoms.  The gate manipulates the target atoms, contingent upon the state of the control atoms.  In practice, this means we engineer the Hamiltonian in Eq.~(\ref{ham}) so that  $\Omega_\alpha\neq0$ only on the target atoms.  Control is through the interaction term $V_{\alpha\beta}$. For state creation in the YC configurations (such as the eye model), we use gates that involve 1 or 2 target atoms.  In the XC configurations (such as the hourglass model), we require a 4-target gate and a 2-target gate which is distinct from the one used in the YC configurations.  Figure~\ref{gates} schematically show the atom configurations for the various gates. {\color{red} Atoms which are not involved in the gate have $\Omega_\alpha=0$ and $\Delta_\alpha=0$??  After the gate is completed one immediately sets $\Omega=0$ and $\Delta=0$ for the affected atoms, to avoid unwanted phases or transitions?? }
In addition to describing the gate actions on the target atoms, it is useful to introduce an extra layer of abstraction.
We group control atoms together which blockade the same transition.  We label the state of that group of control atoms as $|0_c\rangle$ if all of them are in the ground state.  They will then not cause any blockade.  If at least one is excited, we label the state as $|1_c\rangle$.  The state is not uniquely defined by this condition, but for the purposes of our gates, all that matters is the presence or absence of the blockade.  We refer to the two possibilities $|0_c\rangle,|1_c\rangle$ as the \emph{control qubit}.  Fig~\ref{fig:ATOM} gives a schematic representation of a simple case with two control atoms and one target atom.  

Some of our gates will use multiple control qubits.  In that case a given control atom can contribute to the state of more than one control qubit.  The $U_{2c2t}^X$ gate described below, and illustrated in Fig.~\ref{fig:flipy} (d) and \ref{fig:noflipy} (d) is one example.  There the central vertical bond corresponds to an atom which blockades both target atoms.  Physically this behavior is natural, as that control atom is in close proximity to both of the targets. 
%in that geometric arrangement of atoms will result in both tar  is achieved by placing that control atom so that it blockades multiple targets.  
Additionally, in gates with multiple targets, the each target atoms will blockade a select number of other targets, as described below.  Thus only valid dimer coverings appear in the final configurations in Figs.~\ref{fig:flipy} or ~\ref{fig:flipx}.

%It is important to note that our use of the term \textit{control atom} differs from the conventional notion of a \textit{control bit} in quantum computation language. In our system, multiple control atoms may collectively influence a single target atom. Conceptually, the logical control bit corresponds to the joint state of these control atoms.Specifically, we can define the logical control state $\ket{0_c}$ to correspond to the case where all involved control atoms are in the ground state $\ket{g}$. Conversely, if any of the control atoms is in the excited state $\ket{e}$, it corresponds to the $\ket{1_c}$ state of the logical control bit. The notion of the target bit, in contrast, directly maps onto the physical target atom: the state $\ket{g}$ of the target atom is identified with the $\ket{0_t}$ state, and $\ket{e}$ with $\ket{1_t}$. If there is no ambiguity, we omit the subscript $t$ in the following text. See Fig.~\ref{fig:ATOM} for a schematic illustration.

In our protocol the gates always act on target atoms that begin in their ground state.  Thus we only need to define how they act on such states.  This gives us significant flexibility in gate design. Similarly, we only need to consider the control atom configurations which arise during our state preparation protocol.  
Since the gate operations are applied sequentially, some configurations will never appear. 

\begin{figure}[tb]
\begin{equation*}
\begin{array}{ll}
(a)\,\,U^H_{2c2t}&
\raisebox{-0.45\height}{\includegraphics{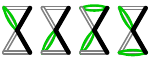}}
\\
(b)\,\,U_{2c4t}
&
\raisebox{-0.45\height}{\includegraphics{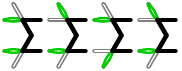}}
\\
\end{array}
\end{equation*}
\caption{Blockaded configurations during XC-$2N$ state preparation.  %Dimer configurations which are unchanged by the gates are shown.  Configurations which do not appear in Figs.~\ref{fig:flipx} or \ref{fig:noflipx} do not occur during the state prepartion.
}\label{fig:noflipx}
\end{figure}

We introduce a total of six {gate operations}: the first four are used for state preparation on the YC cylinder, while the remaining two are used for the XC cylinder.
We use the unified symbol 
$U$ to indicate that these are unitary operations.  Subscripts specifying the number of control and target bits, and (when necessary) superscripts further disambiguate the gates.  The spatial arrangement of atoms in each case is shown in Figs.~\ref{fig:flipy} through \ref{fig:noflipx}.
%, where we present their local configurations on the kagome lattice. 
The control/target atoms are shown as light/dark bonds.  Excited atoms are highlighted in green (control) or red (target).  
Control atoms adjacent to the same target belong to the same control qubit.  Figures~\ref{fig:flipy} and \ref{fig:flipx} show the nontrivial gate actions, corresponding to the cases where some of the target atoms become excited.  Figures~\ref{fig:noflipy} and \ref{fig:noflipx} show the blockaded configurations, where all of the target atoms are blockaded and thus remain unexcited.  These correspond to the control qubits all being in the excited state.

%We only show the gate action for the configurations which arise during the state preparation, as the action on other configurations is irrelevant.  This freedom can be used to more efficiently design the gates.  Implementation is described in Appendix~\ref{}.

The gate operations used in YC-$2N$ state preparations are
\begin{align}
U_{1c1t}^H : 
&
\left\{
\begin{aligned}
U_{1c1t}^H \ket{0_c\,0} &=  \frac{ \ket{0_c\,0} + \ket{0_c\,1} }{\sqrt{2}} \\
U_{1c1t}^H \ket{1_c\,0} &= \ket{1_c\,0}
\end{aligned}
\right.
\label{eq:U_H_action_1c1t}\\
U_{1c1t}^X : 
&
\left\{
\begin{aligned}
U_{1c1t}^X \ket{0_c\,0} &=  \ket{0_c\,1}  \\
U_{1c1t}^X \ket{1_c\,0} &= \ket{1_c\,0}
\end{aligned}
\right.
\label{eq:U_X_action_1c1t}\\
U_{1c2t} :
&
\left\{
\begin{aligned}
U_{1c2t} \ket{0_c\,00} &= 
\frac{
\ket{0_c\,01} + \ket{0_c\,10}}{\sqrt{2}}
 \\
U_{1c2t} \ket{1_c\,00} &= \ket{1_c\,00}
\end{aligned}
\right.
\label{eq:U_H_action_1c2t}\\
U_{2c2t}^X : 
&
\left\{
\begin{aligned} 
U_{2c2t}^X \ket{1_c\,0_c\,00} &= \ket{1_c\,0_c\,01} \\
U_{2c2t}^X \ket{0_c\,1_c\,00} &= \ket{0_c\,1_c\,10} \\
U_{2c2t}^X \ket{1_c\,1_c\,00} &= \ket{1_c\,1_c\,00}
\end{aligned}
\right.
\label{eq:U_X_action_2c2t}
\end{align}
while those for XC-$2N$ geometries are
\begin{align}
U_{2c2t}^H : 
&
\left\{
\begin{aligned}
U_{2c2t}^H \ket{0_c\,0_c\,00} &= 
\frac{
\ket{0_c\,0_c\,10} + \ket{0_c\,0_c\,01}}
{\sqrt{2}} \\
U_{2c2t}^H \ket{1_c\,0_c\,00} &= \ket{1_c\,0_c\,01} \\
U_{2c2t}^H \ket{0_c\,1_c\,00} &= \ket{0_c\,1_c\,10} \\
U_{2c2t}^H \ket{1_c\,1_c\,00} &= \ket{1_c\,1_c\,00}
\end{aligned}
\right.
\label{eq:U_H_action_2c2t}\\
U_{2c4t} : 
&
\left\{
\begin{aligned}
U_{2c4t} \ket{0_c\,0_c\,0000} &=
\frac{\ket{0_c\,0_c\,1010} + \ket{0_c\,0_c\,0101}}{\sqrt{2}}  \\
U_{2c4t} \ket{1_c\,0_c\,0000} &=\frac{\ket{1_c\,0_c\,0010} + \ket{1_c\,0_c\,0001}}{\sqrt{2}}  \\
U_{2c4t} \ket{0_c\,1_c\,0000} &= 
\frac{
\ket{0_c\,1_c\,1000} + \ket{0_c\,1_c\,0100} }
{\sqrt{2}} \\
U_{2c4t} \ket{1_c\,1_c\,0000} &= \ket{1_c\,1_c\,0000}
\end{aligned}
\right.
\label{eq:U_action_2c4t}
\end{align}
As already explained, the subscripts list the number of control and target qubits.  The superscripts $X$ and $H$ distinguish between variants of controlled not and controlled Hadimard gates.  

As demonstrated by their implementation in Appendix \ref{sec:Realization}, these gates can be performed by arranging the atoms in the correct geometric arrangement, and then 
applying the appropriate pulse sequence.
%adiabatically sweeping the control fields.  Alternatively one can use carefully timed pulse sequences.  
In most cases the required atomic arrangement is identical to the spatial arrangement of bonds in Figs.~\ref{fig:flipy} through \ref{fig:noflipx} (i.e. the local configuration of the kagome lattice).  The principle exception is the $U_{2c4t}$ gate in Figs. \ref{fig:flipx} (b) and \ref{fig:noflipx} (b).  There one must engineer a blockade between the target atoms on the top and bottom of the $\Sigma$ shape, for example using the arrangements in Fig.~\ref{fig:U2c4t} % or \ref{fig:U2c4tIM} 
.

\subsection{State Creation for YC cylinders}\label{sec:YC}
\begin{figure}[ht]
\includegraphics[width=0.5\textwidth]{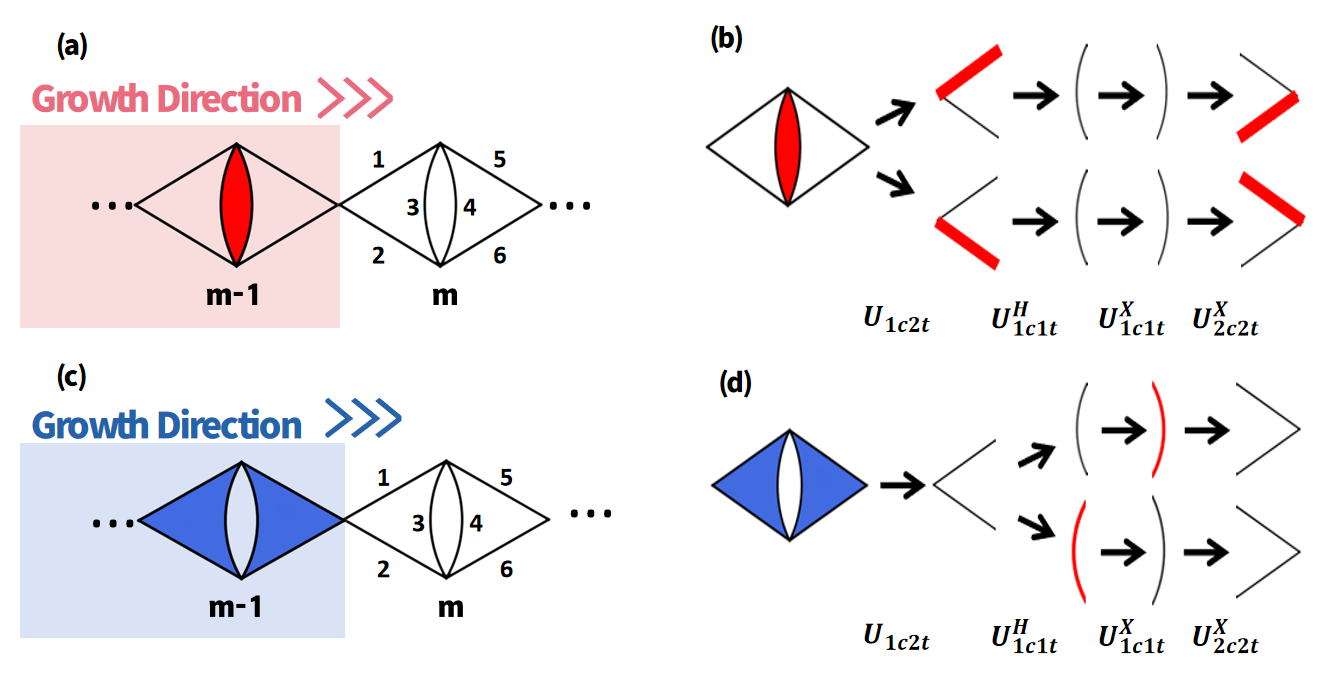}
    \caption{Gate sequence for growing the eye model on a YC-2 cylinder by one unit cell. Panels (a) and (c) show
    two unit cell, labeled $m$ and $m-1$. The atoms within one unit cell are labeled by the numbers 1 through
6. The left cell is in  a coherent superposition of dimer configurations, in one of the two topological sectors [(a) or (c)] or a coherent superposition of the two.  The atoms in right cell are in their ground state.  Panels (b) and (d) show how the states of the atoms in the right cell evolve with each gate, forming a coherent superposition of the bonds in each path.} 
    %The case of the first boundary condition. The atomic states in the shaded region remain unchanged throughout the entire protocol. The dynamics in the unshaded region are activated sequentially, resulting in the growth of the desired state. The site labeling convention $(m, i)$ is explicitly illustrated.(b) State preparation process corresponding to (a). The leftmost state serves as a boundary condition. A sequence of prescribed quantum gates is applied to obtain the state on the right, and this process proceeds iteratively. Vertically aligned arrows represent simultaneous gate operations across branches, while bifurcating arrows indicate equal-weight superpositions of quantum states. The final state is given by the sum over all possible arrow paths.(c) The case of the second boundary condition.
    %(d) State preparation corresponding to (c).
    \label{fig:Eye_growing}
\end{figure}

We illustrate state creation for YC cylinders by first giving our argument for the \emph{eye model} (YC-2), and then generalizing to wider cylinders. 
Figure~\ref{fig:Eye_growing} shows two unit cells of the eye model. We denote the position of a bond by an ordered pair $(m, i)$, where $m$ labels the unit cell, and $i$ indicates the position of the bond within that eye-shaped cell. We imagine that the cell on the left is the right-hand end of a chain corresponding to the Rokhsar-Kivelson state, and the atoms there are in superpositions of the ground and excited states, as described in Sec.~\ref{sec:EyeModel}.  We separately consider the cases that the cell is in the states \eyeA~or \eyeB, and the argument naturally works for a coherent superposition $\alpha\eyeA +\beta\eyeB$.  
The atoms in the cell on the right are all in their ground state.  We wish to apply a set of gates so that we grow the Rokhsar-Kivelson state.  %In the present case, that means converting the cell on the right to \eyeB.

The protocol requires four sequential operations:
%proceeds as follows. The eye units are sequentially grown from smaller to larger $m$. The growth process within each eye unit is further divided into four distinct steps.
\begin{enumerate}
  \item %\textbf{Step 1:}
  {\bf $\mathbf{U_{1c2t}}$ gate:}
  The atoms at positions \((m, 1)\) and \((m, 2)\) are designated as \emph{target atoms}, while the \emph{control bit} is composed of atoms at \((m{-}1, 5)\) and \((m{-}1, 6)\). 
  %The gate operation \(U_{1c2t}\) is applied.
  
  \item %\textbf{Step 2:} 
  {\bf $\mathbf{U_{1c1t}^H}$ gate:}
  The atom at \((m, 3)\) serves as the \emph{target}, with the \emph{control bit} consisting of atoms at \((m, 1)\) and \((m, 2)\). 
  
  \item  
  {\bf $\mathbf{U^{X}_{1c1t}}$ gate:}
  The atom at \((m, 4)\) is set as the \emph{target}, with the \emph{control bit} composed of \((m, 1)\), \((m, 2)\), and \((m, 3)\). 
  %The gate \(U^{X}_{1c1t}\) is applied.
  
  \item 
   {\bf $\mathbf{U^{X}_{2c2t}}$ gate:}
  The atoms at \((m, 5)\) and \((m, 6)\) are designated as \emph{targets}. For \((m, 5)\), the \emph{control bit} consists of atoms \((m, 1)\), \((m, 3)\), and \((m, 4)\); for \((m, 6)\), the \emph{control bit} consists of \((m, 2)\), \((m, 3)\), and \((m, 4)\). 
  %The gate operation \(U^{X}_{2c2t}\) is applied accordingly.
\end{enumerate}

This process is schematically depicted in Fig \ref{fig:Eye_growing}(b) and (d). Each arrow shows the state of subsequent target atoms, after the listed gate.  The chains branch after either the $U_{1c2t}$ or $U^H_{1c1t}$ gates, resulting in equal weight superpositions of the two dimer coverings which are depicted in each of the two cases shown. 
%Here we use arrows to represent the progression of the growing. Specifically, vertically aligned arrows correspond to the same gate operation. The starting point of each arrow can be interpreted as the boundary condition (i.e., the control atoms) for the gate, while the endpoint indicates the resulting state of the target atoms after the gate is applied. In cases where the gate yields an equal superposition of two output states, we represent this with branching arrows that share the same starting point. 
%As illustrated in Fig.~\ref{fig:Eye_growing}, after each step the state grows in a manner which is contingent on the previous step.  
In this way the configuration \eyeA\eye~ is transformed to  \eyeA\eyeB, while \eyeB\eye~ evolves to \eyeB\eyeA.  {  In Sec.~\ref{sec:XC} we explain how in the general case we can relate the branching options to the structure of matrix product states.}

%For the two cases where the left cell is either \eyeA or \eyeB, our protocol can transform the right cell into \eyeB and \eyeA respectively, thereby growing the Rokhsar-Kivelson state. The right panel of Fig.~\ref{fig:Eye_growing} illustrates the working mechanism of our protocol. In particular, each arrow can be interpreted as a gate operation, where a splitting arrow represents an equal-weight superposition of two states, and the final state is obtained by summing over all possible paths.

\begin{figure}[tbh]
\includegraphics[width=0.4\textwidth]{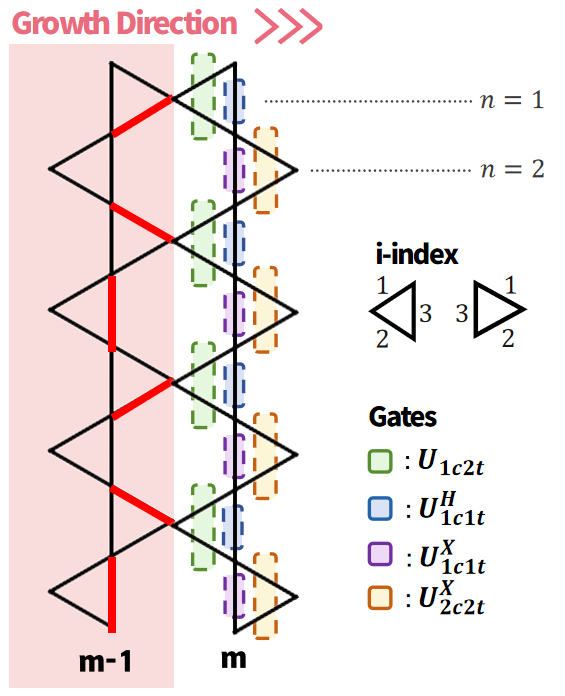}
    \caption{Gate sequence for growing the Rokhsar-Kivelson state on a YC-2N cylinder, here $2N=8$.  Qubits are labeled by integers $(m,n,i)$ -- $m$ labels the annular strip, $n=1,2,\cdots,2N$ labels the triangle, and $i=1,2,3$ labels the bond within each triangle, as depicted in the figure.  The target qubits for each gate is drawn in boxes, whose color denotes the gate type.  All gates of the same color can all be carried out simultaneously. }
    \label{fig:YC_growing}
\end{figure}

The procedure for growing the Rokhsar–Kivelson state on a YC cylinder of arbitrary width can be naturally generalized from the eye model construction. %To systematically 
As shown in Fig.~\ref{fig:YC_growing}, we
label the position of each bond
%, we introduce 
with a triplet index \((m, n, i)\), where \(m\) denotes the index of the annular stripe, \(n\) labels the position of a triangle within the stripe, and \(i = 1, 2, 3\) labels the individual bonds within each triangle. 
%This indexing convention is illustrated in Fig.~\ref{fig:YC_growing}.

%The protocol proceeds as follows. 
The %annular stripes are 
state is 
sequentially grown from smaller to larger $m$. The growth process within each annular stripe is further divided into four distinct steps, which are the generalizations of the same numbered steps used in the eye model:
\begin{enumerate}
  \item  For each odd \(n\), the atoms at positions \((m, n, 1)\) and \((m, n, 2)\) are designated as \emph{target atoms}, colored in green in Fig.~\ref{fig:YC_growing}.  The \emph{control bit} is composed of the atoms located at \((m{-}1, n, 1)\) and \((m{-}1, n, 2)\), which touch the target atoms.  The gate operation \(U_{1c2t}\) is applied. 
  
  \item %\textbf{Step 2:} 
  Again for odd \(n\), the atom at \((m, n, 3)\) serves as the \emph{target atom}, colored in blue.  The control bit is composed of the atoms \((m, n, 1)\) and \((m, n, 2)\), colored in green. The gate \(U^{H}_{1c1t}\) is then applied. 
  
  \item %\textbf{Step 3:} 
  For even \(n\), the atom at \((m, n, 3)\) is selected as the \emph{target atom}, colored in purple.  The control bit consists of atoms at \((m, n{-}1, 2)\), \((m, n{-}1, 3)\), \((m, n{+}1, 3)\), and \((m, n{+}1, 1)\). These are the blue atoms adjacent to the target, as well as the closest green atom on each side.  The  gate operation  \(U^{X}_{1c1t}\) is applied.
  
  \item %\textbf{Step 4:} 
  For even \(n\), the atoms at \((m, n, 1)\) and \((m, n, 2)\) are treated as \emph{targets}, colored in yellow. For \((m, n, 1)\), the control bit is composed of \((m, n{-}1, 2)\), \((m, n{-}1, 3)\), and \((m, n, 3)\); for \((m, n, 2)\), the control bit consists of \((m, n{+}1, 1)\), \((m, n{+}1, 3)\), and \((m, n, 3)\). These are the blue and purple atoms adjacent to the targets, as well as the closest green atoms.  The gate operation \(U^{X}_{2c2t}\) is applied.
\end{enumerate}

In the above steps, the indices \(n{-}1\) and \(n{+}1\) are defined under periodic boundary conditions, which can be experimentally implemented by physically rearranging the Rydberg atoms. Each step of the protocol can be executed in parallel for different values of \(n\):  All of the green gates are performed simultaneously, then all of the blue gates...  Thus the growth time is independent of the width of the cylinder.
This gate sequence produces a uniform superposition of all valid dimer configurations which are consistent with the boundary conditions on the previous strip, growing the Rokhsar-Kivelson state by one annular strip. { Section~\ref{sec:seed} describes how one starts the process, creating the initial strip.}

%The validity of our protocol can be justified as follows. First, each gate operation is designed to ensure that every vertex is occupied by exactly one dimer (i.e., one excited atom). As a result, every component of the resulting quantum state corresponds to a valid dimer configuration. Second, all such quantum configurations appear with equal amplitude, ensuring uniform superposition. Finally, for a cylinder of type YC-\(2N\), our protocol generates \(2^N\) distinct quantum states for each fixed boundary condition, consistent with the structure of the Rokhsar--Kivelson state. Therefore, we conclude that the protocol faithfully constructs the RK state on the YC-\(2N\) cylinder.\footnote{For example, suppose the boundary condition is given by the bitstring 1101. Then, according to the conventions in Sec.~\ref{sec:Larger_cylinder}, the allowed configurations of the newly generated stripe, labeled by the triplet \((L, R, u)\), must satisfy \(L = 0010\), while \(R\) must have the same parity as \(L\), giving \(2^{N}/2\) possible choices for \(R\). Since \(u\) can be either 0 or 1, the total number of allowed configurations is \(2^N\).}

\begin{figure}[tbh]
\includegraphics[width=0.5\textwidth]{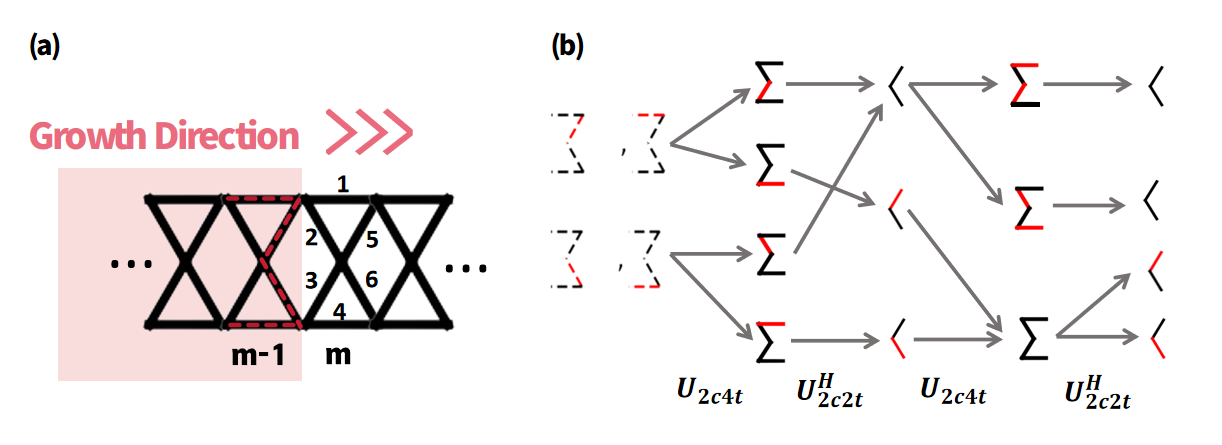}
    \caption{Gate sequence for growing the hourglass model on a XC-4 cylinder by one unit cell. (a) Unit cells are labeled by $m$, and the atoms within one unit cell with the numbers 1 through 6.
    %illustrates a specific step in the growing process, where the dimer covering of the unit cell labeled $m$ is generated based on the boundary condition provided by the neighboring cell labeled $m{-}1$. %The red-shaded region on the left represents RK state, while the atoms in the unshaded region on the right are initially in their ground state. Panel 
    (b) 
    Each gate results in a superposition of excitations, which are contingent on the existing dimer configurations.  The resulting quantum state is a superposition of all paths through this diagram.
    }
    %shows how the quantum states of the atoms in the right unit cell evolve under successive gate operations. The final state is a sum over all the possible arrow paths.\footnote{Although the intermediate weights of different arrows may vary, it turns out that all final paths carry equal total weight.}}
    \label{fig:Hourglass_growing}
\end{figure}

\begin{figure}[tbh]    \includegraphics[width=0.4\textwidth]{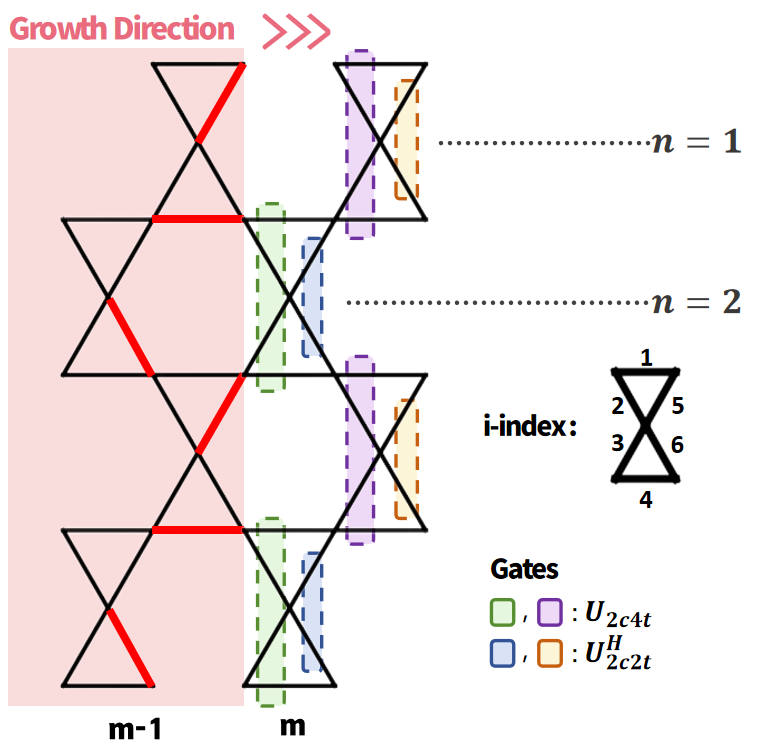}
    \caption{Gate sequence for growing the Rokhsar-Kivelson state on a XC-2N cylinder, here $2N=8$.  Qubits are labeled by integers $(m,n,i)$ -- $m$ labels the annular strip, $n=1,2,\cdots,N$ labels the hourglass unit, and $i=1,2,3,4,5,6$ labels the bond within each hourglass unit, as depicted in the figure.  The target qubits for each gate is drawn in boxes, whose color denotes the gate type.  All gates of the same color can all be carried out simultaneously. }
    \label{fig:XC_growing}
\end{figure}

\subsection{State Creation for XC cylinders}\label{sec:XC}
Similar to
Sec.~\ref{sec:YC}, we illustrate state creation in XC cylinders by first 
%start by giving the state creation protocol for XC cylinders by 
considering the \emph{hourglass model} (XC-4). As shown in Fig.~\ref{fig:XC_growing}, we denote the position of each bond
by an ordered pair $(m, i)$, where $m$ labels the unit cell,
and $i\in\{1,2,3,4,5,6\}$ indicates the location within that hourglass-shaped cell. 
We imagine the cells to the left are in a superposition of all dimer configurations with the chosen topological sector. 
According to Eq.~\ref{hourmps} the two possibilities can be explicitly written as
%to the Rokhsar-Kivelson state, and the atoms there are in superpositions of the ground and excited states. Due to the MPS structure of Eq.~\eqref{hourmps}, the RK state on the left side of the system must take one of the following two forms:
\[
\ket{\Phi_{\text{L}}} = 
\begin{cases}
\frac{1}{2}\left( \ket{\phi_1}\ket{\raisebox{-0.8mm}{\ISigA}} + \ket{\phi_2}\ket{\raisebox{-0.8mm}{\ISigB}} + \ket{\phi_3}\ket{\raisebox{-0.8mm}{\ISigC}} + \ket{\phi_4}\ket{\raisebox{-0.8mm}{\ISigD}} \right), \\
\frac{1}{2}\left( (\ket{\psi_1}+\ket{\psi_2})\ket{\raisebox{-0.8mm}{\ISigE}} + \ket{\psi_3}\ket{\raisebox{-0.8mm}{\ISigF}} + \ket{\psi_4}\ket{\raisebox{-0.8mm}{\ISigG}} \right).
\end{cases}
\]
where $\ket{\phi_i}$ and $|\psi_i\rangle$ are normalized quantum states that specify the dimer configurations of all sites to the left, ending with distinct terminations.
We will focus on the first case, but the reasoning for the second one is identical.  As  in Sec.~\ref{sec:YC}, the algorithm also works for a quantum superpositions of the two states.

%Without loss of generality, we focus on the first case. In this configuration, the boundary condition at the domain wall corresponds to an equal-weight superposition of \(\ket{\ISigA}, \ket{\ISigB}, \ket{\ISigC}, \ket{\ISigD}\). The analysis for the second case proceeds in a completely analogous manner.

The atoms in the cell on the right are all in their ground
state. We  grow
the Rokhsar-Kivelson state
by repeatedly applying the
%The protocol requires 
two sequential operations shown in Fig~\ref{fig:Hourglass_growing}.
\begin{enumerate}
  \item
  {\bf $\mathbf{U_{2c4t}}$ gate:}
  The atoms at positions \((m, 1)\), \((m, 2)\), \((m, 3)\), \((m, 4)\) are designated as \emph{target atoms}, while the \emph{control bit} is composed of atoms at \((m{-}1, 5)\) and \((m{-}1, 6)\). 
  Under this operation the resonating dimer state grows:
  \begin{align}
\ISigA&\to\ISigA\frac{\SigD+\SigE}{\sqrt{2}}&
\ISigC&\to\ISigC\frac{\SigB+\SigC}{\sqrt{2}}\\
\ISigB&\to\ISigB\frac{\SigD+\SigE}{\sqrt{2}}&
\ISigD&\to\ISigD\frac{\SigB+\SigC}{\sqrt{2}}
\end{align}
These superpositions are illustrated in Fig.~\ref{fig:Hourglass_growing}(b) by branching arrows.
  \item
  {\bf $\mathbf{U^H_{2c2t}}$ gate:}
  The atoms at positions \((m, 5)\), \((m, 6)\) are designated as \emph{target atoms}, while the \emph{control bit} is composed of atoms at  \((m, 1)\), \((m, 2)\), \((m, 3)\), \((m, 4)\) . 
\end{enumerate}
After performing these sequential gate operations, the hourglass unit, initially in its ground state, is transformed into a matrix product state
which is one unit cell larger.
% of the form:
% \begin{equation}\label{bd2}
% |\Psi\rangle=\cdots
% \left(
% \begin{array}{cc}
% \XBABB & \XBAAA \\
% \XABBB & \XABAA
% \end{array}
% \right)
% \left(
% \begin{array}{cc}
% \XAAAB & \XAABA\\
% \XBBAB & \XBBBA
% \end{array}
% \right)
% \cdots
% \end{equation}
% , which is equivalent to Eq.~\eqref{hourmps}. In other words, we have effectively grown the wanted RK state.

%It is worth emphasizing that o
Our state creation protocol explicitly leverages the structure of matrix product states. The 
positions of subsequent bonds only depend on those immediately to the left.  The branching diagram in Fig.~\ref{fig:Hourglass_growing}(b),
which describes how our gates grow the quantum state, can be viewed as a representation of the matrix product state in  Eq.~(\ref{hourmps}).  It
is equivalent to the Matrix Product Diagram construction introduced by Crosswhite and Bacon to represent matrix product states and relate them to finite state machines
 \cite{crosswhite}. 
 %(To match their notation, the states should be moved from each vertex to all of the incoming arrows.) 
The diagram in Fig.~\ref{fig:Hourglass_growing}(b) involves subsets of the hourglass shaped unit cell and
%Rather than acting on the entire hourglass shaped unit cell, we implement 
%Our gate sequence, and the diagram in Fig.~\ref{fig:Hourglass_growing}(b), 
corresponds to a decomposition of the matrices in Eq.~(\ref{hourmps}) as
%\begin{widetext}
\begin{align}\label{mat1}
\left(
\begin{array}{cc}
\XABAA&\XABBB\\
\XBAAA&\XBABB
\end{array}
\right)&=
\left(
\begin{array}{ccc}
\SigB&\SigC&\\
&\SigD&\SigE
\end{array}
\right)
\left(
\begin{array}{cc}
\cupC&\\
&\cupA\\
\cupB
\end{array}
\right)\\
\left(
\begin{array}{cc}
\XBBBA&\XBBAB\\
\XAABA&\XAAAB
\end{array}
\right)&=
\left(
\begin{array}{ccc}
\SigA\\
&\SigG&\SigF
\end{array}
\right)
\left(
\begin{array}{cc}
\cupC&\cupB\\
\cupA\\
&\cupA
\end{array}
\right)
\end{align}

 To convert a Matrix product state into a diagram, one begins by drawing the nodes.  There is one node for each matrix element -- and in our notation each node is labeled by that element.  
 One places the nodes corresponding to a given matrix in a vertical line.
 For example, in Fig.~\ref{fig:Hourglass_growing}(b), the four symbols \raisebox{-0.3mm}{\SigB},\,\SigC,
\!\raisebox{-0.1mm}{\SigD},\raisebox{-0.1mm}{\SigE}\, correspond to the
  first matrix in the decomposition in Eq.~(\ref{mat1}).  One then draws arrows connecting nodes in sequential columns.   Nodes are connected if their product would appear in matrix multiplication.  This construction can be applied to any matrix product state.  
  Our growth algorithm amounts to using these diagrams as a  blueprint.  We designed our gates so that at each step we produce the superposition of states prescribed by the diagram.

The procedure for growing the Rokhsar-Kivelson state on an XC cylinder of arbitary width can be naturally generalized from the Hourglass model construction. Here we similarly label the position with a triplet index
$(m, n, i)$, where $m$ denotes the index of the annular stripe, $n$ labels the position of a triangle within the stripe, and
$i = 1, 2, 3, 4, 5, 6$ labels the individual bonds within each triangle. This indexing convention is illustrated in Fig~\ref{fig:XC_growing}.

The state is sequentially grown from smaller to larger $m$. The growth process within each annular stripe is further divided into four steps, which are the generalizations of the same numbered steps used in the hourglass model: 

\begin{enumerate}
  \item For each even $n$, the atoms at positions $(m,n,1)$, $(m,n,2)$, $(m,n,3)$, and $(m,n,4)$ are designated as \emph{target atoms}, marked in purple in Fig.~\ref{fig:XC_growing}. There are two control bits: the first consists of atoms at $(m,n-1,4)$ and $(m,n-1,6)$, controlling the targets at $(m,n,1)$ and $(m,n,2)$; the second consists of atoms at $(m,n+1,1)$ and $(m,n+1,5)$, controlling the targets at $(m,n,3)$ and $(m,n,4)$. The gate operation $U_{2c4t}$ is applied.

  \item For each even $n$, the atoms at $(m,n,5)$ and $(m,n,6)$ are designated as \emph{target atoms}, marked in purple in Fig.~\ref{fig:XC_growing}. There are again two control bits: the first consists of atoms at $(m,n,1)$, $(m,n,2)$, and $(m,n,3)$, controlling the target at $(m,n,5)$; the second consists of atoms at $(m,n,2)$, $(m,n,3)$, and $(m,n,4)$, controlling the target at $(m,n,6)$. The gate operation $U^H_{2c2t}$ is applied.

  \item For each odd $n$, the atoms at positions $(m,n,1)$, $(m,n,2)$, $(m,n,3)$, and $(m,n,4)$ are designated as \emph{target atoms}, marked in purple in Fig.~\ref{fig:XC_growing}. There are two sets of control bits: the first consists of atoms at $(m,n-1,4)$ and $(m,n-1,6)$, controlling the targets at $(m,n,1)$ and $(m,n,2)$; the second consists of atoms at $(m,n+1,1)$ and $(m,n+1,5)$, controlling the targets at $(m,n,3)$ and $(m,n,4)$. The gate operation $U_{2c4t}$ is applied.

  \item For each odd $n$, the atoms at $(m,n,5)$ and $(m,n,6)$ are designated as \emph{target atoms}, marked in orange in Fig.~\ref{fig:XC_growing}. There are again two control bits: the first consists of atoms at $(m,n,1)$, $(m,n,2)$, and $(m,n,3)$, controlling the target at $(m,n,5)$; the second consists of atoms at $(m,n,2)$, $(m,n,3)$, and $(m,n,4)$, controlling the target at $(m,n,6)$. The gate operation $U^H_{2c2t}$ is applied.
\end{enumerate}

\begin{figure}[tbh]
\begin{tabular}{lrll}
    (a)&
    \raisebox{-0.8\height}{\includegraphics[height=4.5cm]{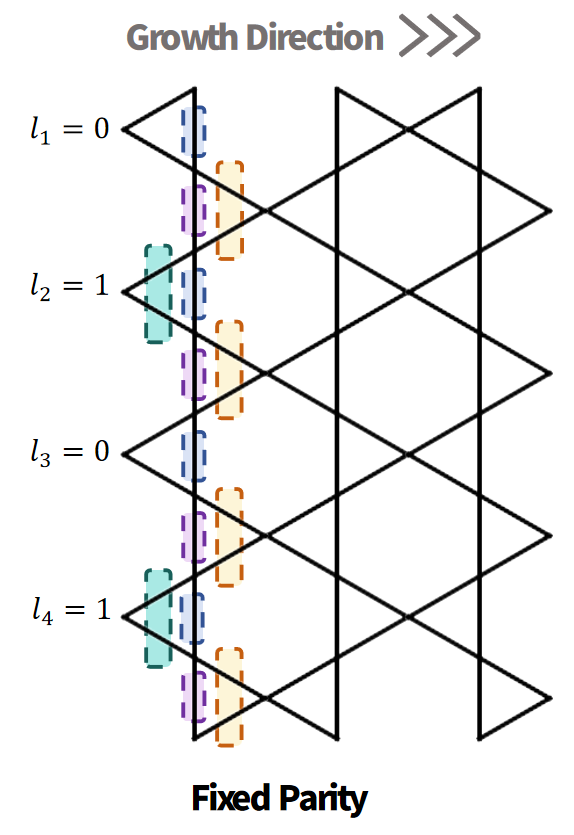}}
    &
    %\hspace{1cm}
    (b)&\hspace{2mm}
    \raisebox{-0.8\height}{\includegraphics[height=4.5cm]{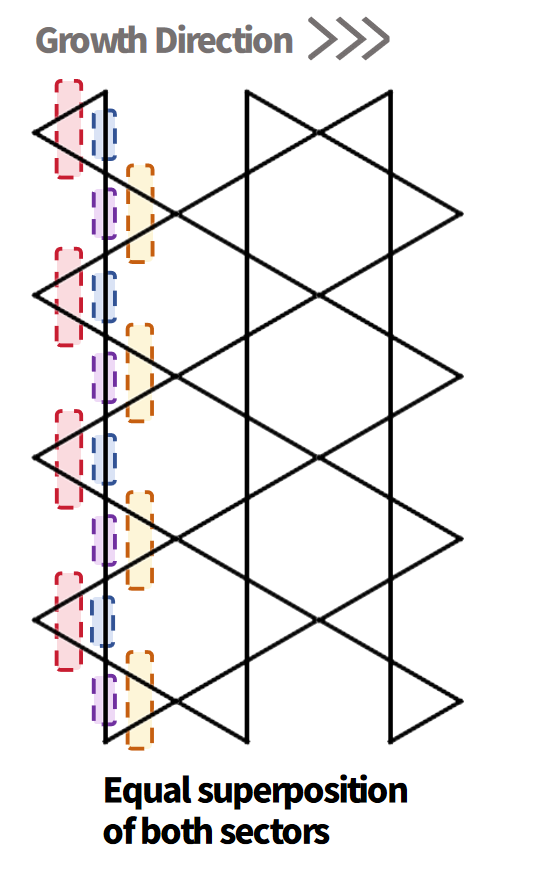}}\\
    %\vspace{5mm}
    \\
    (c)&
    \raisebox{-0.8\height}{\includegraphics[height=4.5cm]{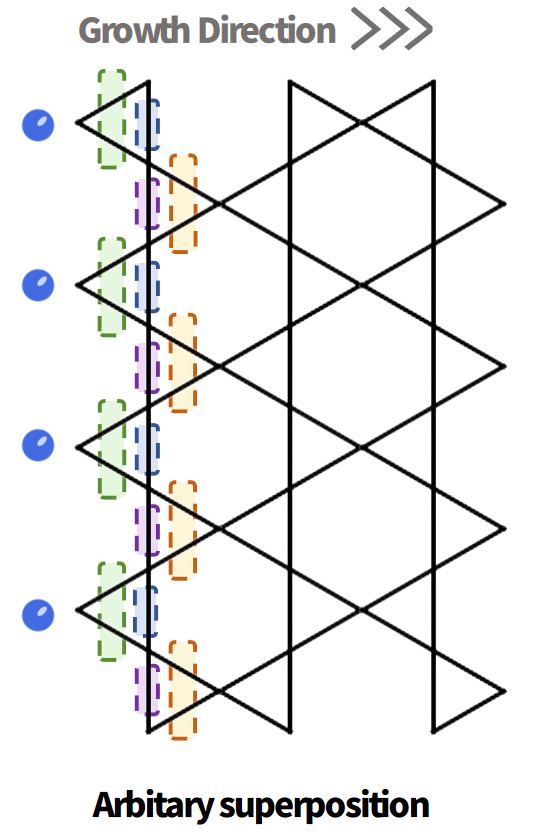}}
%    \hspace{1cm}
&
    (d)&
    \raisebox{-0.8\height}{\includegraphics[height=4.5cm]{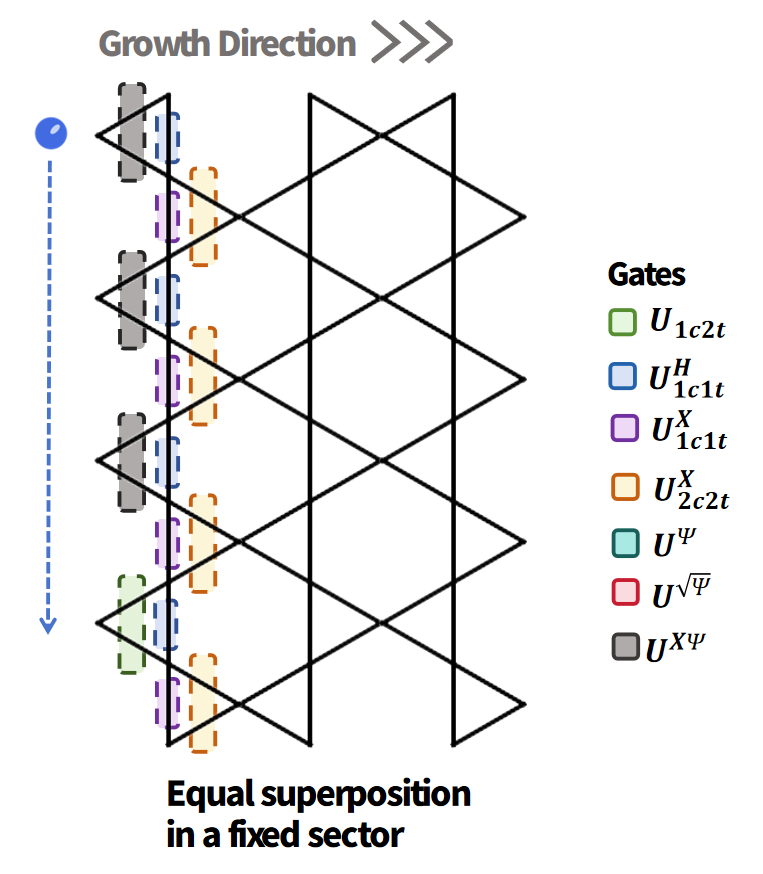}}
\end{tabular}

    \caption{Seeding the dimer covering.
    (a)  For a seed with a fixed edge pattern $L=(l_1,l_2,\cdots l_n)$ one applies $U^\Psi$ gates (cyan) to the pairs of atoms where $l_j=1$. Here we illustrate using $L=(0101)$. Subsequently one applies the same sequence of gates that we use to grow the pattern (shown in blue, purple, and  yellow).  (b) To construct equal weight superposition of  all dimer coverings, one applies $U^{\sqrt{\Psi}}$ gates (red) to all left facing triangles. (c)  Generic superpositions of terminating patterns are formed by placing
    ancilla atoms, shown in blue, to the left of each left-facing triangle. One applies $U_{1c2t}$  gates (green), with the ancilla as control bits. One disentangles the ancilla by targeting each of them with a $U^X_{1c1t}$ gate where the atoms in green form the control bit. (d) One can use a `flying ancilla' to construct a uniform superposition in a single topological sectors. The $U^{X\Psi}$ gate (gray) is used for all triangles except one, where one uses a $U_{1c2t}$ gate, where the ancilla is the control bit, followed by a $U^X_{1c1t}$ gate, where the ancilla is the target.}
    \label{fig:START}
\end{figure}

%\begin{figure}[tbh]
%\includegraphics[width=0.5\textwidth]{start2.png}
%    \caption{
    
%    }
%    \label{fig:START}
%\end{figure}

{
\subsection{Seeding the dimer coverings}\label{sec:seed}
We now describe how to create an initial seed which is used to grow our resonating dimer coverings.  We give our arguments for the YC-$2N$ cylinders, but very similar reasoning applies in the XC-$2N$ case. 
As in the rest of our discussion, we grow from left to right, assuming that initially all atoms are in their ground state.  We introduce several new gates here, whose implementation is described in Appendix~\ref{seedimp}.
%By default, we assume seeding from the left end of the cylinder, and the discussion applies for the right-end case.

As implied by it's matrix-product state representation, the edge of our state naturally has a Hilbert space spanned by $2^N$ basis vectors.  These are labeled by $N$ binary digits, $L=(l_1,l_2,\cdots,l_N)$:  $l_j=1$ if a dimer touches the left-most vertex of the $j$'th left-facing triangle, otherwise $l_j=0$.  We first present a protocol to produce an edge with a fixed pattern, where the binary string $L$ is fixed.  A small change in the protocol allows us to produce a uniform superposition of all possible $L$'s.  We then describe how to make arbitrary superpositions of the various possibilities.  As an important special case, we explain how to generate an equally weighted superposition of all possibilities in one parity sector.
%Finally we pay attention to the two most important cases:  generating an equally weighted superposition of all possible $L$, and generating such a superposition with the constraint that the parity is fixed.

To construct an edge with fixed pattern, the basic strategy is illustrated in Fig.~\ref{fig:START} (a), and requires a new two-qubit gate, $U^{\Psi}$, defined by it's action $U^{\Psi}|00\rangle\to(|10\rangle+|01\rangle)/\sqrt{2}$.  One applies a $U^\Psi$ gate to any pair of atoms in a left-facing triangle for which we want $l_j=1$. No gates are applied to the atoms where $l_j=0$.  One proceeds with the same gate set  that was previously used to grow  the dimer coverings (cf. Fig.~\ref{fig:YC_growing}), shown in blue, purple, and yellow.

To produce a uniform superposition of all dimer coverings, one  applies a $U^{\sqrt{\psi}}$ gate to every pair of atoms in a left-facing triangle.    This gate obeys $U^{\sqrt{\Psi}}|00\rangle\to|00\rangle/\sqrt{2}+(|10\rangle+|01\rangle)/2$. See Fig.~\ref{fig:START} (b).

To produce an arbitrary superposition of terminating patterns we introduce one ancilla atom to the left of each left-facing triangle,
as shown in Fig.~\ref{fig:START} (c).
These ancillae are placed in a quantum state which complements the desired pattern.  For example, if one wants to produce an equal superposition of $L=(1000)$ and $L=(0100)$,  one would take the ancillae wavefunction to be $(|0111\rangle+|1011\rangle)/\sqrt{2}$.
%, one introduces one ancilla for each left-facing triangle, and places those ancillae in the desired superposition. One then applies a $X$ gate to each ancilla, complementing the pattern. 
One applies  $U_{1c2t}$ gates, shown in green, which 
entangle the ancillae with the dimers.  The ancillae act as the control bits, and the atoms in the triangles act as the targets.  At this step there will be an excited dimer on each left-facing triangle if and only if the corresponding ancilla is in its ground state.
Finally, one `erases' the information in the ancillae, by applying  $U^X_{1c1t}$ (controlled-not) gates.  For each gate, the atoms in the triangle act as the control bit, and the corresponding ancilla acts as the target.  This leaves all ancillae in their excited state, disentangled from the dimers. 

There is also a relatively simple gate sequence that we can use to create a uniform superposition of all terminations which have a \emph{fixed parity}.  It involves one `flying' ancilla, which will sequentially interact with each left-facing triangle. See Fig ~\ref{fig:START} (d) for a pictorial illustration.
%It is also useful to be able to generate a seed which is a superposition of 
%all dimer coverings in one topological sector.  The can be done using a `flying ancilla`, that sequentially interacts which each of the left-facing triangles in the leftmost strip.  
We introduce another gate $U^{X\Psi}$,
\begin{align}
U^{X\Psi}&:\left\{
\begin{aligned}
U^{X\Psi} \ket{0,00} &= \frac{1}{\sqrt{2}}  \ket{1,00} + \frac{1}{2}\left(\ket{0,10}
+\ket{0,01}\right) \\
U^{X\Psi} \ket{1,00} &= \frac{1}{\sqrt{2}}  \ket{0,00} + \frac{1}{2}\left(\ket{1,10}
+\ket{1,01}\right) 
\end{aligned}
\right.\nonumber
\end{align}
Here the first bit corresponds to the ancilla, and the other two correspond to the two atoms in the tip of the left-facing triangle,
which we will refer to as the dimer atoms.   
This gate acts similarly to $U^{\sqrt{\Psi}}$, but it entangles the result with the state of the ancilla. The ancilla is flipped when both dimer atoms end up in the ground state.
%Its implementation is given in 
%Appendix~\ref{seedimp}.
%From the definition, the ancilla is flipped when the triangle tip is not touched by a dimer.

%This can be implemented in two steps.  One first places the two dimer atoms within eachother's blockade, and turns on $\Omega$ for a time corresponding to a $\pi/2$ pulse, taking $|\sigma,00\rangle\to |\sigma,00\rangle/\sqrt{2}+\left(|\sigma,10\rangle+|\sigma,01\rangle\right)/2$.  Next one places all three atoms within their blockade radius an apply a $\pi-\Omega$ pulse to the ancilla, which flips its state if and only if neither of the dimer atoms are not excited.

%One starts with the ancilla in a definite state, say $|0\rangle$. 
%For clarity, we assume that $N$ (the total number of triangles) is even, and that we aim to engineer an equal-weight superposition over all configurations with even parity. In this case, the ancilla is initialized in the state $\ket{0}$.
%For clarity, we assume that $N$ denotes the total number of left-facing triangles, and that the ancilla is initialized in the state $\ket{a}$. The value of $a \in \{0,1\}$ is chosen according to the desired topological sector, as explained below.

One begins by placing the ancilla in one of the logical basis states $|a_0\rangle$, with $a_0=0,1$.
The ancilla is moved to the top-most triangle, and a  $U^{X\Psi}$ is applied.  One moves the ancilla the next triangle, and a second $U^{X\Psi}$ gate is applied again. This process is repeated sequentially for all triangles except the bottom one.  After these $N-1$ steps, the system contains a superposition of dimers.  For each dimer pattern, the ancilla is in a state $|a\rangle$ with
%\begin{equation}
$a\equiv(N-1)+a_0+n_d\, ({\rm mod}\, 2),$
%\end{equation}
where $n_d$ is the number of dimers.   
On the last triangle, one  applies a $U_{1c2t}$ gate, where the ancilla acts as the control. This will produce a dimer only if the ancilla is in its ground state.  Thus the system will only contain dimer patterns whose parity is the same as $N+a_0$.
Finally, one places the ancilla in a definite state by applying a  $U^X_{1c1t}$ gate, where the dimer atoms act as the control and the ancilla as the target. This 
%sequence ensures that the ancilla returns to its ground state $\ket{0}$ and is 
disentangles it from the dimer degrees of freedom. The initial state $\ket{a_0}$ is chosen according to the circumference $N$ and the wanted topological sector.

{\color{blue}
%Typically, if one aims to prepare the Rokhsar-Kivelson state as a uniform superposition over both topological sectors, a uniform seed is used. If the goal is to obtain the state within a specific sector, one instead starts from the uniform seed confined to that sector. A fixed edge pattern provides a natural initialization for the matrix product state, and it explicitly encodes the edge degrees of freedom.
}

}

\section{Probes}\label{sec:probes}
After creating the desired state, one would like to perform experiments which confirm that the procedure has been successful, and which probe the exotic properties of these resonating dimer configurations. The  basic strategies were largely developed in \cite{PhysRevX.11.031005}, and experimentally demonstrated in \cite{Semeghini2021}.  There they were not algorithmically generating the dimer configurations, but instead quasi-adiabatically evolving a Hamiltonian into one whose ground state shared the key properties of our superposition of dimer coverings.

Measuring the $Z$-strings is straightforward.  One simply performs a projective measurement on every single atom, determining if it is in the ground state or an excited state.  Averaging over many of these measurements allows one to determine the expectation value of the string operators.  To measure the $X$-strings one first performs a gate which maps $X$ into $Z$ \cite{PhysRevX.11.031005,Verresen2022,Semeghini2021}:  One arranges the atoms so that there is blockade between every set of 3 atoms in each triangle, but no blockade between atoms in other triangles.  One then set $\Delta=0$, and pulses $\Omega(t)$ such that $\int \Omega(t)dt=\frac{4\pi}{3\sqrt{3}}$.  This maps the $Z$ and $X$ segment operators in Fig.~\ref{fig:kagome_lattice} onto one-another.  From measuring $Z$-strings in the new basis one infers the expectation values of the $X$-strings in the original basis.  

These same techniques allow one to apply gates consisting of $Z$-strings or $X$-strings, which create pairs of quasiparticle defects.
Traditionally, the defects formed at the end of $X$-strings are referred to as $e$ anyons, while those formed at the end of $Z$ strings are known as $m$ anyons. The presence of a quasiparticle can be detected by measuring a string operator that encloses it.  A $Z$ loop containing a $e$ particle, or a $X$ loop containing a $m$ particle, gain an extra $-1$.

The $e$ and $m$ defects can be moved around with strings that have one end at the defect, and another at the target location.  These are mutual semions, as moving one about the other multiplies the wavefunction by $-1$.    
It would be particularly exciting to measure these mutual statistics.  Directly measuring this phase is highly nontrivial, as it requires determining the relative phase between two states, $|\Psi_0\rangle$ and $|\Psi_x\rangle$. Here $|\Psi_0\rangle$ is a state which contains both an $e$ and $m$ defect, and $|\Psi_x\rangle$ is the same state after the $e$ defect has been moved along a path encircling the $m$ defect.  The path should contain an even number of sites, so that no phase factor would be acquired in the absence of the $m$ defect.  The mutual statistics correspond to the mathematical statement that $|\Psi_x\rangle=-|\Psi_0\rangle$.

Indirectly one can infer these statistic by simply measuring the $Z$-string which moves the $e$ defect around the $m$ defect.  The state $|\Psi_0\rangle$ should be an eigenstate of this operator, with eigenvalue $-1$ -- while in the absence of the $m$ defect it would have eigenvalue $+1$.  This sign change is proof of the mutual statistics.

As has been demonstrated by a number of related experiments \cite{Satzinger2021,Roberts2024}, a more direct approach to measuring these statistics is to entangle the many-body state with an ancilla.  This requires that one can apply a controlled-$Z$-string.  A controlled-$Z$-string differs from a $Z$-string in that the phase factors are applied  if and only if the ancilla is in its ground state.
%.  If the ancilla is in its ground state the $e$ defect is moved in a circle around the $m$ defect.  If the ancilla is excited, then no operation is performed.
%-string that would move the $e$ defect, but replace the constituent $Z$ gates with controlled-$Z$ gates -- which produce the phase factor contingent on an ancilla atoms being in the ground state.  
Such an operation can be constructed from the
controlled-$Z$ gates which were demonstrated in \cite{Evered2023}.  To perform the statistics measurement, one first places the ancilla in the superposition $\frac{\ket{0}+\ket{1}}{\sqrt{2}}$.  One then moves the ancilla along the path, sequentially applying control-$Z$ gates -- effectively moving a defect contingent on the state of the ancilla.  This produces an entangled state $\ket{\Psi}=(|1\Psi_0\rangle+|0\Psi_x\rangle)/\sqrt{2}$, where the first symbol is the state of the ancilla, and the second is the state of all other atoms.  
One then applies a Hadamard gate to the ancilla, and measure its state.  

To elaborate on this procedure, suppose the mutual phase factor is $e^{i\phi}$, i.e. $|\Psi_x\rangle=e^{i\phi}|\Psi_0\rangle$. After the implementation of the Hadamard gate, the state becomes $H|\Psi\rangle = \frac{1}{2} \left[ (e^{i\phi} + 1)|0\rangle + (e^{i\phi} - 1)|1\rangle \right] |\Psi_0\rangle$. The probability of measuring the ancilla in the $|1\rangle$ state is $\frac{1}{2}(1 - \cos\phi)$. Thus, the result reveals the phase accumulated by moving the defect.  In practice one applies additional gates to the ancilla and determines $\phi$ as the shift of Ramse interference fringes \cite{Satzinger2021,Roberts2024}.
%{\color{red}  After you create the state -- how do you probe it?  How do you prove that you have created something with all of these fun exotic properties? -- Answer: Measure string operators}

\section{Summary}\label{sec:summary}
We have given a protocol for using Rydberg atom arrays to generate the kagome lattice Rokhsar-Kivelson state on a cylinder.  This is an exotic $\mathbb{Z}_2$ spin liquid, 
which is an eigenstate of two types of loop operators.  The state
hosts topological order and quasiparticle excitations which are mutual semions.    We described how to probe this exotic physics.

While Rydberg atom arrays are the most natural platform, our algorithm can be implemented in other physical systems such as transmon arrays or trapped ions. As described in Appendix~\ref{sec:gate-operation}, for the YC-$2N$ geometry one only needs to be able to implement standard single-qubit gates ($X$, $Z$, $H$), standard two-qubit gates (controlled-$X$ and controlled-$H$), and the Toffoli (double controlled not) gate.

Our construction involves `growing' the state along the long axis of the cylinder.  At each stage one implements a series of local gates which extend the Rokhsar-Kivelson state.  The action of these gates are directly related to a matrix product state (MPS) representation of the superposition of dimer covering.  The gates directly create the branching configurations which are encoded in the MPS.  Our construction works for a reconfigurable planar arrangement of atoms:  One does not need to actually construct a 3D cylindrical arrangement of atoms.  {  In Appendix~\ref{sec:Torus} we give an alternative approach, and show how to generalize our construction to a torus.}

 We gain insight into resonating dimer coverings by considering the limit of narrow cylinders. 
Depending on the orientation of the lattice on the narrow cylinder, 
 the state simplifies to either a crystal of resonating bonds (with no long-range entanglement), or an analog of the spin-1 AKLT state.  This small diameter limit is well suited to experiments, as it requires fewer atoms and gates, yet still produces non-trivial physics.

\section*{Acknowledgements}

XCW would like to thank Dong E. Liu for productive suggestions, and JL Dai, RL Li, T Xu, MY Loong, YS Huang for helpful discussions.  We thank Daniel Ranard for bringing to our attention the recent progress on sequential circuits, and Xie Chen for nice discussion on this topic.
This material is based upon work supported by the National Science Foundation under Grant No.  PHY-2409403.

\appendix
\section{Construction of dimer coverings on YC-$2N$ strips}\label{sec:proof_of_modules}

In Sec.~\ref{sec:Larger_cylinder}, we labeled the dimer configurations on a YC-$2N$ strip by specifying $L,R,u$.  The sequence of binary digits $L=(l_1,l_2,\cdots l_N)$ label the left pointing triangles.  If there is a dimer touching $j'th$ triangle point, then $l_j=1$, otherwise $l_j=0$.  The digits composing $R=(r_1,r_2,\cdots r_N)$ similarly label the right pointing triangle,   $u=1$ or 0 depending on  if there is a dimer that touches that point from below.  See Fig.~\ref{fig:module_conn_1} for several examples.  Here we show how to construct a dimer covering from these labels, demonstrating that they uniquely specify the state.  While we focus on YC-$2N$ strips here, the same construction also works for the XC-$2N$ strips.

\begin{figure}[tbh]
\includegraphics[width=0.45\textwidth]{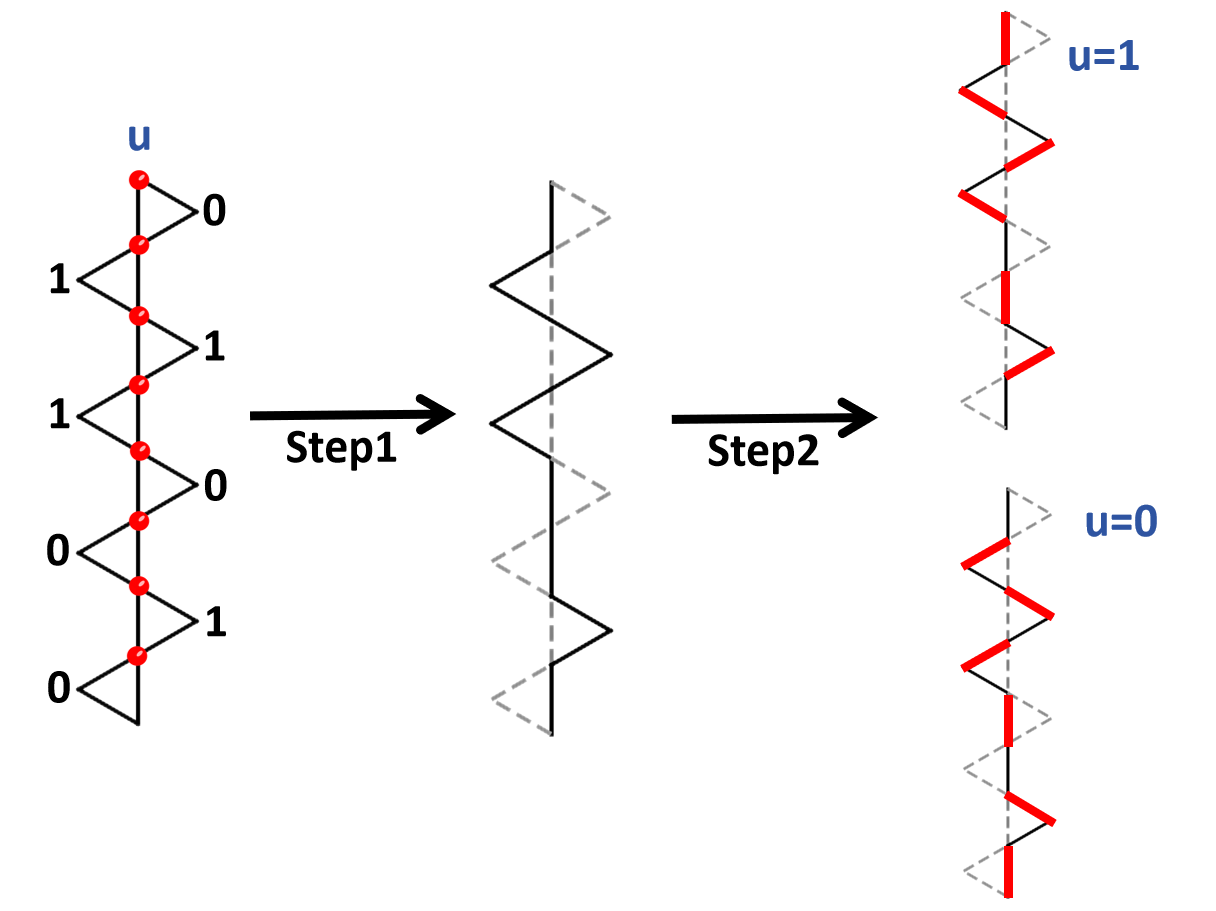}
    \caption{Illustration of how the unique dimer covering configuration is determined from the input $(L, R, u)$.  
We consider the example of a YC-8 strip, with $L = 1100$, $R = 0101$, and $u = 0$ or $1$, as shown in the left panel, where the \emph{middle vertices} are marked with red dots. The bottommost vertex is not marked since it's identified with the topmost one.
\textbf{Step 1} marks the forbidden positions for dimers based on $L$ and $R$; these positions are indicated by gray dashed lines in the middle panel.  
\textbf{Step 2} decorates every other link in this path, depending on the value of $u$. 
%lifts the two-fold degeneracy using $u$: specifically, $u$ determines whether the topmost link in the unmarked string contains a dimer, as shown in the right panel.  
The resulting dimers are represented by thick red lines.
}
    \label{fig:MiddlePoint}
\end{figure}

We begin by showing that for a consistent dimer covering, 
 $L$ and $R$ must have the same parity.  Let ${\cal L}=\sum_{j} l_j$ and ${\cal R}=\sum_j r_j$ be the total number of dimers which touch the left and right vertices, and  let ${\cal M}$ be the number of dimers which do not touch any of the edge vertices -- for the YC model these are all on the vertical bonds.  We can constrain these numbers by noting that each of the $V_m=2N$ middle vertices, shown as red dots in Fig.~\ref{fig:MiddlePoint}, touch exactly one dimer.  Since each left or
right dimer touches one middle vertex, and each middle
dimer touches two middle vertices, we obtain the relationship ${\cal L+R}+2{\cal M}=V_m=2N$.  We therefore deduce that $\cal L+\cal R$ is even.

To generate the dimer pattern we use a two-step process, illustrated in
Figure~\ref{fig:MiddlePoint}.
%illustrates the construction process.  A YC-8 strip is shown with an fixed set of indices.  
In the first step we mark the bonds which are constrained by $L$ and $R$ to not be occupied:  If $l_j=0$, then neither of the edges adjacent to the vertex can support a dimer;
If $l_j=1$, then one of the two adjacent edges must host a dimer, which forces the  edge opposite to the vertex  to remain empty.  As shown in the figure, this leaves a path of potential bonds which snakes from the top to the bottom of the strip.  The length of this path is ${\ell}=\sum_{j}(l_j+1) +\sum_{j}(r_j+1)$, which is even since $L$ and $R$ have the same parity.  In the second step one simply places a dimer on every other bond of this path.  If $u=1$ one begins with a dimer on the top segment, while if $u=0$ one begins with an empty segment.

\section{Realization of Gate Operations}\label{sec:Realization}
In this Appendix, we explain how the six types of \emph{gate operations} introduced in Sec.~\ref{sec:Gates} can be physically realized. In Appendix~\ref{sec:adiabatic} and \ref{sec:Coherent} we present adiabatic and non-adiabatic implementations for these gates  in a Rydberg atom array platform.  In Appendix.~\ref{sec:gate-operation} we give implementations based on digital circuits.  One could also design dissipative gates for this purpose \cite{Harrington2022}.

%We begin by briefly reviewing the two distinct approaches to quantum gate implementation. We illustrate the two different techniques by considering simple $X$ and Hadamard gates on a single qubit, and we introduce the realization of controlled gates.  We then describe the implementation of the many-qubit gates in Sec.~\ref{sec:gate-operation}.

%, and \(V\) accounts for interaction terms when multiple atoms are involved.

\subsection{Adiabatic Gates}\label{sec:adiabatic}

For a time-dependent Hamiltonian \(H(t)\), the system evolves under the unitary operator
\[
U(t) = \mathcal{T} \exp \left( -\frac{i}{\hbar} \int_0^t H(t') \, dt' \right).
\]
If the system is initially in an eigenstate of $H(0)$, and $H$ varies slowly enough, it will evolve into the corresponding eigenstate of $H(T)$, where $T$ is the total gate time.  We describe how this \emph{adiabatic principle} can be used to apply single qubit $X$ and Hadamard gates.  We then describe how to extapolate to the many-qubit gates from Sec.~\ref{sec:Gates}.
%Utilizing the \emph{Adiabatic Theorem}, one can design adiabatic quantum gates. We first introduce the implementation of basic adiabatic gates on the Rydberg atom platform, including single-qubit X gate and Hadamard gate, and controlled gates.

For a single Rydberg atom, labeled $\alpha$, the system is governed by the Hamiltonian
\begin{equation}
H = \frac{\Omega(t)}{2}\sigma^x - \Delta(t) n,
%+ V,
\label{hamatom}
\end{equation}
where \(\Omega(t)\) and \(\Delta(t)\) denote time-dependent Rabi frequency and detuning, \(n=|1\rangle\langle 1|\) is the number operator and $\sigma^x=|1\rangle\langle 0|+|0\rangle\langle 1|$.  For the multi-atom case there will also be an interaction term, as written in Eq.~(\ref{ham}).

\begin{figure}[tbh]
(a)
\raisebox{-0.8\height}
{\includegraphics[height=3.4cm]{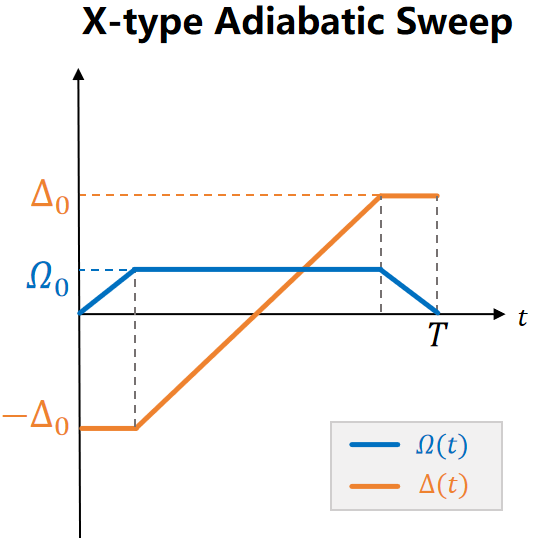}}
(b)
\raisebox{-0.8\height}
{\includegraphics[height=3.4cm]{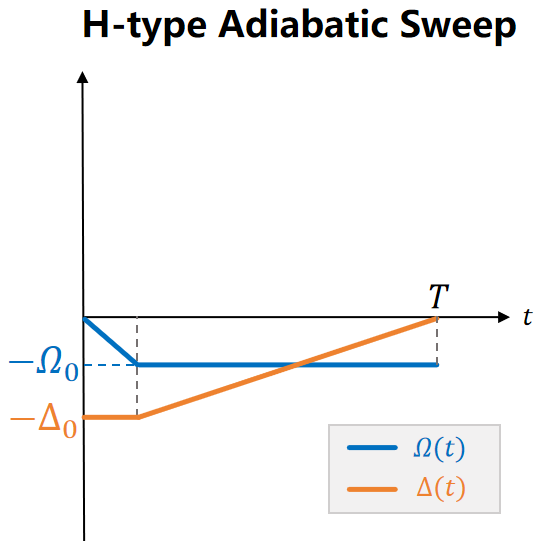}}

    \caption{(a) Illustration for an X-type sweeping pulse. Here we choose a linear variation of the parameter as an example, though other pulse patterns can be used in actual experiments. (b) Illustration for an H-type sweeping pulse.  }
    \label{fig:HXtype}
\end{figure}

To implement a single atom X gate, we consider a time dependent Hamiltonian that starts as $H(0) = -\Delta_i n$ and ends as $H(T) = -\Delta_f n$, with $\Delta_i<0$ and $\Delta_f>0$.  For simplicity we can take $\Delta_i=-\Delta_0$ and $\Delta_f=\Delta_0$. As sketched in  
Fig ~\ref{fig:HXtype} (a),
one first ramps the coupling $\Omega$ to a positive value $\Omega_0$.  One then gradually
sweeps $\Delta(t)$ from $\Delta_i$ to $\Delta_f$.  One finally ramps $\Omega(t)$ to zero,  turning off the dynamics.
This process transfers the atom from $|0\rangle$ to $|1\rangle$, realizing a bit-flip operation. At intermediate times adiabaticity requires that the gap $\Delta_{\rm gap}(t)=\sqrt{\Delta^2+\Omega^2/4}$ must be sufficiently large compared to the rate of change of the Hamiltonian parameters. If $\Delta_0\gg\Omega_0$, one can use the Landau-Zener model to approximate the dynamics \cite{landau1932theorie,zener1932non}, and the probability of a non-adiabatic transition is exponentially small in the ratio ${\Omega_0^2}/{\dot{\Delta}}$, where $\dot{\Delta}=d\Delta/dt\approx \Delta_0/T$. If we take $\Delta_0$ to be a numerical factor times $\Omega$, then adiabacity requires $T\gg 1/\Omega_0$.  There are a number of strategies to speed up these gates or make them more robust against noise 
\cite{STAreview}.
%\cite{Unanyan1997,Berry2009,shortcuts,shortcuts2,shortcut3,Cao2013,Daems}.

To implement the Hadamard gate, we set $H(0) = -\Delta_i n$ and $H(T) = \frac{\Omega_f}{2}\sigma^x$, where $\Delta_i=-\Delta_0<0$, and $\Omega_f=-\Omega_0<0$.  %One could equally well take both $\Delta_i>0$ and $\Omega_f>0$ -- the only requirement is that they have the same sign. {\color{red} WHY? we require the system to start at $\ket{0}$} 
%The sign of $\Omega$ is set by the phase of the drive relative to a reference.
As shown in Fig.~\ref{fig:HXtype} (b), one 
one first ramps $\Omega(t)$ from 0 to $\Omega_f$.  One then ramps
$\Delta(t)$ from $\Omega_i$ to 0.
%the dynamical term to be a negative value $-\Omega_0$, then gradually evolve $\Delta_\alpha(t)$ from $-\Delta_i$ to 0, as shown in Fig ~\ref{HXtype} (b). It 
This process transfers the atom from $\ket{0}$ to $\frac{\ket{0}+\ket{1}}{\sqrt{2}}$. 

We implement controlled gates using the dipole-dipole interaction between Rydberg atoms. By placing the target atoms within the blockade radius of the control atom and  adiabatically evolving the Hamiltonian for the target atoms, a gate is realized. If the control atom is in the excited state $\ket{1_c}$, the target atoms remain in their ground state $\ket{0}$.

%For our example $X$ gate, we start with $\Delta_\alpha(0)<0$ and $\Omega_\alpha(0)=0$ so that the initial ground state is $|0\rangle$.  We end with $\Delta_\alpha(T)>0$ and $\Omega_\alpha(T)=0$ to the final ground state is $|1\rangle$.  At intermediate times adiabaticity requires that the gap $\Delta_{\rm gap}(t)=\sqrt{\Delta_\alpha^2+\Omega_\alpha^2/4}$ must be sufficiently large compared to the rate of change of the Hamiltonian parameters.  In a typical Landau-Zener treatment, the adiabatic parameter is $\lambda=\Delta_\alpha'(t)/\Omega_\alpha^2(t)$, and the probability of a non-adiabatic transition is exponentially small in $1/\lambda$.  Thus the gate requires a time $T\gg 1/\Omega_\alpha$.

The advantage of the adiabatic protocol lies in its robustness to the specific pulse shape; what matters is the initial and final states and the adiabaticity of the evolution process. Moreover, the adiabatic scheme offers a convenient way to design quantum gates involving multiple atoms, without requiring the more complex quantum circuits that might otherwise be necessary. To implement our six gate operations using adiabatic evolution, we  use the two  parameter-sweeping patterns that we introduced in the single atom case, and which are shown in Fig.~\ref{fig:HXtype}.  We refer to these as   $X$ and $H$ sweeps. 
%(This figure is only an example. The actual pulse shapes need not be linear.) 
%Both sweeping types 
In both cases we
start from the same initial conditions $(\Omega(0) = 0, \Delta(0) < 0)$.  We envision that each of the target atoms feel the same $\Omega(t)$ and $\Delta(t)$, while $\Omega=0$ for the control atoms.
In the $X$ sweeps
%, similar to the single-qubit X gate, 
the evolution ends at $(\Omega(T) = 0,\Delta(T) > 0)$, and the $H$ sweeps
%.
%The H-type sweeping, similar to the Hadamard gate, 
end at $(\Omega(T) > 0,\Delta(T) = 0)$.

\begin{figure}[tbh]
(a)
\raisebox{-0.8\height}
{\includegraphics[height=3.5cm]{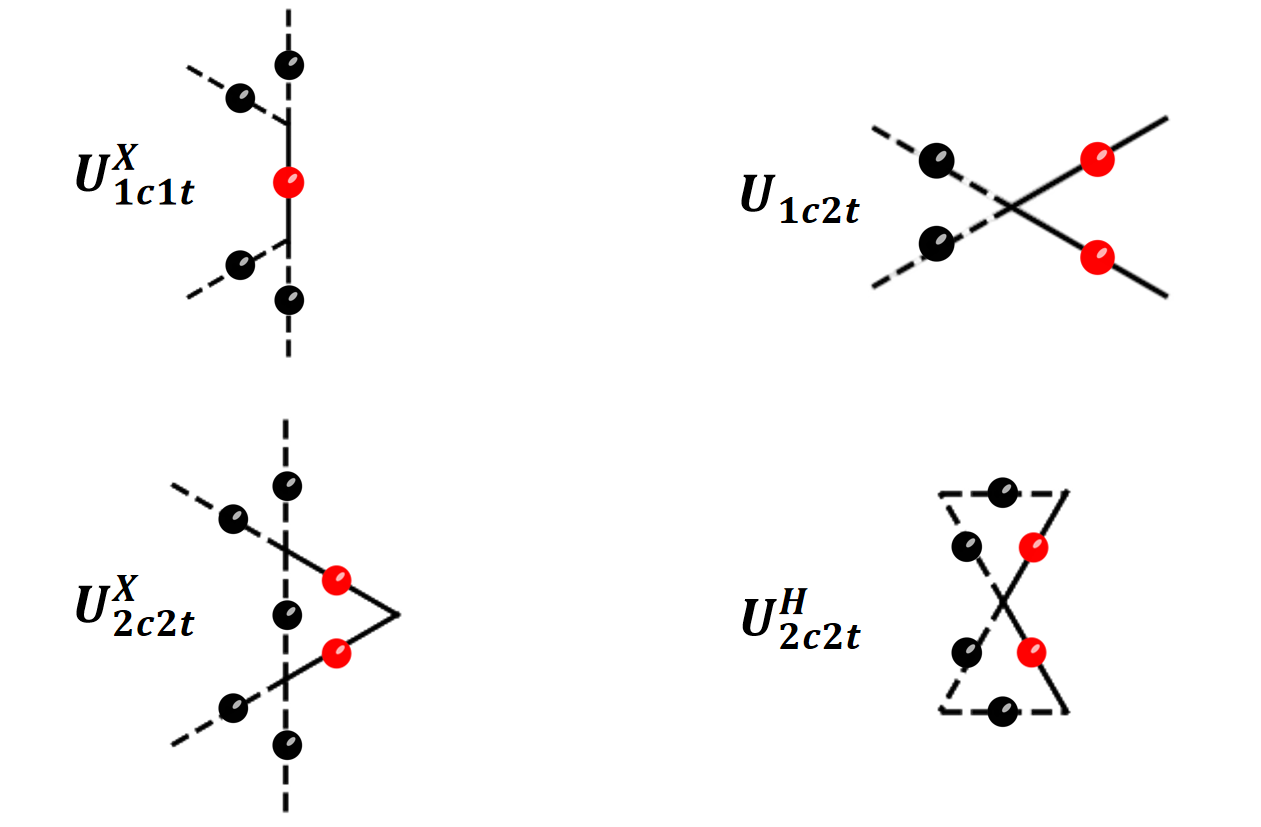}}
(b)
\raisebox{-0.8\height}
{\includegraphics[height=2.2cm]{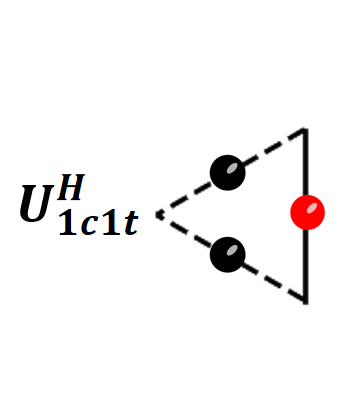}}

    \caption{ Spatial arrangement of atoms for adiabatic implementation of the (a) $U^X_{1c1t}$,$U_{1c2t}$,$U^X_{2c2t}$,$U^H_{2c2t}$ and (b) $U^H_{1c1t}$ gates.  Control atoms are in black, and target atoms in red.
    %are the same as the original configurations in Figs.~\ref{fig:flipy} through \ref{fig:noflipx}. 
    All atoms which share a vertex blockade one-another. 
    Gates in (a) use the $X$ sweep pattern in Fig.~\ref{fig:HXtype} (a), while those in (b) use the $H$ pattern in Fig.~\ref{fig:HXtype} (b). Some of these geometric arrangements can also be used for non-adiabatic gate implementation.   
%    These gate operations are realized by the X-type sweeping. (b) The spatial arrangement of $U^H_{1c1t}$ is the same as the original configuration. It is adiabatically realized by the H-type sweeping.  
    }
    \label{fig:Other4}
\end{figure}

\begin{figure}[tbh]
(a)
\raisebox{-0.8\height}
{\includegraphics[height=2cm]{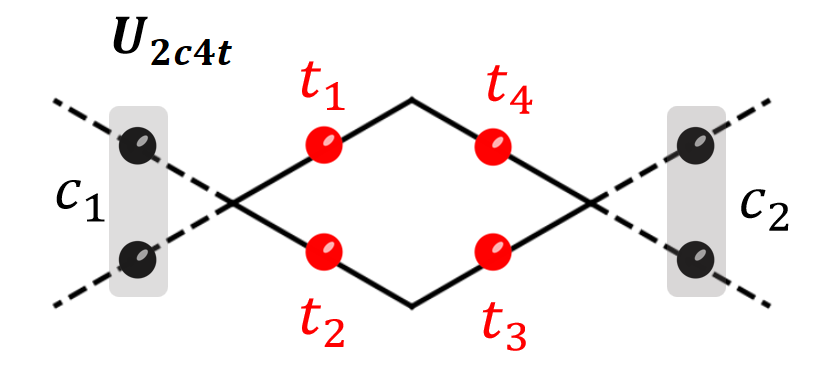}}
(b)
\raisebox{-0.8\height}
{\includegraphics[height=1.8cm]{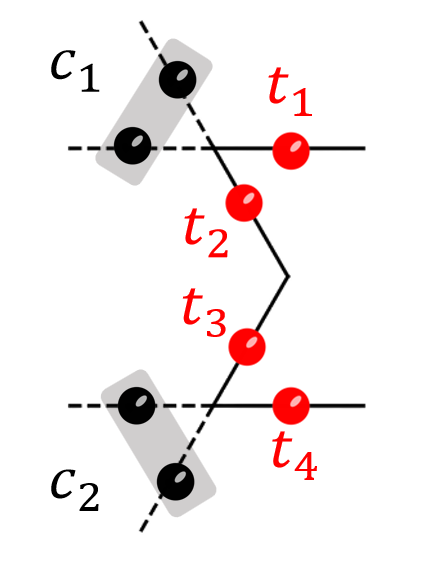}}

    \caption{(a) Arrangement of target and control atoms for adiabatic approach to implementing the $U_{2c4t}$ gate.  All atoms which share a vertex blockade one-another. Note, this configuration is different from the naive locations of the atoms on the bonds of the kagome lattice, shown in (b).  }
    \label{fig:U2c4t}
\end{figure}

The $H$ sweep is used for $U_{1c1t}^H$, and the X sweeps are for  
%the constructions of 
the rest five gate operations. We put the target atoms together with several control atoms in some specific spatial arrangements to achieve wanted blockades.
%, and adiabatically sweep the parameters of all target atoms. 
For most cases, the spatial arrangement is identical to the pattern of bonds in the kagome lattice, as 
%in Figs.~\ref{fig:flipy} through \ref{fig:noflipx}, as 
shown in Fig ~\ref{fig:Other4}. The only exception is $U_{2c4t}$, where instead of the configuration shown in Fig ~\ref{fig:U2c4t} (b), we need a spatial arrangement shown in Fig ~\ref{fig:U2c4t} (a) to impose extra constraints on $(t_1,t_4)$.

An important feature of these pulse sequences is that at all times the gap between eigenstates is of order $\Omega$ (or $\Delta$), and these gaps are independent of the total size of the system.  This feature should be contrasted with adiabatic sweep algorithms that homogeneously drive a many-body system through a continuous quantum phase transition between two phases \cite{Ebadi2021,Semeghini2021}.  At such a phase transition the gap must vanish in the thermodynamic limit.  By manipulating a small number of atoms at a time, we avoid this challenge.

One caution is that during the adiabatic gates the state accumulates both dynamical and geometric phases.  The gates need to be carefully engineered so that these phases do not become imprinted on the superposition of resonating dimers.  The non-adiabatic protocols in Appendix~\ref{sec:Coherent} avoid this challenge.

\subsection{Nonadiabatic Gates}\label{sec:Coherent}

As an alternative to the adiabatic approach, our quantum gates can be implemented via non-adiabatic protocols 
where the pulses $\Delta(t)$ and $\Omega(t)$ are carefully timed so that the system makes a Rabi transition from the initial to final state.  These non-adiabatic gates are typically much faster.

We first introduce the protocol for implementing several fundamental gates that will serve as essential building blocks in the subsequent designs. These basic gates include the single-qubit \( X \) gate, the Hadamard gate, and controlled variants. In addition, we introduce a special gate, which we refer to as the \( \Psi \) gate, as it transforms the \(|00\rangle\) state into the Bell state \(\ket{\Psi} = \frac{1}{\sqrt{2}} ( \ket{01} + \ket{10} )\). Although the \( \Psi \) gate can, in principle, be constructed from \( X \) and Hadamard gates, in the Rydberg atom platform there is  a simpler 
%non-adiabatic protocol for its 
direct implementation.

The \( X \) and \( H \) gates can be understood as specific rotations of the state vector on the Bloch sphere. We begin by setting \(\Delta = 0\), so that the Hamiltonian is $H(t) = \frac{\Omega(t)}{2} \sigma^x$. 
% This leads to the time evolution operator:
% \begin{align}
% U(t) &= e^{-i \int_0^t H(\tau) \, d\tau} 
% = 
% \begin{bmatrix}
% \cos \phi & -i \sin\phi \\
% -i \sin\phi & \cos\phi
% \end{bmatrix}.\\
%\begin{bmatrix}
%\cos\left( \frac{1}{2} \int_0^t \Omega(\tau) \, d\tau \right) & -i \sin\left( \frac{1}{2} \int_0^t \Omega(\tau) \, d\tau \right) \\
%-i \sin\left( \frac{1}{2} \int_0^t %\Omega(\tau) \, d\tau \right) & \cos\left( \frac{1}{2} \int_0^t \Omega(\tau) \, d\tau \right)
%\end{bmatrix}.\\
%\phi&=\frac{1}{2} \int_0^t \Omega(\tau) \, d\tau 
%\end{align}
Under this evolution, the initial state $|0\rangle$ evolves to 
{$|\psi(t)\rangle = \cos(\phi/2) |0\rangle-i\sin(\phi/2) |1\rangle$}, where
{
\begin{align}
\phi&=\int_0^t \Omega(\tau) \, d\tau 
\end{align}
}%
To remove the unwanted phase factor of \(-i\) on the $|1\rangle$ component, one can subsequently set \(\Omega = 0\) and turn on a detuning \(\Delta(t)\) such that $\int \Delta(t) \, dt = \frac{\pi}{2}.$  
%Alternatively, one adjusts the phase of the driving pulse so that $H(t)\propto \sigma^y$.
The $X$ and $H$ gate corresponds to taking {$\phi={\pi}$ and ${\pi/2}$}.  

% Therefore, the full implementation of the gates proceeds as follows:
% \begin{itemize}
%     \item \textbf{\( X \) gate}: First apply a pulse such that $\int \frac{\Omega(t)}{2} \, dt = \frac{\pi}{2}$,
%     followed by a detuning pulse satisfying $\int \Delta(t) \, dt = \frac{\pi}{2}.$ This gives the final state $\ket{1}=(0,1)^T$.
%     \item \textbf{Hadamard (\( H \)) gate}: Apply a pulse such that $\int \frac{\Omega(t)}{2} \, dt = \frac{\pi}{4}$,
%     followed by a detuning pulse satisfying $\int \Delta(t) \, dt = \frac{\pi}{2}$.This gives the final state $\ket{+}=(1/\sqrt{2},1/\sqrt{2})^T$
% \end{itemize}

%The implementation of controlled gates in the same as that in adiabatic protocols. We place the targets inside the blockade range of control atoms, then change the parameters of targets.

Controlled gates are implemented in the same manner as in the adiabatic protocols.  We place the targets inside the blockade range of control atoms, before applying the pulse sequence. 

To implement the \(\Psi\) gate, we place two target atoms within each other's blockade radius to suppress the \(\ket{11}\) state. We set \(\Delta = 0\) and apply a time-dependent Rabi drive \(\Omega_\alpha(t)\). Under these conditions, the Hamiltonian reads $H(t) = \frac{\Omega(t)}{2}(\sigma^x_1+\sigma^x_2) +Vn_1n_2$, where $V$ is extremely large.  The accessible Hilbert space is spanned by the state $|00\rangle$ and the Bell state $\Psi=(|10\rangle+|01\rangle)/\sqrt{2}$.  The Hamiltonian acts as $H\ket{00} =(\Omega/\sqrt{2})\ket{\Psi}$ and $H\ket{\Psi} =(\Omega/\sqrt{2})\ket{00}$.  Hence to the $\Psi$ gate is implemented by a pulse with $
{
\int \sqrt{2}{\Omega(t)} \, dt = { \pi}}$, followed again by a corrective phase pulse $\int \Delta(t) \, dt = \frac{\pi}{2}.$

%It can be verified that under these conditions, the Hamiltonian acts as follows:
%\[
%H(t)\ket{00} = \frac{\sqrt{2}}{2} \Omega(t) \ket{\Psi}, \quad H(t)\ket{\Psi} = \frac{\sqrt{2}}{2} \Omega(t) \ket{00}.
%\]
%
%Since the system is initially in the \(\ket{00}\) state, its dynamics are confined to the two-dimensional subspace spanned by \(\ket{00}\) and \(\ket{\Psi}\). Within this subspace, the Hamiltonian can be written as an effective two-level system,$H_{\text{eff}}(t) = \frac{\sqrt{2}}{2} \Omega(t) \, \sigma_x$. We then only need to implement an X gate with this effective Hamiltonian. Thus:

%\begin{itemize}
%    \item \textbf{\( \Psi \) gate}: Apply a pulse such that $\int \frac{\Omega(t)}{\sqrt{2}} \, dt = \frac{\pi}{2}$,
%    followed by a detuning pulse satisfying $\int \Delta(t) \, dt = \frac{\pi}{2}.$ This gives the final state $\ket{\Psi}$.
%\end{itemize}

With these fundamental gates as building blocks, we construct the six gate operations:
%The $U^X_{1c1t}$ and $U^H_{1c1t}$ gates are controlled X operations -- with the atoms arranged in the manner shown in Fig.~\ref{fig:Other4} (a) and (b).  The $U_{1c2t}$ is a controlled $\Psi$, with the configuration in Fig. \ref{fig:Other4} (a).  The $U

%For example, to execute an $X$ gate on a single qubit we take $\Delta_\alpha=0$, and choose $\Omega_\alpha(t)$ such that $\int_{0}^T (\Omega(t)/2)dt=\pi$. This is typically faster than the adiabatic gate protocol, but requires careful timing and calibration.  
%A Hadamard gate is similar to the X-gate, but with $\int_{0}^T (\Omega(t)/2)dt=\pi/2$.  As with the adiabatic gates, blockade effects are be used to control the process:  If the target is within the blockade radius of an excited control atom, the pulse $\Omega(t)$ cannot change its state.    Consequently, the $U_{1c1t}^H$ and $U_{1c1t}^X$ gates are straightforward to implement.  We briefly discuss the implementation of the multi-target gates below (see Sec.~\ref{sec:Gates} for truth tables).
\vspace{1em}
\noindent
{\bf \boldmath$\ U_{1c1t}^X$:}
%This is %effectively 
%a controlled $X$ gate. 
We arrange the atoms as shown in Fig ~\ref{fig:Other4} (a), then apply an $X$ 
%gate 
pulse to the target.

\vspace{1em}
\noindent
{\bf \boldmath$\ U_{1c1t}^H$:}
%This  is 
%effectively 
%a controlled $H$ gate. 
We arrange the atoms as shown in Fig.~\ref{fig:Other4} (b), then apply %the 
a $H$ gate pulse to the target.

\vspace{1em}
\noindent
{\bf 
\boldmath$U_{1c2t}$:}
%One places the control and target atoms all within each-others blockade radius.  The accessible target Hilbert space is spanned by two states $\psi_i=|00\rangle$ and $\psi_f=(|10\rangle+|01\rangle)/\sqrt{2}$.  If none of the control atoms are excited, one drives Rabi oscillations between the target states by turning on a pulse $\Omega(t)$.  The desired gate ($\psi_i\to\psi_f$) is produced if $\int \Omega(t)dt=\sqrt{2}\pi$.
%This  is  
%a controlled $\Psi$ gate. 
We arrange the atoms as shown in Fig.~\ref{fig:Other4} (a), then apply a $\Psi$ %gate 
pulse to the targets.

\vspace{1em}
\noindent
{\bf \boldmath$\ U_{2c2t}^X$:}
%One arranges the atoms so that $c_1$ blockades $t_1$ and $c_2$ blockades $t_2$, and drives a single-atom $\pi$ pulse on the target atoms, $\int(\Omega(t)/2)dt =\pi$.
%This can be implemented as two simultaneous controlled X gates.  
We arrange the atoms as in Fig.~\ref{fig:Other4} (a), so that $c_1$ blockades $t_1$ and $c_2$ blockades $t_2$.  We then simultaneously apply X gate pulses to each of the targets. 
%The arrangement in Fig ~\ref{fig:Other4} (a) also works, where additionally $t_1$ and $t_2$ blockade one-another can do the job. As mentioned in Sec ~\ref{sec:Gates}, t
This gate is never applied to a state where the control bits are set to $\ket{0_c0_c}$.  Thus the two target atoms are never simultaneously excited and it does not matter if the target atoms blockade one-another.

\begin{figure}[tbh]
\includegraphics[width=0.48\textwidth]{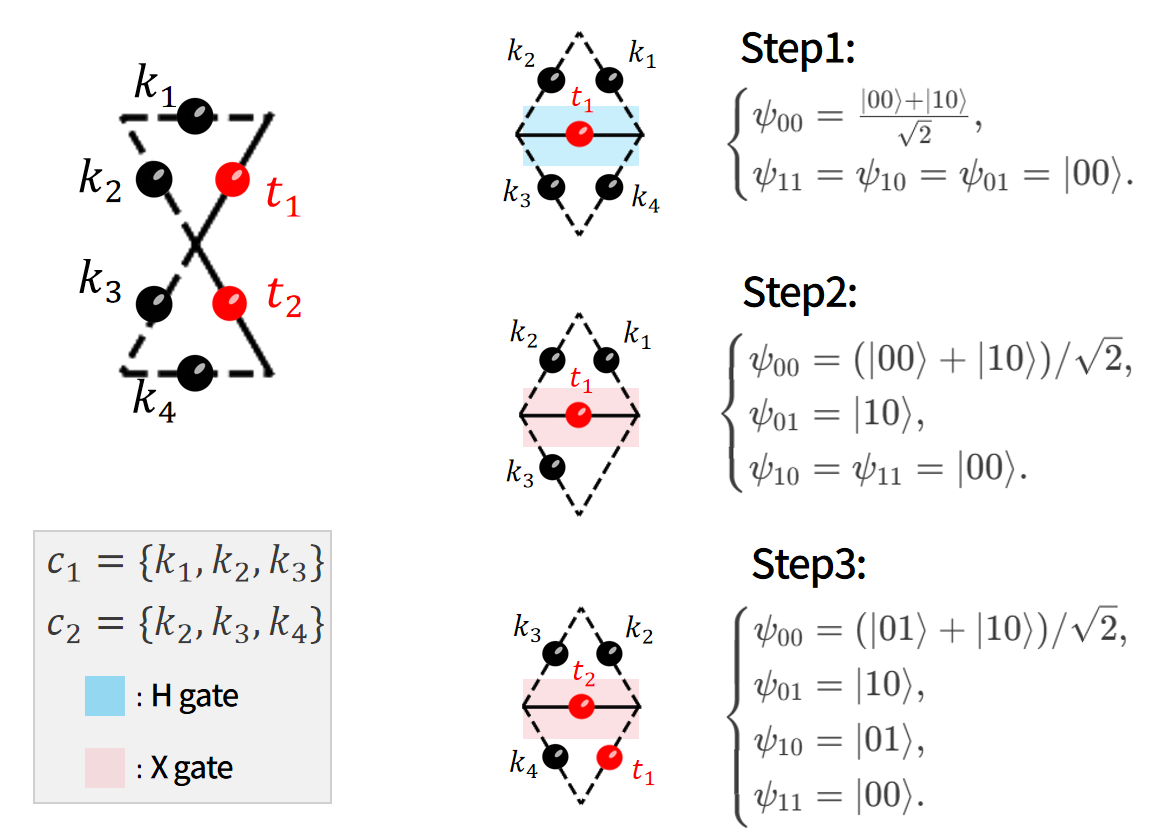}
    \caption{Realizing $U^H_{2c2t}$ with non-adiabatic gates. In the left panel, targets (red) are marked as $t_i$, while control atoms (black) are marked as $k_i$. Control bit $c_1$ is composed of $\{k_1,k_2,k_3\}$ and $c_2$ is composed of $\{k_2,k_3,k_4\}$. The right panel shows the spatial arrangement of atoms at each step, so that atoms on vertex sharing bonds provide blockade. During each step only the atom %located on the solid bond 
    within the shaded area experiences the $\Omega(t)$ and $\Delta(t)$ pulses which changes its state. The state, $\psi_{c_1 c_2}$ at each step is explicitly shown, where $c_j$ is the state of the $j$'th control bit.}
    \label{fig:U2c2tIM}
\end{figure}

\vspace{1em}
\noindent
{\bf 
\textbf{\boldmath$\ U_{2c2t}^H$:}}
This gate can be implemented through a three step process. Control atoms and targets are labeled as in Fig.~\ref{fig:U2c2tIM}, and the spatial arrangement at each step is shown there. For each control qubit state $\ket{c_1c_2}$, the target qubits evolve differently at each step. We denote the corresponding target state as $\psi_{c_1c_2} = \ket{t_1t_2}$.   The wavefunctions after each step of the process are also shown in the figure.
First, one places $t_1$ in the blockade radius of both $c_1$ and $c_2$, and drives an $H$ pulse on $t_1$. Next, one places $t_1$ in the blockade radius of  $c_1$ and applies a $X$ pulse. One finally places $t_2$ in the blockade radius of both $c_2$ and $t_1$, and applies a $X$ gate pulse on $t_2$. 

%Following the systematic

% The arrangement of atoms and the evolution of states at each step is shown in Fig ~\ref{fig:U2c2tIM}. After this process, we get the final state:
% \begin{equation}\label{eq:2C2T}
% \left\{
% \begin{aligned}
% \psi_{00} &= (|01\rangle + |10\rangle)/\sqrt{2}, \\
% \psi_{01} &= |10\rangle, \\
% \psi_{10} &= |01\rangle, \\
% \psi_{11} &= |00\rangle.
% \end{aligned}
% \right.
% \end{equation}
% , which exactly implements the $U_{2c2t}^H$ gate.

%First, one places $t_1$ in the blockade radius of both $c_1$ and $c_2$, and drives a $\pi/2$ pulse.  
%This puts the system into a configuration $\psi_{c_1c_2}=|t_1t_2\rangle$:
%$\psi_{11}=\psi_{10}=\psi_{01}=|00\rangle$ and $\psi_{00}=(|00\rangle+|10\rangle)/\sqrt{2}$.
%  One then places $t_1$ in the blockade radius of $c_1$ and drives a $\pi$ pulse, after which
%  $\psi_{00}=(|00\rangle+|10\rangle)/\sqrt{2}$, $\psi_{01}=|10\rangle$, $\psi_{10}=\psi_{11}=|00\rangle$.
%    Finally, $t_2$ is placed in the blockade radius of both $c_2$ and $t_1$, and a $\pi$ pulse is applied, yielding  $\psi_{00}=(|01\rangle+|10\rangle)/\sqrt{2}$, $\psi_{01}=|10\rangle$, $\psi_{10}=|01\rangle$, and $\psi_{11}=|00\rangle$, as desired.

\vspace{1em}
\noindent{
\textbf{\boldmath$U_{2c4t}$:}}
This gate can be implemented through a three step approach. Control atoms and targets are labeled as in Fig ~\ref{fig:U2c2tIM}, along with the spatial arrangement of atoms and the evolution of the target state as $\psi_{c_1c_2} = \ket{t_1t_2t_3t_4}$.
First, one places $t_1$ and $t_2$ in the blockade radius of $c_1$, and drives a $\Psi$ gate pulse on them. Next, one places $t_3$ and $t_4$ in the blockade radius of $c_1$ and $c_2$. Also, we let $t_1$ blockade $t_4$ and $t_2$ blockade $t_3$. An X  pulse is applied to $t_3$ and $t_4$. One finally puts $t_3$ and $t_4$ in the blockade radius of $t_1,t_2$ and $c_2$, and execute a $\Psi$  pulse on $t_3$ and $t_4$. 
%The arrangement of atoms and the evolution of states at each step is shown in Fig ~\ref{fig:U2c4tIM}. After this process, we get the final state:
% \begin{equation}\label{eq:2C4T}
% \left\{
% \begin{aligned}
% \psi_{00} &= (|1010\rangle + |0101\rangle)/\sqrt{2}, \\
% \psi_{01} &= (|1000\rangle + |0100\rangle)/\sqrt{2}, \\
% \psi_{10} &= (|0010\rangle + |0001\rangle)/\sqrt{2}, \\
% \psi_{11} &= |0000\rangle.
% \end{aligned}
% \right. 
% \end{equation}
% , which exactly implements the $U_{2c4t}$ gate.

%One first applies a $U_{1c2t}$ gate with $c_1$ as the control, and $t_1,t_2$ as the targets.  This puts the system into a configuration $\psi_{c_1c_2}=|t_1t_2t_3t_4\rangle$:  $\psi_{00}=\psi_{01}=(|1000\rangle+|0100\rangle)/\sqrt{2}$ and
%$\psi_{10}=\psi_{11}=|0000\rangle$.
%Next one arranges the atoms so that there is blockade between: $(t_1,t_4), (t_2,t_3), (c_1,c_2,t_3,t_4)$.  One then turns on $\Omega$ for $t_3$ and $t_4$.  Clearly the only state that changes is $\psi_{00}$.  It Rabi flops between $\psi_i=(|1000\rangle+|0100\rangle)/\sqrt{2}$ and
%$\psi_f=(|1010\rangle+|0101\rangle)/\sqrt{2}$.  Choosing $\int \Omega(t)/2 dt=\pi$ transfers the atoms into state $\psi_f$.
%Finally one applies a $U_{1c2t}$ gate where the control atoms are $c_2,t_1,t_2$.  This gate only effects $\psi_{10}$, transitioning it into the desired state, $(|0010\rangle+|0001\rangle)/2$.

\begin{figure}[tbh]
\includegraphics[width=0.48\textwidth]{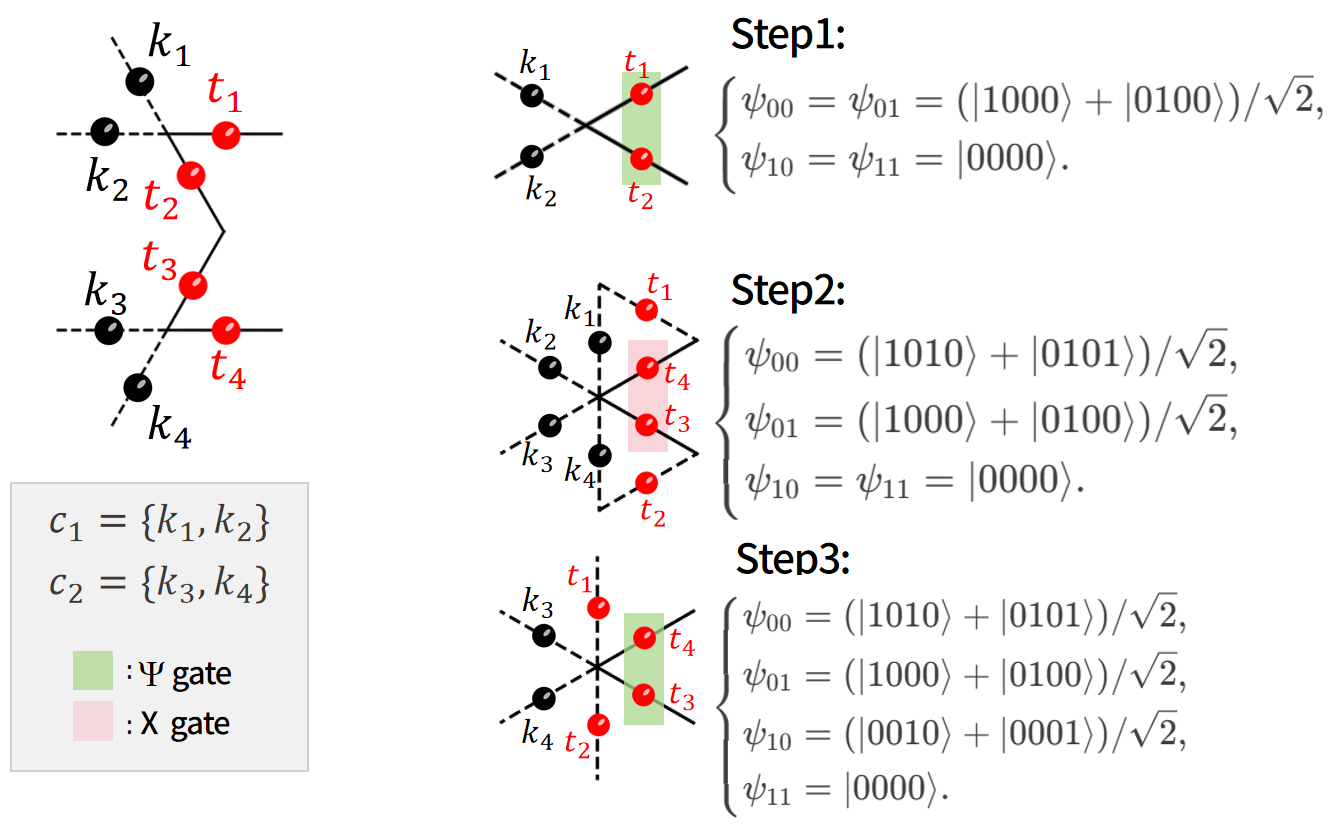}
    \caption{Realizing $U_{2c4t}$ with non-adiabatic gates. 
    Labeling conventions follow Fig.~\ref{fig:U2c2tIM}.
    % In the left panel, targets (red) are marked as $t_i$, while control atoms (black) are marked as $k_i$. Control bit $c_1$ is composed of $\{k_1,k_2\}$ and $c_2$ is composed of $\{k_3,k_4\}$. The right panel shows a feasible spatial arrangement of atoms at each step. Only the atom located on the solid bond within the shaded area is activated. The state, $\psi_{c_1 c_2}$, at each step is explicitly shown.
    }
    \label{fig:U2c4tIM}
\end{figure}

\subsection{Implementation with digital quantum circuits}\label{sec:gate-operation}

The gates in our protocol can also be implemented in digital quantum circuits, enabling the production of the Rokhsar-Kivelson state in other platforms, such as transmon arrays.  The gate sequences are simpler for the YC-$2N$ geometry, making it more suitable for implementing on a digital quantum computer. 

We describe each gate here, which are all low depth circuits built from a small numbers of controlled X and controlled Hadamard gates.  Following the convention in the rest of this paper, which is natural in the setting of Rydberg atoms, our controlled gates are zero-controlled (also known as  open-controlled),  
 meaning the operation is executed only when all control bits are in the $\ket{0}$ state. In our quantum circuit diagrams, we represent this type of control by an open circle. This differs from the usual convention, where gates are \emph{1-controlled}—i.e., activated when the control bits are in the $\ket{1}$ state—typically indicated by a solid dot.

\vspace{1em}  
\noindent
%{\bf 
%Single target controlled Hadamard, 
{\boldmath$ U_{1c1t}^H$}:
%}\\
%\noindent\textbf{\boldmath$\cdot\ U_{1c1t}^H$:} \\
\noindent
This gate is implemented by a controlled Hadamard gate,
\begin{center}
\begin{quantikz}
\lstick{$\ket{c}$} & \octrl{1} & \qw \\
\lstick{$\ket{0}$} & \gate{H} & \qw
\end{quantikz}
\end{center}
which is a standard gate in quantum computation (up to a possible relabeling of the logical states).  It is readily implemented with most hardware.
%As already explained, in Rydberg atoms one implements this gate by placing the ground state target atom within the blockade radius of the control atoms.  One then drives a $\pi/2$ pulse on the target atom, defined by $\Delta_\alpha(t)=0$ and

%If any of the control atoms are excited

%Alternatively, it can be implemented through an adiabatic quantum gate. First, the Rabi frequency \(\Omega_\alpha\) is turned on, and then \(\Delta_\alpha\) is slowly tuned from a large negative value to zero. Finally, \(\Omega_\alpha\) is turned off. This adiabatic process ensures the system remains in the instantaneous eigenstate, transitioning to the desired operation in a controlled manner.

\vspace{1em}  
\noindent
%{\bf Single target controlled not, \boldmath
{\boldmath$U_{1c1t}^X$}:
%}\\
%\vspace{1em}
%\noindent\textbf{\boldmath$\cdot\ U_{1c1t}^X$:} \\
%\noindent
This gate is implemented by a controlled X (CNOT) gate,
\begin{center}
\begin{quantikz}
\lstick{$\ket{c}$} & \octrl{1} & \qw \\
\lstick{$\ket{0}$} & \targ{} & \qw
\end{quantikz}
\end{center}
%Alternatively, it can be implemented adiabatically. First, the Rabi frequency \(\Omega_\alpha\) is turned on, and then \(\Delta_\alpha\) is slowly tuned from a large negative value to large positive value. Finally, \(\Omega_\alpha\) is turned off, and $\Delta_\alpha$ is also turned off to avoid unwanted phase factors. 
which is also a standard gate.

\vspace{1em}  
\noindent
%{\bf Two targets controlled $\Psi$ gate,
{\boldmath$U_{1c2t}$}:
%}\\
This operation can be implemented as a controlled-Hadamard gate followed by a Toffoli gate:
\begin{center}
\begin{quantikz}
\lstick{$\ket{c}$} & \octrl{1} & \octrl{2} & \qw \\
\lstick{$\ket{0}$} & \gate{H}\gategroup[2,steps=2,style={dashed,rounded
corners,fill=blue!10, inner
xsep=2pt},background,label style={label
position=below,anchor=north,yshift=-0.2cm}]{{$\Psi$
gate}}  & \octrl{1} & \qw \\
\lstick{$\ket{0}$} & \qw       & \targ{}  & \qw
\end{quantikz}
\end{center}
%The alternative adiabatic gate can be designed as follows. First, the Rabi frequency \(\Omega_\alpha\) is turned on, and then \(\Delta_\alpha\) is slowly tuned from a large negative value to large positive value. Finally, \(\Omega_\alpha\) is turned off, and $\Delta_\alpha$ is also turned off to avoid unwanted phase factors. 
which is effectively a controlled $\Psi$ gate, as marked in the blue shaded area.

\vspace{1em}  
\noindent
%{\bf Two targets controlled  not,
{\boldmath$\ U_{2c2t}^X$}:
%}\\
%\vspace{1em}
%\noindent\textbf{\boldmath$\cdot\ U_{2c2t}^X$:} \\
%\noindent
%
This minimal implementation of this gate is simply two CNOT gates, between one of the control bits and one of the target bits:
\begin{center}
\begin{quantikz}
\lstick{$\ket{c_1}$} & \octrl{2} & \qw & \qw \\
\lstick{$\ket{c_2}$} & \qw & \octrl{2}  & \qw \\
\lstick{$\ket{0}$} & \targ{}  & \qw  & \qw \\
\lstick{$\ket{0}$} & \qw   & \targ{}   & \qw
\end{quantikz}
\end{center}

\vspace{1em}
\noindent
%{\bf Two target two control gate with target blockade,
{\boldmath$\ U_{2c2t}^H$:}
%}\\
Here the two targets blockade one-another.  One implementation is with 
a double controlled-Hadamard gate, a CNOT gate, and a Toffoli gate:
\begin{center}
\begin{quantikz}
\lstick{$\ket{c_1}$} & \octrl{1} & \octrl{2} & \qw & \qw \\
\lstick{$\ket{c_2}$} & \octrl{1} & \qw       & \octrl{2} & \qw \\
\lstick{$\ket{0}$} & \gate{H}  & \targ{}  & \octrl{1}       & \qw \\
\lstick{$\ket{0}$} & \qw       & \qw       & \targ{}  & \qw
\end{quantikz}
\end{center}
%The alternative adiabatic gate can be designed as follows. First, the Rabi frequency \(\Omega_\alpha\) is turned on, and then \(\Delta_\alpha\) is slowly tuned from a large negative value to large positive value. Finally, \(\Omega_\alpha\) is turned off, and $\Delta_\alpha$ is also turned off to avoid unwanted phase factors. 
which corresponds to the steps shown in Fig.~\ref{fig:U2c2tIM}.

\vspace{1em}
\noindent
%{
%\bf Two control four target, 
{\boldmath$U_{2c4t}$}:
%}}\\
A depth six circuit can implement the desired gate,
\begin{center}
\begin{quantikz}
\lstick{$\ket{c_1}$} & \octrl{2} & \octrl{2}\slice{} & \octrl{1} & \octrl{1}\slice{} &\qw &\qw &\qw \\
\lstick{$\ket{c_2}$} & \qw & \qw & \octrl{2} & \octrl{1} &\octrl{1} &\octrl{1} &\qw \\
\lstick{$\ket{0}$} & \gate{H} & \octrl{1} & \qw & \octrl{3} &\octrl{1} &\octrl{1} &\qw \\
\lstick{$\ket{0}$} & \qw & \targ{} & \octrl{1} & \qw &\octrl{1} &\octrl{1} &\qw \\
\lstick{$\ket{0}$} & \qw & \qw & \targ{} & \qw &\gate{H} &\octrl{1} &\qw \\
\lstick{$\ket{0}$} & \qw & \qw & \qw & \targ{} &\qw &\targ{} &\qw 
\end{quantikz}
\end{center}
which has been divided into three sections by red dashed lines.  These correspond to the three steps in Fig.~\ref{fig:U2c4tIM}.

{

{
\section{Gate Operations for Seed}
\label{seedimp}
In Appendix~\ref{sec:Realization} we gave realizations of the gates used to grow the Rokhsar-Kivelson state on a cylinder.  Here we briefly explain how to implement the gates introduced in Sec.~\ref{sec:seed} to seed the dimer configurations.

The gate defined by $U^\Psi|00\rangle=(|01\rangle+|01\rangle)/\sqrt{2}$ is the $\Psi$ gate introduced in Appendix~\ref{sec:Coherent}.  As described there, it is implemented by placing the two atoms within their blockade radius, and turning on a driving pulse $\Omega(t)$ with { $\int \sqrt{2}{\Omega(t)} \, dt =  \pi$}, followed by a corrective $\Delta$ pulse, $\int \Delta(t) \, dt = \frac{\pi}{2}$.  The gate $U^{\sqrt{\Psi}}|00\rangle=|00\rangle/\sqrt{2}+(|01\rangle+|01\rangle)/2$ is similar, excepts one takes {$\int \sqrt{2}{\Omega(t)} \, dt = { \pi/2}$}.  It is natural to call this a $\sqrt{\Psi}$ gate, since applying it twice gives the $\Psi$ gate.

The $U^{X\Psi}$ gate involves one ancilla atom and two dimer atoms.  One first applies a $\sqrt{\Psi}$ gate to the dimer atoms.  One then applies a $U_{1c1t}$ (control-$X$) gate where the dimer atoms form the control bit, and the ancilla is the target.
}

{
\section{Gluing cylinders together}
\label{sec:Torus}
Here we give a protocol to `glue' two dimer coverings together.  We explain how this procedure allows us to make superpositions of dimer coverings on a torus.  We then show how it can be used in an alternative approach to state preparation.  It allows us to produce the Rokhsar-Kivelson state on a cylinder (or torus) in a time which scales as circumference, rather than the length.  We give our argument for a YC-$2N$ cylinder, but a similar argument works for a XC-$2N$ geometry.

Consider the situation illustrated in Fig.~\ref{fig:Torus}, where one has three annular strips, labeled $m-1$, $m$, and $m+1$.  The strips $m\pm1$ strips are in superpositions of dimer coverings, of the same parity, while all the atoms in the $m$ strip are in their ground state.  Our procedure entangles the central strip with the others in such a way that it contains a superposition of all dimer coverings consistent with the surrounding strips.

\begin{figure}[tbh]
\includegraphics[width=0.4\textwidth]{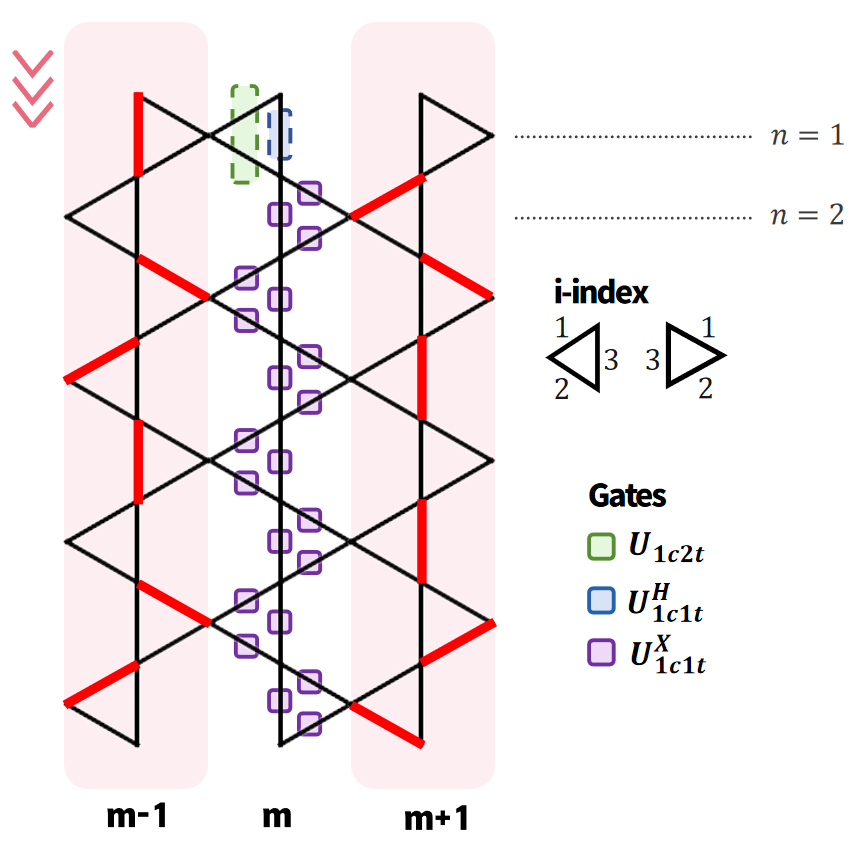}
    \caption{
    Gate sequence to glue together two previously prepared superpositions of dimer coverings, shaded in red.  The dimer superposition on the left and right should have the same parity.  The atoms in the central region, marked $m$, begin in their ground state.  One sequentially applies gates, sweeping from the top  to the bottom. 
    % as the   
    % The bonds on the left and right, shaded in red, are in a superposition of dimer c 
    % for preparing the toric RK state on a YC-8 cylinder. Qubits are labeled by integers $(m,n,i)$, and different gates are sequentially acted from top to bottom, following the convention in Fig ~\ref{fig:YC_growing}. The red shaded region stands for the prepared region, i.e. the two ends of a cylinder.
    }
    \label{fig:Torus}
\end{figure}

As in our growth algorithm, we label the atoms in the central strip by integers $(m,n,i)$. 
It is also useful to imagine one particular pattern in the pre-existing superposition of dimer patterns.  We wish to produce a superposition of dimer patterns on the central bonds where $L=(l_1,l_2,\cdots l_N)$ and $R=(r_1,r_2,\cdots r_N)$ are consistent with the neighboring strips.  The figure shows one example, where $L=(1,0,1,0)$ and $R=(0,1,1,0)$.  
%Our protocol is illustrated in Fig.\ref{fig:Torus}. As in Sec.\ref{sec:YC}, we use the array $(m, n, i)$ to label atom positions.
As suggested by the green and blue rectangles  in the figure, we first apply the gate $U_{1t2c}$ to bonds at $(m,1,1)$ and $(m,1,2)$,
followed by $U_{1t2c}^H$ on $(m,1,3)$.
If $l_1=0$, this results in coherent superposition of each of the two atoms in the green box being excited.  Conversely, if $l_1=1$ the atom in the blue box is placed in a superposition of $|0\rangle$ and $|1\rangle$, while the green atoms remain in their ground state.  In each case we have 
generate a coherent superposition which will be seeds for the $u = 0$ and $u = 1$ patterns. Subsequently, we sequentially apply controlled-X gates to the remaining atoms and grow the state downward, triangle by triangle. Within each triangle, the gates are applied, top to bottom, in the sequence $i = 1 \rightarrow 3 \rightarrow 2$. For each step, the control bits are composed of the atoms spatially adjacent to the target atoms, as shown in Fig.~\ref{fig:Torus}.

As a first application, we present a protocol for preparing a uniform superposition of all dimer converings on a torus.  We begin by producing a YC-$2N$ cylinder of length $L_x-1$, consisting of either a uniform superposition of all configurations in one topological sector, or an arbitrary superposition of the two possible sectors. 
The final torus will be formed from
$L_x$ total number of annular strips, which must be even.
%To produce a defect free dimer covering, we require either $N$ or $L_x$ to be even, or both.  
%That way each dimer configurations will have the same parity on strip 1 as it will on strip $L_x-1$. 
We then glue the two ends together, as illustrated in Fig.~\ref{fig:Twoways} (a).
\begin{figure}[tbh]
\includegraphics[width=0.48\textwidth]{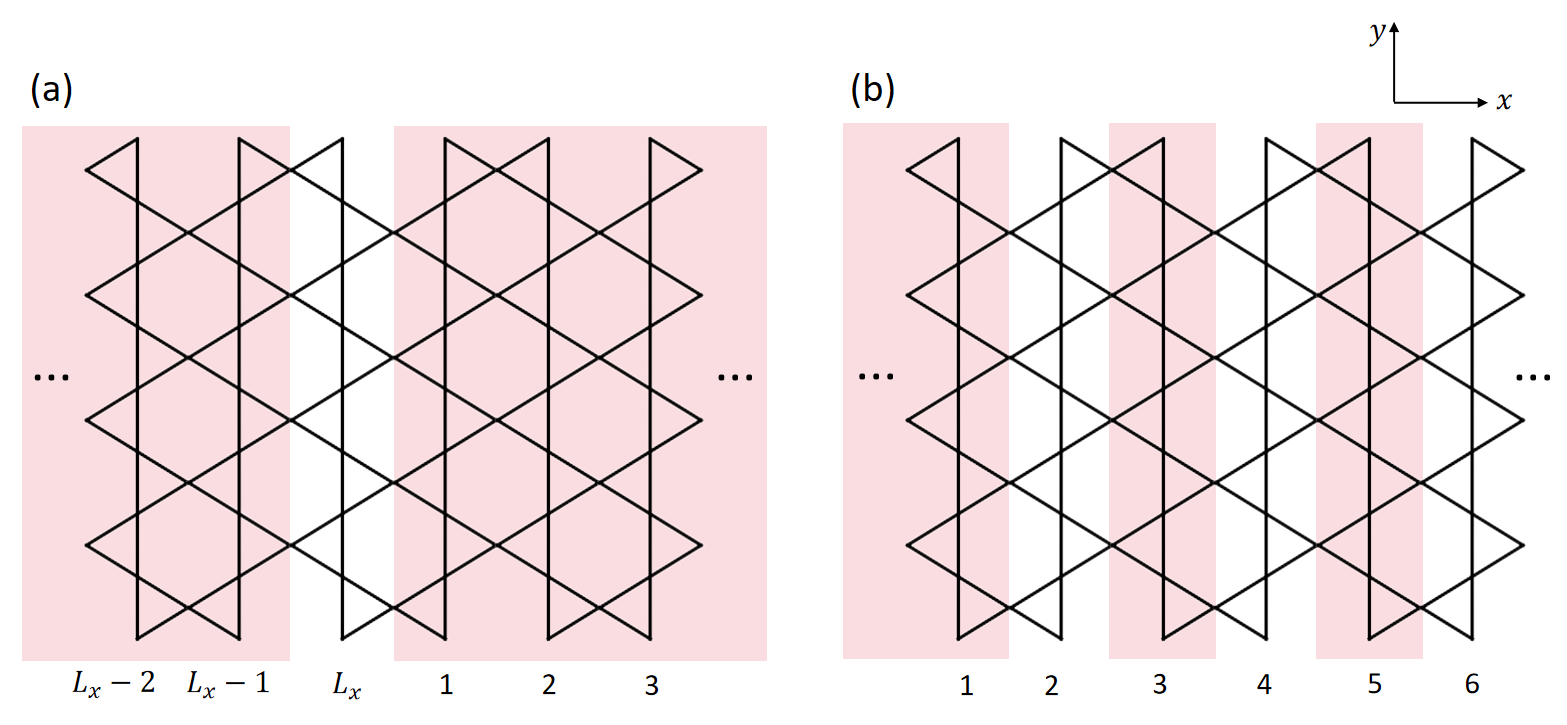}
    \caption{
    (a) Gluing two ends of a cylinder to make a torus. The red shaded area corresponds to a cylinder formed by the procedure in Sec.~\ref{sec:StateCreation}.  Atoms in the white region  begin in their ground state.  (b)
    One can also produce a superposition of dimer coverings by gluing together a collection of independently prepared annular strips.  The resulting topology can be a cylinder or a torus.
    %Making half of the annular strips, and then glue them up. The red shaded area is in superpositions of dimer coverings, and the blank area is to be glued up.
    }
    \label{fig:Twoways}
\end{figure}

%{\color{blue}The following might need modification}
For a torus there are two distinct classes of non-contractable loops, that either wrap around in the $x$-direction (horizontal in Fig.~\ref{fig:Twoways}) or the $y$-direction (vertical in Fig.~\ref{fig:Twoways}).
There are four topologically distinct dimer coverings, corresponding to the expectation values of those non-contractable $Z$-strings, that can be labeled $|m_x,m_y\rangle$, with $m_x,m_y=0,1$.  These are isomorphic to the logical states of the Kitaev toric code \cite{kitaev2003fault}.  The quantum number $m_y$, corresponding to eigenvalue of the $Z$-loops in the $y$ direction on alternate strips, is determined by the state of the cylinder before the ends are glued together.  One can readily produce states with $m_y=0,1$, or a superposition.
%As discussed in Appendix~\ref{seed}, one can readily produce either topological sector,  by changing the states of the ancilla used as a seed.  
The other quantum number, $m_x$, can be understood as the parity of the sum of the $u$ index of every ring. Our approach yields a uniform superposition of these indices, and hence a superposition $|m_x=0\rangle+|m_x=1\rangle$.  This corresponds to an eigenstate of a $X$-loop in the $y$ direction.  One can use $X$-loop and $Z$-loop operators to manipulate these states.

As a second application of the ability to glue strips together, we present an alternative approach for preparing the Rokhsar-Kivelson state on a cylinder or torus, of size \( L_x \times L_y \).  
%within a specified topological sector. We label the \( L_x \times L_y \). 
The procedure begins by preparing \( L_x/2 \) annular strips, each initialized in the desired topological sector using the method introduced in Sec.~\ref{sec:seed}.
One then arranges them in an alternating pattern, as illustrated by the red-shaded regions in Fig.~\ref{fig:Twoways}(b). These strips are then glued together using the procedure described in Fig.~\ref{fig:Torus}. This growth procedure is dual to the one described in the main text.  There the growth takes a time proportional to $L_x$, and independent of $L_y$.  This alternative scheme instead takes a time which  scales linearly with \( L_y \), and is independent of \( L_x \). 
%A limitation of this method, however, is that it can only produce states within a fixed topological sector and does not allow the preparation of an equal-weight superposition of multiple sectors.

%Similar to the well-known Kitaev toric code~\cite{kitaev2003fault}, the RK state on the torus, as a \(\mathbb{Z}_2\) spin liquid, encodes two logical qubits. In our preparation scheme, the logical qubit associated with the cylinder's longitudinal direction is initialized in the superposition state \( (\ket{0} + \ket{1})/\sqrt{2} \), while the logical qubit associated with the circumferential direction can be set to \( \ket{0} \), \( \ket{1} \), or their equal-weight superposition, depending on the choice of the initial boundary condition/topological sector. One could use X-loop operators to manipulate those logical states.

}

%\newpage

%\bibliographystyle{apsrev4-2}
%\bibliography{bib}


\end{document}